\pdfoutput=1
\documentclass[letterpaper,12pt]{article}
\usepackage{setspace}
\usepackage{arxivstyle}
\usepackage{physics}
\usepackage{bbold}
\usepackage{tikz}
\usetikzlibrary{arrows.meta}

\usepackage{subcaption}

\DeclareCaptionLabelFormat{panelnumber}{\thefigure(#2)}
\newcommand{\R}{\mathbb{R}}
\newcommand{\Z}{\mathbb{Z}}
\newcommand{\id}{\mathbb{1}}
\newcommand{\AdS}{\text{AdS}}
\newcommand{\vare}{\varepsilon}
\newcommand{\SL}{\text{SL}}
\newcommand{\U}{\text{U}}
\newcommand{\mO}{\mathcal{O}}
\newcommand{\tpiint}{\int_0^{2\pi}}

\newcommand{\pl}{\partial}

\def\longeq#1{
\begin{equation}
    \begin{aligned}
    #1
    \end{aligned}
\end{equation}
}

\def\lceq#1{
\begin{equation}
\begin{gathered}
    #1
\end{gathered}
\end{equation}
}

\def\mx#1{
\left(
\begin{matrix}
#1
\end{matrix}
\right)
}

\newcommand{\BH}{\mathrm{BH}}

\newcommand{\phys}{\text{phys}}

\newcommand{\mL}{\mathcal{L}}

\newcommand{\figTopoCylinderH}{%
  \begin{tikzpicture}
    \node[inner sep=0,outer sep=0] (cylinder)
      {\includegraphics[width=4.5cm]{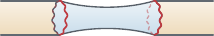}};
    \draw[black,line width=0.69pt,-{Stealth[length=3.5pt,width=3.0pt]}]
      ([xshift=-5.0mm,yshift=-1.125mm]cylinder.center)
      -- ([xshift=5.0mm,yshift=-1.125mm]cylinder.center);
    \node[inner sep=0,outer sep=0,font=\footnotesize,text=black]
      at ([yshift=0.625mm]cylinder.center) {$h$};
  \end{tikzpicture}%
}

\usepackage{placeins}

\def\la{\label}
\def\nref#1{(\ref{#1})}
\def\half{{1\over2}}
\newcommand{\be}{\begin{equation}}
\newcommand{\ee}{\end{equation}}
\newcommand{\bea}{\begin{eqnarray}}
\newcommand{\eea}{\end{eqnarray}}

\title{The low energy limit of black holes coupled to a conformal field theory}

\author[a,b]{Ahmed Almheiri}
\author[c]{Bingxin Lao}
\author[d]{Juan Maldacena}

\affiliation[a]{New York University Abu Dhabi, Abu Dhabi, UAE }
\affiliation[b]{Simons Center for Geometry and Physics, Stony Brook University, Stony Brook, NY, USA }
\affiliation[c]{Jadwin Hall, Princeton University, Princeton, NJ, USA}
\affiliation[d]{Institute for Advanced Study, Princeton, NJ, USA}

\emailAdd{almheiri@nyu.edu}
\emailAdd{bingxin.lao@princeton.edu}
\emailAdd{malda@ias.edu}

\abstract{We consider the low energy limit of near extremal black holes interacting with a CFT.  We conjecture that the low energy limit can be described in terms of a boundary CFT, or BCFT, whose Affleck--Ludwig boundary entropy is set by the black hole entropy $S_0$.  We present some evidence for this picture by considering a toy model of JT gravity coupled to a CFT and performing several explicit computations at the lowest orders in the topological expansion.  We also explain how this toy model applies to the description of near extremal magnetic black holes in the Standard Model, in a regime where they could lead to a tabletop test of quantum gravity.
}

\date{}

\begin{document}
\maketitle
\section{Introduction}
\label{sec:introduction}
In quantum gravity, we often expect that a black hole can ultimately be described by a quantum system in Hilbert space whose dimensionality is set by the black hole entropy.   This presupposes that we can isolate the black hole from the external system, an assumption that is approximately true in most cases. This separation works better for near-extremal black holes; at very low temperatures, or very close to extremality, the coupling to external fields is very suppressed \cite{Brown:2024evaporation}.

In this paper, we study a situation where the coupling to external fields {\it cannot} be ignored, even at very low energies. This situation arises when we have degrees of freedom described effectively by a two dimensional conformal field theory propagating along the radial and time directions \cite{Almheiri:2014cka, Almheiri:2019qdq, Engelsoy:2016xyb, Almheiri:2019psf, Goto:2020wnk}. The two dimensional CFT can be very simple, such as free fermions.

We propose that the low energy dynamics can then be described as a conformal invariant boundary condition in the CFT, i.e., the boundary conformal field theory (BCFT). Schematically, we claim
\be
 {\rm B + CFT = BCFT ~.
 }
 \ee
The black hole degrees of freedom (B)  are incorporated into the boundary condition seen by the
exterior conformal field theory (CFT). Our aim is to identify the part of this description that
follows from the low-energy gravitational path integral, and to understand what further input
is needed to obtain a consistent boundary CFT (BCFT).

An analogy is an impurity with some localized degrees of freedom interacting with a gapless medium. For example, in the Kondo model, the spin degrees of freedom of the impurity interact with the s-wave degrees of freedom of a Fermi liquid, and the whole system flows in the IR to a BCFT \cite{Affleck:1990by,Affleck:1995ge}.

An explicit four-dimensional black hole where these ideas are relevant is a magnetically charged black hole in the Standard Model. In that case, the electrically charged fermions at the lowest Landau level on the sphere provide a large number of fields behaving as two dimensional massless fields \cite{Maldacena:2020skw, Maldacena:2018gjk}\footnote{In the regime we consider, we can ignore some very small masses induced by interaction terms \cite{MiranskyShovkovy2015}.}.
\begin{figure}[t]
  \centering

  \begin{subfigure}[b]{0.48\linewidth}
    \centering
    \includegraphics{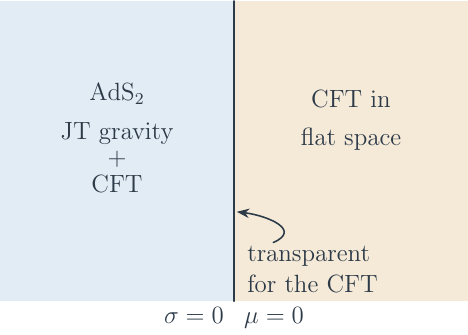}
    \caption{Lorentzian spacetime}
    \label{fig:setup-lorentzian}
  \end{subfigure}\hfill
  \begin{subfigure}[b]{0.48\linewidth}
    \centering
    \includegraphics{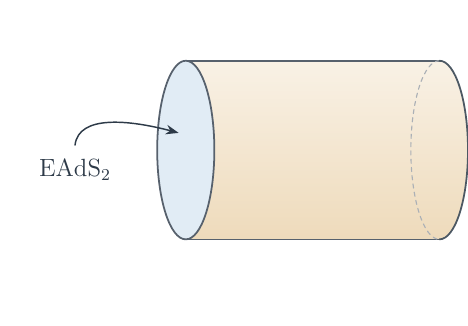}
    \caption{Euclidean spacetime}
    \label{fig:setup-euclidean}
  \end{subfigure}

  \caption{ (a) Lorentzian description of JT gravity
  coupled to a flat CFT bath. The light blue region is the gravitating region; the light yellow region on the right is flat space. The CFT lives in both regions and has transparent boundary conditions at the interface. (b) The same setup at finite temperature and Euclidean time.}
  \label{fig:lorentzian-setup}
\end{figure}

\paragraph{The setup.} We consider a toy model consisting of JT gravity minimally coupled to a matter CFT. At the $\AdS_2$ boundary, we connect the gravitating region to a flat space region without gravity where the same  CFT lives. At the interface, we have transparent boundary conditions for the CFT. See figure \ref{fig:lorentzian-setup}. This model has been studied extensively \cite{Almheiri:2014cka,Engelsoy:2016xyb,Almheiri:2019psf,Almheiri:2019qdq}. This is not an arbitrary toy model; JT gravity is a good approximation for the physics of near extremal black holes in higher dimensions, and the CFT can be present for some of them.

\paragraph{The boundary reparametrization mode and its low energy dynamics.}
In the gravitational region, the metric is locally a rigid $\AdS_2$ space, but there is an important degree of freedom: a reparametrization that relates the $\AdS_2$ time coordinate to the bath time coordinate \cite{Maldacena:2016upp,Engelsoy:2016xyb,Jensen:2016pah}. In our situation, the action for this mode consists of two terms. One is the familiar Schwarzian term coming from the JT gravity action \cite{Maldacena:2016upp,Engelsoy:2016xyb,Jensen:2016pah} and the other is a term induced by the matter CFT. For the dominant black hole geometry, which in Euclidean space is a disk, the induced action is given just in terms of the conformal anomaly of the CFT \cite{Polyakov:1981rd}, which is proportional to the central charge $c$, where $c=c_L = c_R$ in order to have no gravitational anomaly. At low energies, this term dominates, and its coupling is weak when $c\gg 1$. This is an example of ``induced gravity,'' in which the dynamics of gravity is induced by the matter theory. It is important that, in contrast to the Schwarzian action, this induced gravity action for an infinite bath is scale invariant and, for large $c$, it can remain weakly coupled at low energies.

For large $c$ and small fluctuations around the trivial reparametrization,  $\theta = \varphi + \epsilon(\varphi)$,  the action for quadratic fluctuations takes the form
\be \la{ActionINtro}
I_2 = \sum_{n=2}^\infty n (n^2-1) \left[ { 2 \pi \phi_{r,{\rm phys} } \over \beta } n + { c \over 12 }  \right] |\epsilon_n|^2
\ee
where $\phi_{r, {\rm phys} } $ is the Schwarzian coupling, and $\epsilon_n$ are the Fourier modes of $\epsilon(\varphi)$.   We see that at low energies, or as $\beta/n   \gg \phi_{r, {\rm phys}}/c $, we can drop the Schwarzian term, yielding a temperature independent action.

\paragraph{The black hole as an energy mirror.}
If we send some matter into the black hole and increase its temperature relative to the bath, then it will Hawking radiate the degrees of freedom of the CFT. This Hawking radiation involves a two dimensional version of the black body formula,  $\dot E \propto c (T^2- T^2_{\rm bath} ) $, where $T$ is the temperature.  The energy of the black hole coming from the Schwarzian action is $ E \propto \phi_{r,\rm{phys}} T^2 $, where $ \phi_{r,\rm phys}$ is the coefficient of the Schwarzian term in the action and has dimensions of time, defined in~\eqref{eq:sec2_phi_phys}. This implies that the energy decreases exponentially back to its equilibrium value with a characteristic time
\lceq{
\label{DefTc}
\tau_c \propto \frac{\phi_{r, \text{phys}}}{c} \, .
}
This implies that we return to equilibrium after a time $\tau_c$. In the low energy limit, we are considering times $t \gg \tau_c$, and therefore the Hawking radiation can be viewed as instantaneous. At low energies, the energy that goes into the black hole immediately comes out since the time $\tau_c$ is approximated by zero.
\nref{DefTc} is the inverse temperature scale where we transition  from the Schwarzian-dominated regime to the CFT-dominated regime in the dynamics of the reparametrization mode \nref{ActionINtro}.

This means that, at low energies, we have the following  condition for the stress tensor at the boundary
\be
\la{TTBAR} T = \bar T
\ee
This physically states that no energy can flow through the interface between the bath and the gravity region, and that the energy gets reflected.

A direct way to derive~\nref{TTBAR} is to use the equations of motion for the reparametrizations. A small variation of the CFT induced action leads to
\be
\delta I_{\text{CFT}} \propto  \int dt { \delta \theta \over \dot \theta } T_{tn} ~,~~~~~~~T_{t n} \propto T - \bar T ~,
\ee
as we show in detail in section \ref{sec:disk_stress_tensor}. Here $n$ is the direction normal to the boundary and $\theta$ is a standard $\AdS_2$ time coordinate. Of course, \nref{TTBAR} is one of the basic conditions for a conformal boundary condition in a CFT.

\paragraph{The thermal ensemble in the disk approximation.}
At a very low but nonzero temperature, the thermal ensemble is described by a cylinder in the bath region attached to a gravitational disk, at the leading order in the topological expansion, see figure \ref{fig:lorentzian-setup}(b).

At finite temperature, we can view the thermal circle as space and the original spatial direction as Euclidean time. This is the so called ``closed string channel''. Then the entire gravitational path integral can be viewed as preparing a state in the closed-string channel. Due to \nref{TTBAR}, this state obeys
\be
(L_n - \bar L_{-n} ) | \text{BH} \rangle =0 ~,~~~~~n\in  \mathbb{Z}
\ee
States that obey this condition can be expanded in terms of so called Ishibashi states \cite{Ishibashi:1988kg,Cardy:1989ir}. Each Ishibashi state is a particular linear combination of a Virasoro primary with $h= \bar h$ and its descendants. We denote it by $|h \rangle \rangle$.

The Euclidean disk partition function produces just the vacuum family. So the disk produces the state
\be \la{DiskIshi}
| \text{BH}\rangle_{\text{disk}} = e^{ \tilde S_0 } |0 \rangle \rangle
\ee
where $|0 \rangle \rangle $ is the identity Ishibashi state and $\tilde S_0$ includes all constants that multiply the state, and its main contribution is from the coefficient $S_0$ of the topological term in the gravity action.    We explicitly check this statement by performing several perturbative computations to leading order in the $1/c$ expansion by writing the reparametrizations as $\theta(\varphi) = \varphi + \varepsilon(\varphi) $, expanding the induced action to quadratic order, and then performing the path integral.

  One very explicit way to analyze the system is to transform the CFT problem on a general reparametrized disk, with $\theta(\varphi)$, into a flat disk via two separate holomorphic changes of coordinates, one for the bath and one for the gravitating disk. This is the so called ``conformal welding problem'' \cite{SharonMumford2006}.  See figure~\ref{fig:disk}. Once we have these maps, we can easily compute the action using the conformal anomaly.  Using this method, we can also perform a few other related computations, such as correlation functions.  The same mathematical problem has been studied for other reasons in \cite{TakhtajanTeo2003,TakhtajanTeo2004,Wang2019LoewnerEnergy, AlekseevShatashviliTakhtajan2024}.

An alternative way to think about it is to realize that, in the closed string channel, we can view the problem with a general reparametrization, $\theta(\varphi) $, as follows. We start with the CFT with a trivial reparametrization $\theta = \varphi$, and then we insert an operator $U[\theta] $, written in terms of the stress tensor and the reparametrization $\theta(\varphi)$ that transforms the CFT state coming from the left side of the boundary into another on the right side of the boundary.  This operator has an explicit form displayed in \nref{UOperator}.  Then the induced action from the CFT is expressed in terms of its vacuum expectation value, $\langle 0 | U[\theta] | 0 \rangle$. Curiously, such expectation values were discussed in  \cite{Fewster:2018} as a tool to investigate probability distributions for the energy density in CFTs.

\paragraph{Correlation functions in the bath region and the OTOC.}
Of course, if we are willing to accept the fact that we obtain the vacuum Ishibashi state \nref{DiskIshi}, then we can also perform some computations at any central charge. For example, consider a two point function of two operators inserted at general points on the bath. As we mentioned, the induced CFT action is scale invariant. In fact, it is invariant under the $\SL(2,\R)$ symmetries of the infinite bath that preserve the interface. For this reason,  the two point function depends non-trivially on only one conformal cross ratio. We can compute the two point function as follows. Imagine reflecting the left moving parts of the operator so that now we have a problem involving only right movers. The correlator is just given by the conformal block of the identity in the OPE channel where the two original right-moving parts of the two operators are fused to the identity.  We have checked this by an explicit computation at large $c$, though we expect the answer to hold for all $c$.   The large $c$ computation also allows us to compute the out of time ordered correlator (OTOC). Interestingly, it continues to exhibit maximal chaotic behavior, growing as $c^{-1} e^{ 2 \pi t/\beta }$, in the regime before it becomes of order one. This is despite the induced action being non-local and not given by Einstein gravity. Also, by considering the OPE in the boundary channel, we can extract some boundary operators. Their dimensions determine the long-time behavior, and we can read off quasinormal frequencies that exhibit gravitational corrections, or corrections in a $1/c$ expansion.

\paragraph{Short bath and the induced Schwarzian.}
Another interesting computation involves considering a finite bath of length $L$ with another known local boundary state on the other side. In this case, we already obtain the expected identity character at quadratic order. This is the answer we expect on the basis of \nref{DiskIshi}. This means that all higher orders in $1/c$ can only change this result by an overall constant.  This suggests an interesting non-renormalization theorem for this particular quantity.  For a short bath $L/\beta \ll 1 $, we again induce a new Schwarzian action with a coefficient proportional to  $c L$.

\paragraph{Irrelevant deformations in the gravity region produce other Ishibashi states.} It is natural to wonder whether a small deformation of this problem can lead to other Ishibashi states. Indeed, this is the case. If we deform the CFT in the gravitational region by an operator of conformal dimensions $h = \bar h$, then we get the $h$ Ishibashi state in the closed string channel. Namely, we get an additional contribution of the form $e^{\tilde S_0 } \lambda |h \rangle \rangle $ on the right hand side of \nref{DiskIshi}, where $\lambda$ is related to the small coefficient of the operator insertion.  Note that we do not deform the CFT in the bath region, so we continue to expect a conformal invariant BCFT in this case as well. In fact, in a complete theory of quantum gravity it is natural to think that our toy model will arise only as a low energy approximation.  Therefore, assuming that the relevant or marginal deformations are not present, it is natural to expect some irrelevant deformations of the CFT parametrized by the insertions of operators $\mathcal O_h$ in the gravitating region, with $h>1$. In the bath region, these can be made small by taking a low energy limit, or imagining that they become effectively zero in the same way that the gravitational coupling becomes zero.  But a small irrelevant deformation in the gravitational region continues to have an effect in the low energy description outside.

\begin{figure}[!htbp]
 \centering
\includegraphics[width=0.5\linewidth]{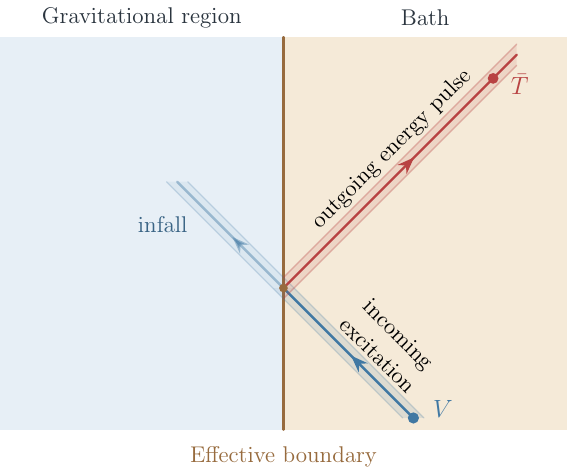}
 \caption{An excitation coming from the right leads to an energy pulse that is denoted in red. Other quantum numbers of the excitation continue through the interface unhindered, in the disk or usual black hole topology. }
 \label{fig:ReflectionIntro}
\end{figure}

\paragraph{Information loss at the disk level.}
In summary, at this order in the topological expansion, the black hole is behaving as an energy mirror. For example, vacuum correlators of insertions of the left and right moving parts of the stress tensor can be computed by applying a reflection trick, by viewing all the right moving insertions as left moving ones at the reflected points on the other side of the boundary.   It is important to note that other information about the excitations is free to go through the boundary. For example, if the CFT is given by a large number of free fermions, then the information of the fermion species can go through the interface in this approximation. It is ``leached'' of its energy, which gets reflected at the boundary, but it continues otherwise unhindered into the gravitating region. See figure \ref{fig:ReflectionIntro}.  In this approximation, the information about infalling particles is lost, and all that emerges is a pulse of energy created by the stress tensor and its descendants. We have an energy mirror but a perfect absorber of all other information not carried by the stress tensor.

\paragraph{A proper BCFT or Cardy state in the full gravity theory.}
Even though we have information loss at the disk level, it is natural to conjecture that in a full theory of quantum gravity, the black hole should become a proper boundary condition for the CFT, which is fully local and obeys all the relevant Cardy conditions; see \cite{Cardy:2004hm}. In this full theory, information would be preserved.
In other words, at the disk level we have only the identity Ishibashi state, but in the full theory, we expect to obtain an appropriate superposition of Ishibashi states that gives rise to a proper Cardy state, i.e., a proper BCFT. We have already said that a more complete bulk theory is expected to contain irrelevant deformations of the CFT which lead to new Ishibashi states already at the disk level.

\paragraph{Higher topologies.}
With this philosophy in mind, it is natural to include the effects of higher topologies. The first difficulty that one encounters is that they diverge. The divergence comes from the thin handles, which propagate the vacuum of the CFT, whose negative energy gives rise to an exponential divergence \cite{Saad:2019lba, Jafferis:2022wez, Moitra:2021uiv}. At this point one is tempted to give up and say that we have to consider the problem in a full quantum theory of gravity which presumably will not lead to this divergence.  For our purposes, we will assume that this divergence is cut off by introducing a minimum cutoff on the length of closed geodesics.  Of course, this introduces some arbitrariness. However, we will still find some interesting computations to discuss.

\paragraph{The cylinder.}
The simplest non-trivial topology for the gravitating region is that of a cylinder or double trumpet. The path integral involves two independent boundary reparametrization integrals as well as an integral over a modulus $b$ that specifies the minimal size of the double trumpet circle.

We attach two separate baths at the two ends, and we perform the integrals. We impose a small cutoff $b \geq b_c$ for the $b$ integral to avoid the divergence due to small handles.
A separate divergence arises when the vacuum propagates in the closed string channel. This one arises due to the fact that the induced action for the reparametrization mode is $\SL(2,\R)$ invariant, but the path integral is over ${\rm Diff}(S^1)/\U(1)$ rather than ${\rm Diff}(S^1)/\SL(2,\R)$, as it was on the disk. So, while we had no problem with the disk, we now have a divergence. This divergence can be cured by remembering that we had the original Schwarzian action. This results in a residual breaking of the precise conformal symmetry, which is simple to incorporate for this diagram.

\paragraph{Open-closed consistency and Cardy state condition.}
For a general  BCFT,  one of the consistency conditions is that the thermal partition function on a strip with two boundaries can also be interpreted as the propagation of the CFT on the closed circle \cite{Cardy:2004hm}. This is the open-closed duality

\lceq{ \la{ClOp}
\langle \text{BH} | e^{ -  \ell H_{\rm closed}} | \text{BH} \rangle = \Tr  \left[ e^{ - \tau_o H_{\rm open} } \right], \quad \tau_o = {2 \pi^2 \over \ell } , ~~~~ \ell = { 2 \pi L \over \beta}
}
where $H_{\rm closed} = L_0+\bar{L}_0 -\frac{c}{12}$ and $H_{\rm open} = L_0 - \frac{c}{24}$.
The trace on the right hand side is over the Hilbert space on the interval. This is the Hilbert space with two local boundary conditions. This is the same as the space of boundary operators of the BCFT.  Furthermore, if we have an elementary BCFT, we have a unit coefficient for the identity contribution on the right hand side,  which is the dominant term as $\ell \to 0$.
 Note that, in addition to $\ell \to 0$, we also need $\tau_c \ll L $ in order to neglect the original Schwarzian contribution, with $\tau_c$ in \nref{DefTc}.

 \begin{figure}[t]
  \centering
  \includegraphics{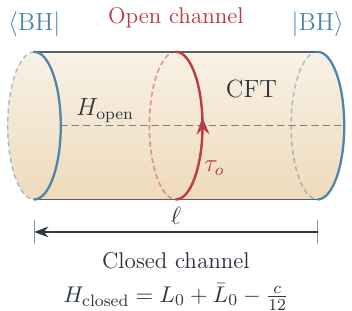}
  \caption{Open- and closed-channel descriptions of the same cylinder.  In the closed channel, the evolution between the two boundary states
  proceeds along the cylinder axis, with Hamiltonian
  $H_{\rm closed}=L_0+\bar L_0 -{c \over 12 }$ and evolution length $\ell$.
  In the open channel, $H_{\rm open}$ acts on the spatial interval
  indicated by the horizontal dashed line, while Euclidean evolution
  winds around the red circumference of length $\tau_o$.
 }
  \label{fig:open-closed-Intro}
\end{figure}
\begin{figure}[t]
  \centering
  \begin{minipage}[t]{0.48\linewidth}
    \centering
    \includegraphics{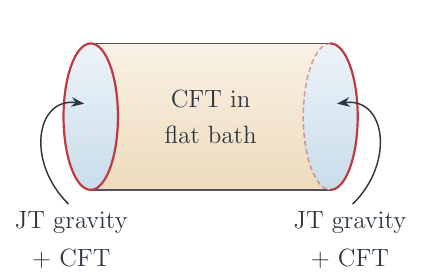}
    \par\smallskip
    {\small (a) Two disks}
  \end{minipage}\hfill
  \begin{minipage}[t]{0.48\linewidth}
    \centering
    \includegraphics{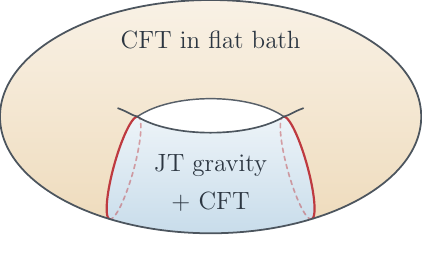}
    \par\smallskip
    {\small (b) Double trumpet}
  \end{minipage}
  \caption{Two gravitational fillings of the same finite bath cylinder, colored in yellow. In (a), two disconnected gravitational disks (blue) cap its ends, giving a sphere. In (b), a connected double trumpet joins the two ends, giving a torus. The CFT lives throughout both geometries. The red curves represent the interfaces between the gravitational region and the flat space.}
  \label{fig:open-closed-geometries}
\end{figure}

In our case, this problem translates to a situation in which we have a finite bath interval with a gravitating region attached to each of its two sides. We can consider a couple of topologies for the gravitating region. The most obvious one is to consider two separate black holes, or two disks at the two ends, see figure \ref{fig:open-closed-geometries}(a). Here, we get only the contribution of the vacuum Ishibashi state \nref{DiskIshi} in the closed channel. For small $\ell$, this does not yield a contribution that resembles the identity in the open channel.

However, there is an interesting contribution from a new topology: a connected contribution in which the gravitational region joins the two sides, so that the total Euclidean spacetime has the topology of a torus; see figure \ref{fig:open-closed-geometries}(b). For large $c$, there is a saddle point in the integral over $b$ and the reparametrizations that can be viewed as an eternally traversable wormhole, similar to the one in \cite{Maldacena:2018lmt}. Curiously, the length of this wormhole (related to the value of $b$) is such that a signal that starts from one of the boundaries takes the same time to go via the wormhole as it takes to go via the bath.
Therefore, the configuration is barely consistent with causality from the perspective of the BCFT interpretation.  Since the CFT is on a torus, we can easily take the small $\ell $ limit where the partition function is essentially that of a torus with a very small spatial cycle and a long Euclidean time circle.
The Gaussian path integral around the saddle point produces a unit overall coefficient.
This looks like a miracle from this perspective. It looks natural when we realize that one expects this full gravitational problem to compute a trace along the Euclidean time direction.

The conclusion is that this new topology gives the expected contribution so as to be consistent with \nref{ClOp}. Namely, the leading term in the  partition function goes as
\be \la{IdCo}
Z \sim   \exp\left( { 2 \pi^2 \over \ell } { c \over 24 } (2-1) \right) + \cdots ~,~~~~~~ {\rm for } ~~~\ell \ll 1
 \ee
 The factor of 2 arises from the CFT on a flat torus; the subtraction arises due to the warp factor of the gravitating region via the conformal anomaly. Together, they give the expected contribution from the identity on the strip. Note that the two disk solution in figure \ref{fig:open-closed-geometries}(a) goes as $e^{ 2 S_0} e^{ \pi^2 \over 12 \ell }$, but the torus diagram in figure \ref{fig:open-closed-geometries}(b) goes as \nref{IdCo} and it dominates for $\ell $ small enough, $\ell \ll c/S_0 $.

 To obtain this result \nref{IdCo}, it is important that we obtain negative energy, which is consistent with the fact that these traversable wormholes typically have negative energy \cite{Gao:2016bin,Maldacena:2018lmt}. This ensures that the connected contribution grows for small $\ell$, eventually overwhelming the contributions of the two separated disks. In particular, we have a large but fixed cutoff dependent contribution from the small $b$ region in the integral, which is overtaken at small $\ell $ by  \nref{IdCo}. These comments also imply that we expect to get \nref{IdCo} in a more general gravity theory that is  approximately described by our toy model at low energies.

It is very significant that the coefficient of \nref{IdCo} is just one. This is consistent with the idea that the black hole corresponds to a single elementary BCFT, rather than to $e^{S_0}$ separate boundaries. In particular, this means that the black hole entropy, appearing in \nref{DiskIshi}, can be interpreted as the Affleck--Ludwig boundary entropy of the BCFT \cite{Affleck:1990by}.

\begin{figure}[htbp]
  \centering
  \includegraphics{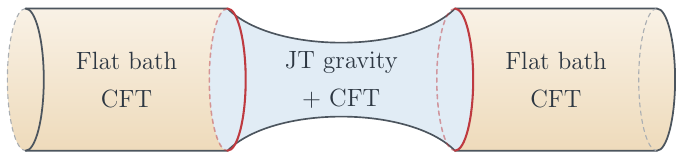}
  \caption{The double-trumpet gravitational region (blue)
  joined to two finite flat CFT baths (yellow). The red circles
  mark the interfaces, which are transparent to the CFT.}
  \label{fig:double-trumpet-setup-intro}
\end{figure}
\paragraph{The cylinder connecting two baths.} Another interesting configuration consists of a gravitating cylinder, or double trumpet,  connecting two separate baths; see figure \ref{fig:double-trumpet-setup-intro}. In this case, the integral over reparametrizations produces effective reflecting boundary conditions for the stress tensor on each interface. We can expand the diagram in the closed channel in terms of primaries with dimensions $h_i= \bar h_i$. For dimensions $h_i> c/24$, and $c> 25$, the integral over all the moduli is finite. As for other similar geometries \cite{Saad:2018bqo}, this can be interpreted as an average. In this case, the natural interpretation is an average over boundary conditions for a given CFT. If we expand the full black hole state in terms of  Ishibashi states
\be
|\mathrm{BH} \rangle = \sum_i \alpha_i |h_i \rangle \rangle
\ee
then the expansion of the cylinder diagram can be interpreted as giving us the average values
\be
 \overline{ \alpha_i \alpha^*_j}_{\rm con} = K(h_i,c) \delta_{ij} ~,~~~~~~~~{\rm for } ~~~~ h_i > { c \over 24 }
 \ee
with $K(h_i,c)=A(h_i,c)^2F(h_i,c)$ given in \nref{eq:proposal_cylinder_match}. For smaller values of $h_i$, the computation exhibits a small handle divergence, so we would need to cut it off, which means we do not have a concrete prediction.

\paragraph{Higher topologies and their interpretation.}

Connected geometries can be naturally interpreted as arising from an average over the coefficients of the boundary state, in analogy to the ensemble interpretation of JT gravity \cite{Saad:2019lba}. More precisely, we imagine a fixed CFT and average over possible boundary states in it. The disk implies that we have a large average coefficient for the vacuum Ishibashi state. The cylinder gives the size of the Gaussian fluctuations,  with non-Gaussian corrections from connected geometries with more boundaries. Of course, the qualitative features do not guarantee the existence of the ensemble. See \cite{Kusuki:2022boundaries} for a closely related discussion.

This picture is clouded by two issues; one is the small handle divergences. We assumed that these can be cut off without a major change, but this is not clear.
The other issue is the breaking of conformal symmetry when the identity primary flows through a trumpet. This gives rise to corrections that scale like powers of the inverse temperature, which could lead to a further RG flow at very long times. In other words, at the disk level we approach a BCFT-looking result at $\beta \gg \tau_c$ \nref{DefTc}, but higher topologies could induce further RG flow at values of $\beta$ which go as some power of $e^{ S_0}$.  Similar features were observed in other computations such as \cite{Okuyama:2019xbv, Engelhardt:2020replica, Janssen:2021entropy, HernandezCuenca:2024geometry,Antonini:2025airy}.   Then we would need to solve this RG flow to find a truly local BCFT and the appropriate value of the boundary entropy.

\paragraph{Magnetic black holes.} We can apply this discussion to magnetic black holes in the Standard Model \cite{Maldacena:2020skw}, particularly in the regime of charges where the electroweak symmetry is unbroken in the near horizon region. Then the fermions are all massless and are charged under the hypercharge magnetic field. Their lowest Landau levels give rise to massless fermions moving along the radial and time directions. This means that their physics can be approximated by our toy model. The region where the fermions are massless stops at the corona, the region separating the near horizon region from the usual Standard Model vacuum. We can say that this provides a boundary condition for all the Standard Model fermions and acts as an effective wall or end of the bath for the massless fermions. As we lower the temperature, or the energy scales we consider, we transition between regimes dominated by the original Schwarzian term, to a CFT dominated regime, to a new effective Schwarzian term at even lower energies. Interestingly, we could have relatively light black holes of the order of a few kilograms where the quantum gravity effects associated with strongly coupled reparametrizations become important at eV-type energies, which are tabletop masses and energies.
 Therefore they could lead to a tabletop test of quantum gravity, {\it if nature was kind enough to produce such magnetically charged black holes! } (A very big {\it if}.)

\paragraph{Outline.} In section \ref{sec:disk}, we discuss in more detail the disk computations.
In section \ref{sec:cylinder}, we discuss the cylinder topology, including the connection to the strip computation discussed above.
In section \ref{sec:bh_as_bd_state}, we discuss more general aspects of higher topologies.
In section \ref{sec:Magnetic}, we discuss the case of magnetically charged black holes in the Standard Model.

\section{The disk}
\label{sec:disk}

\subsection{The basic setup}
We consider JT gravity in AdS$_2$ coupled to a two-dimensional CFT. The CFT extends beyond the AdS boundary into a non-gravitating region with a fixed flat spacetime metric, which we call the \textit{bath}. The interface between the gravitational region and the bath is transparent to the CFT. The Lorentzian and Euclidean setups are shown in figures~\ref{fig:lorentzian-setup}\subref{fig:setup-lorentzian} and~\ref{fig:lorentzian-setup}\subref{fig:setup-euclidean}, respectively. On the two dimensional surface $\mathcal{M}$, the gravity action has the usual JT gravity form
\be \la{JTAct}
 -I_{\rm grav} = \frac{S_0}{ 4 \pi } \int_{\mathcal{M}} \sqrt{g } R + \frac{1}{4 \pi}
 \int_{\mathcal{M}}  \sqrt{g} \phi (R+2)  + \frac{1}{2\pi}\int_{\partial \mathcal{M} } \sqrt{h} K   (S_0 + \phi)
\ee
with the usual boundary terms. Here we set $4G_N = 1$ and the AdS radius to one. Though we have a CFT that lives on $\mathcal{M}$, it does not couple to the dilaton field $\phi$ explicitly. The path integral over $\phi$ along a contour parallel to the imaginary axis imposes the condition that the metric is locally $\AdS_2$ (or $H_2$ in Euclidean signature). The CFT matter will then be important for finding the form of $\phi$ in a solution.

In this section, we will focus on the case where the geometry of the gravitating region is a disk (see figure~\ref{fig:lorentzian-setup}\subref{fig:setup-euclidean}), or its black hole continuation.

In Euclidean signature, the metric $g_w$ of the disk can be written as
\begin{equation}
  \label{Diskme}
  \vcenter{\hbox{\includegraphics{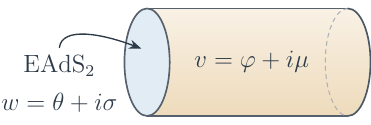}}}
  \qquad
  \begin{gathered}[t]
    ds^2 = \frac{d\theta^2+d\sigma^2}{\sinh^2\sigma}
    =4\frac{dr^2+r^2d\theta^2}{(1-r^2)^2},
  \end{gathered}
\end{equation}
where $\theta\sim\theta+2\pi$ and $-\infty<\sigma<\sigma_b<0$. The bath is a thermal cylinder with period $\beta$. The metric $g_v$ of the outside cylinder can be written as
\be \la{Cylme}
 ds^2 = { 1 \over   \epsilon^2 } ( d\varphi^2 + d   \mu^2 ) = { 1 \over  \epsilon_{\phys}^2 } ( d \varphi_{\phys}^2 + d \mu_{\phys}^2 ) ~~~  , \quad \mu > 0,
 \ee
 with
 \be
 \label{eq:sec2_phys}
 \varphi = { 2 \pi \over \beta } \varphi_{\phys} ~,~~~~~~~~ \mu = { 2 \pi \over \beta } \mu_{\phys} ~,~~~~~~~  \epsilon = { 2 \pi \over \beta }  \epsilon_{\phys}
 \ee
 Here $\varphi_{\phys}$ is the physical Euclidean time outside and $\varphi $ is a rescaled version using the physical inverse temperature $\beta$. Similarly, $\mu_{\phys}$ is the physical distance and  $ \epsilon_{\phys}$ the physical cutoff. We will mostly use the rescaled variables, since the conformal symmetry implies that the physical temperature drops out of most of our computations.
 The two metrics match along a curve $(\theta(\varphi), \sigma_b(\varphi) )$.
 This means setting $ds = d\varphi_{\phys}/ \epsilon_{\phys} = d\varphi/\epsilon$, leading to
 \be
 \label{eq:sec2_displacement}
 \frac{\theta'(\varphi)^2 + \sigma_b' (\varphi)^2}{\sinh^2 \sigma_b(\varphi)} = \frac{1}{\epsilon^2},~~~~~~~ \sigma_b = -\epsilon \theta'(\varphi ) + O(\epsilon^3)
  \ee
As is well known, the JT gravity path integral reduces to an integral
over a boundary mode described by a reparametrization of the circle,
$\theta(\varphi)$. More precisely, the integration is over
$\mathrm{Diff}(S^1)/\mathrm{PSL}(2,\mathbb{R})$. The JT action for this boundary reparametrization is controlled by the boundary value of the
dilaton $\phi$. We impose
\begin{equation}
\label{eq:sec2_phi_phys}
\left.\phi\right|_{\partial\mathcal{M}}
=\frac{\phi_r}{\epsilon}
=\frac{\phi_{r,{\phys}}}{\epsilon_{\phys}},
\qquad
\phi_r=\frac{2\pi}{\beta}\phi_{r,{\phys}},
\end{equation}
where $\phi_{r,{\phys}}$ is the renormalized dilaton parameter in physical units, which we hold fixed when varying the temperature. After subtracting the standard boundary divergence, the gravity action \eqref{JTAct} reduces, apart from the topological term, to the Schwarzian action
\begin{equation}
\label{eq:sec2_Sch}
I_{\rm JT}[\theta]
=-\frac{\phi_r}{2\pi}\int_0^{2\pi}d\varphi\,
\left\{\tan\frac{\theta(\varphi)}{2},\varphi\right\},
\qquad
\{f,\varphi\}
=\frac{f'''}{f'}-\frac{3}{2}\left(\frac{f''}{f'}\right)^2.
\end{equation}
Its coefficient is $\phi_r/(2\pi)=\phi_{r,{\phys}}/\beta$, so this
contribution becomes small at low temperatures.

In the coupled system, integrating out the CFT generates an additional
contribution to the boundary action,
\begin{equation} \label{IndAct}
I_{\rm eff}[\theta]
=I_{\rm JT}[\theta]+I_{\rm CFT}[\theta],
\qquad
I_{\rm CFT}[\theta]=-\log Z_{\rm CFT}[\theta].
\end{equation}
Here $Z_{\rm CFT}[\theta]$ is the CFT partition function on the disk-plus-bath geometry with boundary identification $\theta(\varphi)$. As we will see, the CFT contribution dominates the boundary dynamics in the infrared. We will primarily study this matter-dominated regime, neglecting the Schwarzian term when this approximation is justified.

\subsection{Equations of motion and the boundary condition for the stress tensor}
\label{sec:disk_stress_tensor}

In this section, we discuss the equations of motion for the action $I_{\rm CFT} = - \log Z_{\rm CFT}[\theta]$  induced by the CFT. We anticipate that the result is to set $T= \bar T$ at the boundary of the gravitating region.
This was already done in appendix A of~\cite{Almheiri:2019qdq}, and we review and slightly simplify the derivation here for completeness.

Consider a small variation $\theta (\varphi) \rightarrow \theta(\varphi) + \delta \theta (\varphi)$.
We can shift the bath coordinate $v \rightarrow \tilde{v} = v + \zeta^v$ ($\bar{v} \rightarrow \tilde{\bar{v}} = \bar{v} + \zeta^{\bar{v}}$), such that to first order we have
\lceq{ \label{DefZet}
\theta (\varphi + \zeta^\varphi) = \theta(\varphi) + \delta \theta (\varphi) = \theta(\varphi) + \theta'(\varphi) \zeta^\varphi  +\cdots  \, , \quad \zeta^\varphi = \frac{\delta \theta}{\theta'} \, , ~~\zeta^{\mu} =0
}
where $\zeta^\varphi $ is the component of the vector along the $\varphi $ direction, and we set the component normal to the boundary to zero. Both components are evaluated at the interface in \nref{DefZet}.  Therefore,  the gluing has its original functional form in the shifted coordinate,
while the bath metric acquires a variation. Originally we have  $ds^2=\epsilon^{-2}dv d\bar v$. In the shifted coordinate
\lceq{
ds^2 = \epsilon^{-2} dv d \bar{v} = \epsilon^{-2} \left(d \tilde {v} d \tilde{\bar{v}} - 2 \pl_a \zeta_b d v^a d v^b\right), \quad \delta g_{ab} = - \left(\nabla_a \zeta_b + \nabla_b \zeta_a\right) \, .
}
In other words, a small change in the gluing function $\theta(\varphi) $ can be equivalently parameterized as a change in the metric of the outside cylinder. Note that \nref{DefZet} only defines the vector at the boundary; we are free to choose it arbitrarily in the rest of the cylinder. We choose it so that it rapidly goes to zero away from the boundary.
By the definition of the stress tensor, the change in the action is
\longeq{
\delta I_{\rm CFT}
&=\frac12\int_{D_v}d^2v\,\sqrt g\,
       T_{ab}\,\delta g^{ab}=\int_{D_v}d^2v\,\sqrt g\,
       T_{ab}\nabla^a\zeta^b =+\int_{\partial D_v}ds\,n^aT_{ab}\zeta^b \\
&=-\int_0^{2\pi}d\varphi\,
       T_{\mu\varphi}\frac{\delta\theta}{\theta'} =-i\int_0^{2\pi}d\varphi\,
       \left(T_{vv}-T_{\bar v\bar v}\right)
       \frac{\delta\theta}{\theta'}.
\label{eq:disk_stress_metric_response}
}
where we integrated by parts, used the stress tensor conservation, picked up a boundary term, and used \nref{DefZet}.
Note that \nref{eq:disk_stress_metric_response} is the full variation of (minus the log of) the CFT partition function, including the conformal anomaly part.

Since
$\delta\theta$ is arbitrary and $\theta'>0$, the equation of motion is therefore
\begin{equation}
 \left. T(v)\right|_{v=\varphi}
 =\left.\bar T(\bar v)\right|_{\bar v=\varphi},
 \label{eq:disk_stress_eom}
\end{equation}
where we use the normalization $T(v) = -2\pi T_{vv}$ and $\bar{T}(\bar{v}) = -2\pi T_{\bar{v}\bar{v}}$.   In the closed string channel, this means that there is no momentum along the boundary. In the Lorentzian boundary interpretation, this says that no energy can penetrate the gravitating region from the exterior.

It is worth noting that we also obtain a similar equation from the interior perspective by setting $T_{ww} = T_{\bar w \bar w}$ at the boundary of the gravitating $\AdS_2$ region, since the Schwarzian term is real and cancels in $T- \bar{T}$. In other words, we can also rewrite \nref{eq:disk_stress_metric_response} as
\be \la{VarInThe}
\delta I =  - i  \int d\theta (T_{w w } - T_{\bar w \bar w } ) { \delta \theta }
\ee
Of course, for us it will be very important that \nref{eq:disk_stress_eom} is exactly the condition of vanishing energy flux across the boundary in the context of boundary conformal field theory (BCFT) \cite{Cardy1984}.

\subsection{Recovering the action from its small variation}

Using ideas similar to the ones described in \cite{Fewster:2018,AlekseevShatashviliTakhtajan2024}, we can now integrate
this small variation to write the action in terms of an operator expression.
We write a one parameter family of reparametrizations interpolating between the identity and the final reparametrization
\be
\theta_s(\varphi) \equiv\varphi+s(\theta(\varphi)-\varphi) ~,~~~~~~ 0 < s < 1
\ee
and a small change in $s$ results in the small variation $\partial_s \theta_s(\varphi) $.   Therefore, if we call $U(s)$ the operator that implements the reparametrization $\theta_s$, with the initial condition $U(0) = \id $ we find that
\be \la{UDer}
 \partial_s U(s) = U(s) Q[\xi_s], \quad ~~~~~~\xi_s(\varphi) =   { \partial_s \theta_s(\varphi ) \over \theta_s'(\varphi ) }
\ee
where
\lceq{
Q[\xi_s] =  \frac{1}{2\pi i} \int d \varphi \xi_s(\varphi) \left[ T(\varphi) - \bar T(\varphi)\right]=i \sum_n\xi_s^{(n)}(L_n-\bar L_{-n}) \, ,
}
and $\xi_s^{(n)}$ are the Fourier components of $\xi_s(\varphi)$.  Integrating this equation we find that the operator that implements the reparametrization is
\be \label{UOperator}
 U = U(1) = {\cal {A P }} \exp \left( \int_0^{1} ds Q[\xi_s] \right)
 \ee
where ${\cal A P } $ is the anti-path ordered exponential.
The effective action therefore satisfies
\be
e^{ - I_{\rm CFT} }  \propto \langle 0 | U | 0 \rangle
\ee
It is straightforward to expand this in small fluctuations. We find
\begin{equation}
1+ \int_0^1 ds\,Q[\xi_s]
+\int_0^1 ds_1\int_0^{s_1}ds_2\,Q[\xi_{s_2}] Q[\xi_{s_1}]+\ldots
\end{equation}
  Setting $\theta = \varphi + \vare(\varphi) $, and expanding
$\xi_s\approx\vare-s\vare\vare'$, we get
\begin{equation}
-\sum_{m,n}\frac{\vare_n\vare_m}{2}
\langle0|(L_n-\bar L_{-n})(L_m-\bar L_{-m})|0\rangle
=-\frac{c}{12}\sum_{n\geq2}|\vare_n|^2 n(n^2-1)
\end{equation}
where we used
$\langle0|L_nL_m|0\rangle=\frac{c}{12}n(n^2-1)\delta_{n,-m}$
for $n\geq2$ and the matrix element vanishes for $n =0, \pm 1$. It is also useful to note that if a Virasoro primary of weight $h= \bar h$ is propagating along the closed string channel, we then get a similar result, but now with
\begin{equation}
\langle h | U | h \rangle = 1
 -\frac{c}{12}\sum_{n\geq1}|\vare_n|^2 n \left( n^2 + { 24 h \over c } -1 \right) + \cdots
\end{equation}
We will reproduce both of these quadratic actions by deriving the conformal anomaly and solving for conformal welding, which is the approach we follow in the subsequent computations.

\subsection{The disk and the vacuum Ishibashi state}

\label{subsec:Ishibashi}

In the previous section we discussed the effect of a reparametrization on the CFT. Now we consider the disk case and view the integral over reparametrizations as creating a state in the closed string channel for the CFT.
We can say that we are interested in
\lceq{
\label{eq:sec2_BH}
\bra{\text{BH}}_{\rm disk}  = \int_{\text{Diff($S^1$)/PSL(2,$\R$)}} d \mu[\theta] \, \,  \langle 0 | U[\theta]    \, ,
}
It is important that the measure is invariant under reparametrizations. With the equations of motion in mind, we perform a change of variables in the integral via an infinitesimal reparametrization.  Then, for the variation of  $U$, we can use an equation similar to \nref{UDer}, but with a $\xi$ given by a small reparametrization, which can be an arbitrary function which then leads to
\be \la{IshCon}
\bra{\text{BH}}_{\rm disk} \left[T(\varphi) - \bar T(\varphi) \right]= 0 ~,~~~~~~~{\rm or } ~~~~~~~\bra{\text{BH}}_{\rm disk} (L_n - \bar L_{-n} ) =0
\ee
This argument assumes that the functional integral decays sufficiently fast at the boundaries of the integration domain.
This is one of the conditions that a boundary state in a BCFT obeys. Given a (Virasoro)$^2$ representation of weight $h=\bar h$, we can construct a solution of \nref{IshCon} which is called an ``Ishibashi state'' $|h \rangle \rangle $. It is a particular linear combination of the primary and its descendants that essentially acts as a reflection on all descendant states \cite{Ishibashi:1988kg, Cardy:2004hm}
\lceq{
\label{Ishiexp}
|h \rangle \rangle  = \sum_{q=0}^{\infty} \sum_{A,B} \left(M_q^{-1}\right)_{AB} |h;q,A\rangle \otimes \overline{|h;q, B\rangle} \, ,
}
where $M_q^{-1}$ is the inverse Gram matrix of the basis $|h;q,A\rangle $ for the given descendant level $q$. In particular, we can expand the vacuum Ishibashi state based on order $1/c$, and to leading order we have
\be \label{eq:sec2_BH_vac}
|0 \rangle \rangle = |0 \rangle + \sum_{n=2}^\infty { 12  \over c n(n^2-1)} { L_{-n} \bar L_{-n} }|0 \rangle + \cdots
\ee

Now, since we obtained the state in \nref{eq:sec2_BH} by acting on the vacuum via the stress tensor operators in $U$, we see that the final state should be a linear combination of identity descendants. Since the state should also obey \nref{IshCon}, we conclude that it is proportional to the vacuum Ishibashi state.
In other words, at the disk level we have
\be
\bra{\text{BH}}_{\rm disk} = e^{\tilde{S}_0} \langle \langle 0 | \, ,
\ee
  where $\tilde{S}_0$ includes all constant contributions to the normalization of the disk state.

\subsection{Deriving the action induced by the CFT}

We now provide an alternative formula for  $I_{\rm CFT}[\theta]$. The idea is that the CFT lives on a geometry conformally equivalent to a reference cylinder. If we can write the physical metric as $ds^2=e^{2\rho}d\hat{s}^2$, where $d\hat{s}^2=|dz|^2$ is the metric of a flat reference cylinder, the partition function of the CFT on the physical metric is given by
the conformal anomaly
\begin{equation}
\label{ConfAn}
\log\left(\frac{Z_{\rm CFT}}{Z_{\rm flat}}\right)
=
\frac{c}{24\pi}
\left\{
\int_\Sigma d^2z\,\sqrt{\hat g}\,
\left[(\hat\nabla\rho)^2+\rho\hat R\right]
+
2\int_{\partial\Sigma}d\hat{s}\,\rho\hat K
\right\}.
\end{equation}
Here $c$ is the central charge, $Z_{\rm flat}$ is the partition
function on the reference cylinder, and all quantities carrying a
hat are defined with respect to the reference metric $\hat g$.
In particular,
$(\hat\nabla\rho)^2
=\hat g^{ab}\partial_a\rho\,\partial_b\rho$.

The task is therefore to find a common reference cylinder for the disk and the bath and determine the corresponding Weyl factor $\rho$. At first sight, this looks very simple, since the two regions are individually conformal to half-cylinders, \eqref{Diskme} and \eqref{Cylme}. However, we have a nontrivial
gluing between them, specified by the function $\theta(\varphi)$.
In order to take this into account, it is convenient to perform a further coordinate transformation to a new cylinder parametrized by
$z=x+iy$, with $x \sim x + 2 \pi$,
\lceq{
\label{CFP}
\begin{cases}
    z = G(w) & w = \theta + i \sigma \\
    z = F(v) & v = \varphi+i \mu
\end{cases}, \quad \theta = \theta (\varphi) \quad \text{at} \quad \sigma = \mu  = 0 \, ,
}
Here $G$ is holomorphic for $\sigma<0$ and $F$ is holomorphic for $\mu>0$. Notice that we are working at the leading order in the cutoff $\epsilon$ and we have set $\epsilon=0$ in writing \nref{CFP}. This is enough for our purposes, since we will see that this leads to finite terms in the effective action. Requiring the conformal map to be continuous across this interface
gives
\begin{equation}
\label{eq:sec2_welding}
G\bigl(\theta(\varphi)\bigr)=F(\varphi).
\end{equation}
The problem of determining the holomorphic functions $G$ and $F$ for a given $\theta(\varphi)$ is called the \textit{conformal welding problem}. For the smooth, orientation-preserving circle reparametrizations considered here, existence follows from the classical conformal welding theorem for quasisymmetric homeomorphisms \cite{SharonMumford2006,Bishop2007ConformalWelding,TakhtajanTeo2003}. The solution is unique after fixing the conformal normalization.

Once these maps are known, the metrics on the inside and outside
half-cylinders can be written as
\begin{equation}
\label{MetrConf}
ds^2=\frac{|dz|^2}{|G'(w)|^2\sinh^2\sigma},
\qquad
ds^2=\frac{1}{\epsilon^2}
     \frac{|dz|^2}{|F'(v)|^2},
\end{equation}
respectively. Both are related by a Weyl transformation to the same
flat reference cylinder, $d\hat{s}^2=|dz|^2=dx^2+dy^2$.
Equation~\eqref{MetrConf} therefore determines the Weyl factor
needed in~\eqref{ConfAn}.

\begin{figure}[t]
  \centering
  \includegraphics[width=0.95\linewidth]{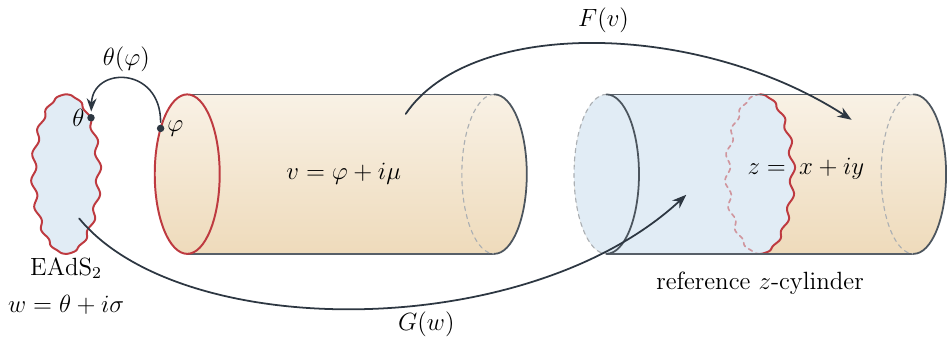}
  \caption{The gravitational disk coupled to the bath and its image under
  conformal welding. The shaded region is mapped by $G(w)$, and the bath
  by $F(v)$.  }
  \label{fig:disk}
\end{figure}

Due to the flatness of the reference metric, $\hat R=0$. The extrinsic curvature is generically nonzero because the gluing surface is a wiggly curve in the $z$ cylinder, see figure~\ref{fig:disk}. However, since the scale factor is continuous and the extrinsic curvature has the opposite sign with respect to the two regions, there is no contribution from the last term in \nref{ConfAn}. The integral within the bulk is conformal invariant so we can evaluate it either in the $z$ coordinates or in the $w$ (or $v$) coordinates.

We must also include the reference partition function
$Z_{\rm flat}$. For the infinite reference cylinder, only the
vacuum state propagates, and hence
\begin{equation}
\label{ZFlat}
\log Z_{\rm flat}
=\frac{c}{12}(y_{\max}-y_{\min}),
\end{equation}
where $y_{\max}$ and $y_{\min}$ regulate its two ends.
To obtain the effective action, we must therefore evaluate both
this reference contribution and the conformal anomaly, keeping
track of how the welding maps relate the original and reference
coordinates.
\paragraph{Relative radial scale after welding.}
An important point is the following. We choose the origin of the $y$ component of the $z$ coordinate,  the reference coordinate, by matching it to the bath, i.e., taking the boundary condition $F(v) - v \rightarrow 0$ at $\mu \rightarrow \infty$. One would be tempted to think that one could also set similar boundary conditions for $G$. However, once we make the choice for $F(v)$, we should allow for a potential shift
\be \la{CoordShift}
  y_{\min} = \Im G(w) = \sigma_{\min} - \Delta y ~,~~~{\rm as} ~~~~ \sigma_{\min} \to - \infty \, .
\ee
The $\Delta y$, which depends on the $\theta$ implicitly, measures the relative radial scale of the disk and the bath after welding.
The constant shift between the two coordinates \nref{CoordShift} is captured by the zero frequency term in the functions $F(\varphi)$ and $G(\theta)$. More precisely, by the imaginary value of that zero frequency term
\lceq{
\label{eq:sec2_delta}
\Delta y  = \frac{1}{2 \pi} \left[\int d \varphi \Im F(\varphi) - \int d \theta \Im G(\theta) \right] = \frac{1}{2 \pi}  \int d \varphi ( 1 - \theta')    \Im F(\varphi)  \, .
}
In appendix~\ref{app:relative_radial_scale}, we calculate this shift for some functions $\theta(\varphi)$ in detail.

\paragraph{Evaluation of the action.}
In the gravitational region, the metric $g_w$ in~\eqref{MetrConf} gives
the Weyl factor
\begin{equation}
\rho=\rho_0+\rho_G,
\quad
\rho_0=-\log|\sinh\sigma|,
\quad
\rho_G=-\frac12\log|G'(w)|^2.
\end{equation}
Here $\rho_0$ comes from the hyperbolic metric \nref{Diskme}, while $\rho_G$
comes from the conformal map. We can evaluate the integral of the square of the gradient in either the $z$ or $w$ coordinates, due to the conformal symmetry the overall factors in the metric drop out.
Expanding the square of the first term of~\eqref{ConfAn} gives
\begin{equation}
\begin{aligned}
\label{eq:sec2_split}
\int_{\mathcal D_z} d^2z \sqrt{\hat{g}}(\hat{\nabla} \rho)^2
&=
\int_{\mathcal D_w} d^2w \sqrt{g_w}
\Big[
(\nabla \rho_0)^2
+(\nabla \rho_G)^2
+2\nabla \rho_0\cdot\nabla \rho_G
\Big] \\
& = - \left(\mL_{\rm b} + \mL_G + \mL_\times\right),
\end{aligned}
\end{equation}
where the first term $\mL_{\rm b}$ depends only on the hyperbolic Weyl factor, the second $\mL_G$ depends only on $G'$, and the third $\mL_\times$ is their cross term.  We begin with the term involving only $\rho_0$
\begin{equation}
\label{eq:bulk_int}
\begin{aligned}
  -\mL_b & = \int_{\mathcal D_w} d^2w \sqrt{g_w} (\nabla \rho_0)^2 = 4 \int d^2 w | \partial_w \log( \sinh \sigma )|^2 = \int d \theta \int_{\sigma_{\min}}^{\sigma_b } d\sigma { 1 \over \tanh^2 \sigma }\\
   &= \int_0^{2 \pi} d \theta \left[\sigma - \coth \sigma\right]_{\sigma_{\min}}^{\sigma_b} =   2 \pi \left(\frac{1}{\epsilon} - \sigma_{\min} - 1\right) + O(\epsilon) \, .
\end{aligned}
\end{equation}
We neglect the term going like $1/\epsilon $ because it is proportional to the length of the boundary once we express it in terms of the physical cutoff $\epsilon_{\phys}$ \nref{eq:sec2_phys}. The $-1$ is an extra contribution to the entropy of the disk that should be added to $S_0$.\footnote{The finite constant $-1$ in~\eqref{eq:bulk_int} originates from the hyperbolic area. More generally, for a smooth hyperbolic surface $\Sigma$ with $R=-2$ and asymptotic boundaries $\pl \Sigma_\epsilon$, Gauss--Bonnet gives the area $A_\epsilon=\int_{\partial\Sigma_\epsilon}K\,ds-2\pi\chi(\Sigma)$, where we have introduced the asymptotic cutoff $\epsilon$. Notice that $K=1+O(\epsilon^2)$, the integral $\int_{\pl \Sigma_\epsilon} K ds = 2\pi/\epsilon$ giving the first term in~\eqref{eq:bulk_int}. Then the rest of the term is simply $-2\pi\chi(\Sigma)$. Such an area contribution to $\log Z_{\rm CFT}$ is therefore $-c\chi(\Sigma)/12$, which can be absorbed consistently across topologies by defining $S_0^{\rm ren}=S_0-c/12$.}
 The remaining contribution requires keeping an explicit regulator $\sigma_{\min}$ near $\sigma = -\infty$, until all terms have been combined. To interpret this term, recall that the cylindrical coordinate sends the smooth disk center to $\sigma\rightarrow -\infty$. Introducing a cutoff $\sigma_{\min}$, we can describe the metric in terms of a long cylinder in the coordinates \nref{Diskme} up to $\sigma_{\rm min} $, and then this cylinder has an endcap which is described by the metric $ds^2 = 4 e^{  2 \sigma_{\rm min} } ({ d r^2 + r^2 d \theta^2 } )$. Switching from this metric to a unit disk gives a contribution to the conformal anomaly from the extrinsic curvature term in \nref{ConfAn} which is  $\frac{c}{6} \sigma_{\min}$, plus a constant that we absorb in the definition of $\tilde S_0$.
Combined with the reference partition function~\eqref{ZFlat} and the shift~\eqref{CoordShift}, we have the final contribution $\frac{c}{12} (\sigma_{\min} - y_{\min}) = \frac{c}{12} \Delta y$.

Using that $G$ is holomorphic, and so $\rho_G$ is harmonic ($ \nabla^2 \rho_G = 0$), we can do an integration by parts in the cross term and get
\longeq{
\label{eq:sec2_crossterm}
-\mL_\times &= 2 \oint_{\pl \mathcal{D}_w} \rho_0 \pl_n \rho_G d s - 2 \int_{\mathcal{D}_w}  d^2 w \sqrt{g_w} \rho_0 \nabla^2 \rho_G \\
&= i \int_0^{2 \pi} d  \theta \log (\epsilon \theta') \left[\frac{G''(\theta)}{G'(\theta)} - \frac{\bar{G}''(\theta)}{\bar{G}' (\theta)}\right] + O(\epsilon) \, ,
}
where we use the radial derivative
\lceq{
(\pl_n f) ds = i \left(\pl_w f d w - \pl_{\bar{w}} f d\bar{w}\right) ,
}
and only keep the terms up to zeroth order in the cutoff. In addition, we used the expression of $\sigma_b$ in~\eqref{eq:sec2_displacement} to compute
\lceq{
\rho_0 |_{\pl \mathcal{D}_w} = -\log (\epsilon \theta') + O(\epsilon^2), \quad d w |_{\pl \mathcal{D}_w} = d \theta + O(\epsilon) \, .
}
Also, in ~\eqref{eq:sec2_crossterm}  we have dropped the boundary term at $\sigma = \sigma_{\rm min}$ for the following reason.
When $\sigma_{\min} \rightarrow -\infty$, we have
\lceq{
\rho_0  = \sigma_{\min} + \log 2 + O(e^{2\sigma_{\min}}), \quad G(w) = w + g_0 + O(e^{\sigma}), \quad \pl_\sigma \rho_G = O(e^{\sigma}),
}
Then the integral along the regulated boundary simplifies to
\lceq{
\oint_{\sigma = \sigma_{\min}} \rho_0 \pl_n \rho_G d s = -  \rho_0 \int_0^{2 \pi} d \theta \pl_\sigma \rho_G = O\left(|\sigma_{\min}| e^{\sigma_{\min}}\right) \rightarrow 0 \, .
}
Moreover, due to the harmonic property of the $\rho_G$, we have the identity
\lceq{
0 = \int_{\mathcal{D}_w} d^2 w \sqrt{g_w} \nabla^2 \rho_G = - \oint_{\sigma = \sigma_{\min}} \pl_\sigma \rho_G d\theta  + \oint_{\pl \mathcal{D}_w} \pl_n \rho_G ds \, \Rightarrow \,  \oint_{\pl \mathcal{D}_w} \pl_n \rho_G ds = 0 \,.
}
With this identity, we can see that the logarithmic term $\log \epsilon$ actually vanishes. In conclusion, the cross term is simply
\lceq{
-\mL_\times  = i \int_0^{2 \pi} d  \theta \log \theta' \left[\frac{G''(\theta)}{G'(\theta)} - \frac{\bar{G}''(\theta)}{\bar{G}' (\theta)}\right] \, . \
}
For the last term, $-\mL_G$, we apply integration by parts again to obtain
\longeq{
-\mL_G = \int_{\mathcal{D}_w} d^2 w \sqrt{g_w} &(\nabla \rho_G)^2 = \oint_{\pl \mathcal{D}_w} \rho_G \pl_n \rho_G ds \\
&= \frac{i}{4} \int_0^{2\pi} d \theta \log |G'(\theta)|^2 \left[\frac{G''(\theta)}{G'(\theta)} - \frac{\bar{G}''(\theta)}{\bar{G}' (\theta)}\right]  \, .
}
For the integration within the bath region, the computation procedure is roughly the same. In the bath region with coordinate $v$ we split the Weyl factor as $\rho_0 = -\log \epsilon$ and $\rho_F = -\frac{1}{2} \log |F'|^2$. $\rho_0$ is a constant, which brings a significant simplification and
\longeq{
-\mL_b &= \int_{\mathcal{D}_v} d^2 v \sqrt{g_v} (\nabla \rho_0)^2 = 0 \, , \\
-\mL_\times  &= 2 \int_{\mathcal{D}_v} d^2 v \sqrt{g_v} (\nabla \rho_0) \cdot (\nabla \rho_F)  = 0  \, ,\\
-\mL_F &= \oint_{\pl \mathcal{D}_v} \rho_F \pl_n \rho_F ds = -\frac{i}{4} \int d \varphi \log |F'(\varphi)|^2 \left[\frac{F''(\varphi)}{F'(\varphi)} - \frac{\bar{F}''(\varphi)}{\bar{F}'(\varphi)}\right] \, .
}
In summary, after integrating out the CFT, we obtain the following effective action
\longeq{
\label{eq:sec2_sumI}
-I =  -& E \Delta  y + \frac{i c}{24 \pi} \tpiint d \theta \log \theta' \left[\frac{G''}{G'} - \frac{\bar{G}''}{\bar{G}' }\right]  \\
& +\frac{i c }{96 \pi} \left\{\int_0^{2\pi} d \theta \log |G'|^2 \left[\frac{G''}{G'} - \frac{\bar{G}''}{\bar{G}' }\right] - \tpiint d \varphi \log |F'|^2 \left[\frac{F''}{F'} - \frac{\bar{F}''}{\bar{F}'}\right]\right\},
}
where we subtract the reparametrization-independent divergence $\frac{c}{12} (y_{\max} +1/\epsilon)$ and all the primes are derivatives with respect to the natural argument of the function. The term involving $E \Delta y$ is related to the mismatch of the coordinates \nref{CoordShift} and it captures the energy of the state. In this case, this energy is just the vacuum Casimir energy along the cylinder $E = -c/12$. Note that a positive $\Delta y$ makes the reference cylinder longer.

We can further simplify the expression for the action as follows. The crucial relation we can use is the conformal welding condition~\eqref{eq:sec2_welding}. Starting from~\eqref{eq:sec2_welding} and taking derivatives on both sides we obtain
\begin{align}
&  G'(\theta) \theta' = F'(\varphi) ~,~~~~~~~ G''(\theta) {\theta'}^2 + G'(\theta) \theta'' = F''(\varphi) \, .
 \\ \la{FpGpThp}
 &  {F'' \over F' } = { G'' \over G' } \theta' + { \theta'' \over \theta' } ~,~~~~~~~ |G'|^2 {\theta'}^2 = |F'|^2 \, .
  \\
 &   {F'' \over F' } - {\bar F'' \over \bar F' } = \left( { G'' \over G' }- { \bar G'' \over \bar G' }\right)  \theta' \, .
\end{align}
where all functions are evaluated at their common boundary.
Note that $G$ and $F$ are not required to be real.
One can then show that the total action takes the form
\be
\begin{aligned}
 \la{ActDisk}
-I
&=  -E \Delta y + \frac{i c}{48 \pi} \tpiint d \theta \log \theta' \left(\frac{G''}{G'} - \frac{\bar{G}''}{\bar{G}'}\right) \\
&=  -E \Delta y + { i c \over 48 \pi } \int_0^{2\pi} d\varphi \log \theta' \left( { F'' \over F' } - {\bar F'' \over \bar F' } \right) \, .
\end{aligned}
\ee
We are free to express the action only in terms of the conformal map $G$ or $F$. Notice that this action is independent of the physical temperature $\beta$, which implies that it dominates at low temperatures. Since it is proportional to $c$, it can be analyzed perturbatively for large $c$, $c\gg 1$.
In other words, at very high temperatures, the system is dominated by the Schwarzian action \nref{eq:sec2_Sch} but as we lower the temperatures, {\it before reaching strong coupling}, the action \nref{ActDisk} becomes dominant and it can remain weakly coupled. This transition happens when
\be
{ \phi_{r, {\rm phys}} \over \beta } \sim c
\ee

 As a side comment, note that the equation of motion of \nref{ActDisk} gives, as we expected from section   \ref{sec:disk_stress_tensor},
\lceq{
\label{eq:sec2_eom}
 0= \Im \left[-   \frac{E}{2}
  F'(v)^2 + \frac{c}{12} \left\{F(v),v\right\} \right]  \propto T - \bar T  \, .
}
where we computed the stress tensor starting from a constant one in the $z$ coordinates,  $T(z) = \bar T(\bar{z})=  - E/2$, and conformally transforming it with the map $F$.

\paragraph{Comment on the relation with the math literature.} In the case that the bath has infinite length, we find that $E= -c/12$ and this action agrees with the actions considered in
\cite{TakhtajanTeo2004,AlekseevShatashviliTakhtajan2024} where the connection of this action with the structure of the Virasoro group was explored\footnote{ In other words, the action \nref{ActDisk} for $E=-c/12$ has the same form as the one in equation (1.1) of \cite{TakhtajanTeo2004}, up to integrations by parts, see appendix \ref{TTAction}. Also, \cite{Wang2019LoewnerEnergy} showed a connection to the Loewner energy. }. This functional is invariant under inversion of the boundary reparametrization $\theta(\varphi)$, so that the action evaluated on $\theta(\varphi)$ equals that evaluated on its inverse $\varphi(\theta)$. It is also invariant under independent circle-preserving M\"obius transformations of the interior and exterior boundary coordinates, corresponding to left and right composition of the welding map. These properties give a $\mathrm{PSL}(2,\mathbb{R})\times\mathrm{PSL}(2,\mathbb{R})$ symmetry of the induced action. Here, we provide a concrete physical realization of this mathematical object in JT gravity coupled to a CFT.

\paragraph{The bath at finite length.} It is interesting that we can also generalize this discussion to the case that the length of the bath is finite, so that $\mu \in [0, \ell]$, with $\ell = \frac{2 \pi L}{\beta}$, where $L$ is the physical length of the bath. In that case we require that the function $F$ maps the circle at $\mu=\ell$ to a straight circle at the right boundary of the $z$ cylinder. In this case, the computation of the action is similar. The only change is that the integrals over $v$ run over a finite region. The straight circle at the right boundary of the $z$ cylinder has vanishing extrinsic curvature,  so we can still ignore the term involving $\hat{K}$ in~\eqref{ConfAn}, even in the presence of this extra boundary $\Sigma_\ell$. The computation we need to modify is the evaluation of $-\mL_F$, which becomes
\lceq{
-\mL_F = -\frac{i}{4} \int d \varphi \log |F'|^2 \left[\frac{F''}{F'} - \frac{\bar{F}''}{\bar{F}'}\right] + \oint_{\Sigma_\ell} \rho_F \pl_n \rho_F d s \, .
}
The straight circle boundary condition means that the imaginary part of $F$ is constant, therefore
\lceq{ \la{BCbc}
\Im F |_{\mu = \ell} = \frac{F - \bar{F}}{2 i} |_{\mu = \ell } = \text{constant} \quad \Rightarrow \quad F' =  \bar{F}' \quad \text{and} \quad  F'' = \bar{F}'' \, .
}
This gives
\bea
{ F'' \over F' } - { \bar F'' \over \bar F' } = 0 \quad \Rightarrow \quad  \partial_n \rho_F \propto \pl_\mu \log |F'|^2 = 0
\eea
along the boundary $\Sigma_\ell$.
The conclusion is that for a finite cylinder we get the same expression for the action in terms of $F$ and $G$. The functions $F$ and $G$ are different because now $F$ obeys a different boundary condition \nref{BCbc}.

During the derivation of the above actions, the only approximation we make is to expand the quantities to the leading order of the UV cutoff $\epsilon$, which should be valid for arbitrary reparametrization fluctuations that are fixed and smooth in the $\epsilon \to 0 $ limit.

\subsection{Partition function}
In general it is difficult to perform the whole Euclidean path integral using the full action~\eqref{ActDisk}, since the dependence on $\theta$ is highly nonlinear. We therefore expand to quadratic order in fluctuations around a saddle and compute the partition
function to one-loop order. It is instructive to solve the problem for a finite external cylinder, since it is not much more difficult.

\paragraph{Expansion of the action around the saddle.}
It is obvious that the round configuration $\theta = \varphi$, while $F(v) = v$ and $G(w) = w$, is a solution to the conformal welding condition~\eqref{eq:sec2_welding} and the equation of motion~\eqref{eq:sec2_eom}. Perturbing around this saddle, we can write
\lceq{
\label{eq:sec2_vare}
\theta = \varphi + \vare(\varphi) \, , \quad \vare(\varphi) = \sum_{n \in \Z} \vare_n e^{i n \varphi} \, , \quad \vare_n^* = \vare_{-n} \, .
}
Since $\theta \sim \theta +2\pi$ and $\varphi \sim \varphi + 2\pi$, $\vare(\varphi)$ is a real periodic function which leads to the last two equations above. Since
$F$ and $G$ are holomorphic and admit the boundary conditions
\lceq{
\label{eq:sec2_bdy}
\begin{cases}
 \Im F(v) = \ell, & v = \varphi+ i \ell, \\
 G(w) - w \rightarrow \text{constant} & \sigma \rightarrow -\infty, \\
\end{cases}
}
the corrections to the maps have Fourier expansions
\begin{equation}
 \begin{aligned}
 F(v)&=v+f(v),&
 f(v)&=\sum_{n\in\mathbb Z}f_n e^{inv},\\
 G(w)&=w+g(w),&
 g(w)&=g_0+\sum_{n=1}^{\infty}g_{-n}e^{-inw}.
 \end{aligned}
 \label{eq:sec2_Fourier}
\end{equation}
We fix the arbitrary axial translation of the reference cylinder by placing the outer bath boundary at $y = \ell$. The Fourier coefficients $f_n$ and $g_n$ are of order $\vare$. In terms of these Fourier coefficients one can expand the action \eqref{ActDisk}  and the expression for $\Delta y$ \eqref{eq:sec2_delta}   to quadratic order in $\vare$ to obtain
\longeq{
\label{eq:sec2_expand_I}
E \Delta y =  \frac{E}{2\pi} \tpiint (1-\theta') \Im F &= -\frac{E}{2 \pi} \tpiint d \varphi \vare' \Im f + O(\vare^3) \\
\frac{i c}{48 \pi} \tpiint d \varphi \log \theta' \left(\frac{F''}{F'} - \frac{\bar{F''}}{\bar{F}'}\right) &= \frac{ic}{48 \pi} \tpiint d \varphi \vare' (f'' - \bar{f}'') + O(\vare^3) \, .
}
\paragraph{Solving conformal welding problem.}
It is enough to solve the conformal welding problem to the linear order of $\vare$ to get the quadratic action. We substitute~\eqref{eq:sec2_vare} and~\eqref{eq:sec2_Fourier} into the conformal welding condition~\eqref{eq:sec2_welding} and the first condition in~\eqref{eq:sec2_bdy}. By matching the coefficients for each mode $e^{i n \varphi}$, we get
\begin{equation}
 f_n=\varepsilon_n+g_n \, \,  (n \le 0),\quad f_n = \vare_n \, \,  (n \ge 1), \quad
 f_n e^{-n\ell}=f_{-n}^*e^{n\ell}\, \, (n \in \Z) \, .
 \label{gnvals}
\end{equation}
For $n = 0$, $f_0 = \vare_0 + g_0$ is real and for $n\ge 1$ we have
\begin{equation}
 f_n=\varepsilon_n,\quad
 f_{-n}=q^n\varepsilon_{-n},\quad
 g_{-n}=-(1-q^n)\varepsilon_{-n},\quad
 q=e^{-2\ell},
 \label{eq:sec2_disk_welding_sol}
\end{equation}
where we also used the reality condition for $\vare$, i.e., $\vare_n^* = \vare_{-n}$. Inserting this in~\eqref{eq:sec2_expand_I} gives
\lceq{
-I^{(2)} = - \frac{c}{12} \sum_{n=1}^\infty \left(n^3+ \frac{12 E}{c} n\right) (1-q^n) |\vare_n|^2 \, .
}
If we use the cylinder vacuum energy $E=-c/12$, the two contributions combine to
\begin{equation}
  -I^{(2)}
 =- \frac c{12}\sum_{n=2}^{\infty}
       n(n^2-1)(1-e^{-2n\ell})|\varepsilon_n|^2
       +O(\varepsilon^3).
 \label{ZShF}
\end{equation}
For $\ell \rightarrow \infty$, the quadratic action is of a simple form
\lceq{
\label{eq:sec2_inf_I2}
-I^{(2)}
 =- \frac c{12}\sum_{n=2}^{\infty}
       n(n^2-1)|\varepsilon_n|^2 \, .
}
This is clearly a nonlocal action, it has one less power of $n$ than the Schwarzian action expanded to quadratic order.

We can use \nref{eq:sec2_inf_I2} to calculate the two point function
\lceq{
\label{eq:sec2_varesq}
\langle \vare_n \vare_{-m} \rangle = \frac{12 \delta_{nm}}{c n (n^2-1)}, \quad n, m \ge 2 \, ,
}
or, in position space,
\lceq{
\langle \vare(\varphi) \vare (0)\rangle = \frac{12}{c} \sum_{|n|>1} \frac{e^{i n \varphi}}{|n|(n^2-1)} = \frac{12}{c} \left[-1+ \frac{3}{2} \cos \varphi + 2 \sin^2 \frac{\varphi}{2} \log \left(4 \sin^2 \frac{\varphi}{2}\right)\right] \, .
}

As a side remark, in the infinite-bath limit, the quadratic action and propagator coincide with those of the boundary reparametrization mode on the $\AdS_2$ string, upon matching $c/12=4\pi T_s$, where $T_s$ is the string tension in AdS units \cite{Giombi:2022pas, Giombi:2023zte}.

\paragraph{Comments on the zero modes.}
It is obvious that we have three zero modes in~\eqref{ZShF}, i.e., the constant mode $\vare_0$ and $\vare_{\pm 1}$, if we choose the vacuum to propagate along the reference cylinder. This is a consequence of the $\rm PSL(2,\R)$ symmetry acting on the disk. In particular, the constant mode $\vare_0$ corresponds to the rigid rotation of the disk boundary, $\theta \rightarrow \theta + \vare_0$. This is the $\rm U(1)$ subgroup of the full disk isometry group $\rm PSL(2,\R)$.

Let us check that the full action is indeed invariant under   $\rm PSL(2,\R)$. Under an infinitesimal  transformation
\lceq{
\delta \theta  = \eta (\theta)  =   \vare_0 +  \vare_{1} e^{ i \theta} +  \vare_{-1} e^{ - i \theta}
}
Using~\eqref{VarInThe} we get
\lceq{
\delta I = - \frac{1}{2 \pi i} \tpiint d \vartheta \eta (\vartheta) \left[T(w) - \bar{T} (\bar{w})\right]_{w = \bar{w} = \vartheta} \, .
}
The assumption that the vacuum propagates along the cylinder implies that the expansion has only terms involving powers of $e^{ i n w} $ with $|n|> 1$ so that this integral vanishes.

\paragraph{The Gaussian integral and its measure.}
The gravitational path integral uses the measure inherited from JT
gravity on $\operatorname{Diff}(S^1)/\operatorname{PSL}(2,\mathbb R)$. The JT measure
is the symplectic measure associated with the disk orbit
\cite{Stanford:2017thb,Saad:2019lba}. Up to an overall normalization
and orientation, its symplectic form is
\begin{equation}
 \Omega\ \propto\
 \int_0^{2\pi}d\varphi\,
 \left[
   \frac{\delta\theta'\wedge\delta\theta''}{\theta'^2}
   -\delta\theta\wedge\delta\theta'
 \right].
 \label{eq:disk_loop_symplectic}
\end{equation}
At one loop only the measure evaluated at the saddle is needed. Therefore we are using the measure
\begin{equation}
 d\mu
 =\mathcal A \prod_{n=2}^\infty
       n(n^2-1)\,d^2\varepsilon_n,\qquad
 d^2\varepsilon_n
 =d(\operatorname{Re}\varepsilon_n)\,
  d(\operatorname{Im}\varepsilon_n),
 \label{eq:disk_loop_measure}
\end{equation}
where we package all the normalization factors in $\mathcal{A}$. The disk topology contributes $e^{S_0}$ and the reference partition function contributes $e^{\frac{c}{12} \ell}$ (see~\eqref{ZFlat} and we fix $y_{\max} =\ell$). Therefore, the one loop partition function is
\longeq{
 Z^{(1)}(\ell)
 &=e^{S_0} e^{c\ell/12}\mathcal A
   \prod_{n=2}^\infty n(n^2-1)
   \int d^2\varepsilon_n\,
   \exp\!\left[
    -\frac c{12}n(n^2-1)(1-q^n)|\varepsilon_n|^2
   \right]\\
 &=e^{S_0} \mathcal A \left(\frac{c}{12 \pi}\right)^{3/2}
   e^{c\ell/12} \prod_{n=2}^\infty(1-q^n)^{-1}  \, ,
\label{eq:sec2_Z1}
}
Here we use the zeta function regularization for the infinite product $\prod_{n=2}^{\infty} \left(12 \pi/c\right) = (12 \pi/c)^{-3/2}$ since $\zeta(0) = \sum_{n=1}^\infty 1=-1/2$ through analytic continuation. Interestingly we can write the one loop partition function in terms of the vacuum Virasoro character $\chi_0 (q)$ for $c > 1$
\lceq{
\label{eq:sec2_Z1chi}
Z^{(1)} (\ell) = e^{S_0} \mathcal A \left(\frac{c}{12 \pi}\right)^{3/2} \chi_0 (q), \quad \chi_0 (q) = q^{-c/24} \prod_{n=2}^{\infty} (1-q^n)^{-1} \, .
}

This result supplies a check of the idea that the integral over reparametrizations produces the vacuum Ishibashi state.   One interesting aspect of this formula is that its $\ell$ dependence is precisely the one we expect
for $c>1$. In other words, all higher $1/c$ corrections that depend on $\ell $ vanish. In principle, we can have a $c$ dependent prefactor that is independent of $\ell$. This suggests a non-renormalization theorem or some localization formula\footnote{For the unitary minimal models with $0<c<1$, the vacuum has an additional independent null vector besides the level one null vector $L_{-1}\ket{0}$, and it has a different vacuum Virasoro character. It is interesting to understand how to see it in our framework.
}.

\paragraph{The short-bath limit.}
For small $\ell$, \nref{ZShF} plus the vacuum energy on the bath cylinder becomes
\begin{equation}
\label{ESchCou}
 {c \ell \over 12 } -I^{(2)}
 ={c \ell \over 12 } -\frac{c\ell}{6}\sum_{n=2}^{\infty}
       n^2(n^2-1)|\varepsilon_n|^2
       +O(\ell^2).
\end{equation}
Recall that the Schwarzian action~\eqref{eq:sec2_Sch} expanded to quadratic order is
\lceq{
-I_{\rm Sch} =  \frac{\phi_r}{2} - \phi_r \sum_{n=2}^\infty n^2 (n^2-1) |\vare_n|^2 + O(\vare^3) \, .
}
The induced quadratic action~\eqref{ESchCou} is exactly of the same form if we define the effective Schwarzian length
\lceq{
\label{eq:sec2_phic}
\phi_c = \frac{c \ell}{6}, \quad ~~~~~~~\ell  = \frac{2 \pi L}{\beta}, \quad ~~~~~~\phi_{c, \text{phys}} = \frac{\beta}{2 \pi} \phi_c = \frac{c L}{6}.
}
The theory effectively becomes Schwarzian theory with physical length scale $\phi_{c, \text{phys}}$. Here we derived this to the linearized order, but we show in appendix~\ref{app:short_bath}, that the same is true to all orders as we take the $\ell \to 0$ limit. Of course, this is not surprising since we can view the short bath limit as a type of local boundary, and the Schwarzian is the generic term that we get with a local boundary condition \cite{Maldacena:2016upp}. Note that the small $\ell$ limit of \eqref{eq:sec2_Z1chi} does indeed look like the Schwarzian disk answer.

For $\ell \ll1$, retaining $c \ell$ in the exponent, the vacuum Virasoro character has the asymptotic form
\longeq{
\label{eq:sec2_chi_smalll}
\chi_0 (e^{-2 \ell}) = \frac{2}{\sqrt{\pi}} e^{\frac{\pi^2}{12 \ell}} \ell^{3/2} e^{\frac{c \ell}{12}} \left [1 + O(\ell)\right] \, ,
}
where we have used the modular property of the Dedekind eta function to get the asymptotics. Restoring the $\beta$ dependence gives the one loop partition function as
\lceq{
\label{eq:sec2_short_bath_Z}
Z^{(1)} (\ell)  \propto e^{S_0 - \beta E_0} \left(\frac{\phi_{c, \text{phys}}}{\beta}\right)^{3/2} \exp \left(\frac{\pi \phi_{c, \text{phys}}}{ \beta}\right) \, .
}
The first factor $e^{\frac{\pi^2}{12 \ell}}$ in~\eqref{eq:sec2_chi_smalll} can be interpreted as a shift in the ground state energy $E_0$, $\delta E_0 = - { \pi \over 2 4 L} $. After this shift, the remaining temperature
dependence is that of the Schwarzian disk. The power
$(\phi_{c, \text{phys}}/\beta)^{3/2}$ is the one-loop factor, while
$e^{\pi \phi_{c, \text{phys}}/\beta}$ is the classical saddle
contribution. The proportionality constant is independent of $\beta$
and $L$ in the fixed measure convention. This agrees with the vacuum-character construction in~\cite[Sec.~3.1]{Mertens:2017mtv},
who obtained the Schwarzian partition function by taking $c\to\infty$
and $q\to1$ with $q^{c/24}$ fixed, after absorbing the zero-point
energy and overall normalization. The short bath therefore provides a geometric realization
of the same character-to-Schwarzian limit, consistent with the
nonlinear boundary action derived in
appendix~\ref{app:short_bath}.\footnote{The character limit keeps
$c\ell$ fixed, while semiclassical control of our Gaussian calculation
additionally requires this fixed value to be large. The comparison
does not by itself establish one-loop exactness of the full finite-bath path integral.}

Finally, note that we have reflecting boundary conditions for the energy or the degrees of freedom created by the stress tensor. These degrees of freedom can be roughly viewed as those of a single scalar field contributing the $\chi_0$ factor in \nref{eq:sec2_Z1chi}. At small $\ell$, this part contributes the term $e^{\frac{\pi^2}{12 \ell}}$, which can be viewed as the ground state energy on the strip of a $c=1$ degree of freedom. On the other hand, the rest of the degrees of freedom of the CFT contribute with the factor $e^{ c \ell \over 12} $ which corresponds to the ground state energy in the closed channel. These degrees of freedom feel as if they are in an infinite cylinder, since they can pass through the interface without impediment. Though the degrees of freedom live in a semi-infinite cylinder, it is only the bath region that ends up contributing to the partition function.

\paragraph{Restoring the JT Schwarzian action.}
The preceding calculation shows that the CFT alone generates a Schwarzian action when the bath is short. In the full JT gravity coupled to CFT, the
induced contribution coexists with the Schwarzian term already present in the gravitational
action. Restoring the latter allows us to determine how the competition between the two contributions affects the temperature
dependence of the disk partition function.

We first keep the physical length $L$ finite and evaluate the one loop partition function, then obtain the infinite bath result by taking $L\rightarrow \infty$. With  $\phi_{r,\rm{phys}}$ fixed, we can write the full action up to the quadratic order
\begin{equation}
\label{eq:sec2_Itot}
 -I_{\rm tot}^{(2)} =  \frac{\phi_r}{2} + \frac{c \ell}{12}
 - \sum_{n=2}^{\infty}n(n^2-1)
       \left[\phi_r n+\frac c{12}(1-q^n)\right]
       |\varepsilon_n|^2.
\end{equation}
It is useful to define
\lceq{
\lambda = \frac{c}{12 \phi_r} = \frac{c \beta}{24 \pi \phi_{r,\rm{phys}}} \, , ~~~~~~q = e^{ - 2 \ell }
}
and the ratio of the matter contribution and JT contribution for each Fourier mode is
\lceq{
\frac{I_{\text{CFT}}^{(n)}}{I_{\text{Sch}}^{(n)}} = \frac{\lambda}{n} \left(1 - q^n\right) \, .
}
For an infinite bath $q \rightarrow 0$, the CFT matter dominates the modes $n\ll\lambda$,
whereas the Schwarzian term dominates the others with $n\gg\lambda$.
In physical frequency units, $\omega_n=2\pi n/\beta$, this gives
\begin{equation}
 \omega_c=\frac{2\pi\lambda}{\beta}
          =\frac{c}{12\phi_{r,\rm{phys}}},
 \qquad
 \tau_c=\omega_c^{-1}=\frac{12\phi_{r,\rm{phys}}}{c}.
 \label{eq:disk_loop_crossover}
\end{equation}
For large $c$, this is a shorter time scale than the time scale $\phi_{r,\rm{phys}}$ where the Schwarzian action on its own becomes highly quantum. This means that for large $c$, the matter contribution suppresses the gravitational fluctuations. So, for large $c$ and infinite bath length,  the theory can remain semiclassical at all energies! For a finite bath length, the theory goes back to the Schwarzian action, which becomes strongly coupled at sufficiently low energies.

It is straightforward to compute the   partition function up to one loop from ~\eqref{eq:sec2_Itot}
\begin{equation}
 Z^{ (1)} (\ell)
 =\mathcal{A} \pi^{-3/2} e^{S_0}  \exp\left( \frac{\phi_r}{2 } + \frac{c\ell}{12}\right)
     \prod_{n=2}^\infty
     \frac{1}{\phi_r n+(c/12)(1-q^n)} \, .
 \label{eq:jtapp_gaussian}
\end{equation}
Let us set $\ell \to \infty $ first to obtain\footnote{Let us give some details. One encounters the product  $\prod_n { n \over n+ \lambda } \propto \Gamma(2 + \lambda) N^{-\lambda}$ with $N \propto \beta/\epsilon_{\rm phys} $. And $N^{-\lambda } = \lambda^{-\lambda } e^{ - \lambda \log( N/\lambda) } $, and the last term is related to the last term in the ground state energy \nref{GSEnS}. }
\begin{equation}
    \begin{aligned}
    \la{GFn}
\left.  Z^{ (1)}(\ell) e^{ - { c \ell \over 12} } \right|_{\ell = \infty} &=  \mathcal{A} e^{S_0}  \exp\left( \frac{\phi_r}{2 }\right) \left(\frac{\phi_r}{\pi}\right)^{3/2} \prod_{n=2}^{\infty } { 1 \over n + \lambda } \\
&= e^{ - \beta E_0}  e^{\tilde S_0} \exp \left( { c \over 24 \lambda } \right)    \lambda^{-3/2} \lambda^{ -\lambda } { \Gamma(2 + \lambda) \over \sqrt{ 2 \pi } } e^{\lambda} \, ,
    \end{aligned}
\end{equation}
where $E_0$ is
\lceq{ \label{GSEnS}
E_0 = { d_1 \over \epsilon_{\rm phys} } + d_2 { c  \over \phi_{r,\rm{phys}}}  + d_3   { c \over \phi_{r,\rm{phys}}}    \log \left(\frac{\phi_{r,\rm{phys}}}{\epsilon_{\rm phys}}\right)  \, ,
}
and it includes UV divergent terms, with $d_i$ being some numerical constants. In the second equality of~\eqref{GFn}, we have included all overall constants into the definition of $\tilde S_0$. The term  $\exp \left( { c \over 24 \lambda } \right)   $ is the classical Schwarzian contribution.  The following terms are the one loop expression, which transitions between its Schwarzian form at small $\lambda $ and  the CFT dominated form at large $\lambda$. This transition occurs for $\lambda \sim 1$. Note that at this transition point the classical term involving $\exp \left( { c \over 24 \lambda } \right)   $ is still large, even though the dynamics of small fluctuations is dominated by the CFT.

\begin{figure}[t]
    \centering
    \includegraphics[width=0.8\linewidth]{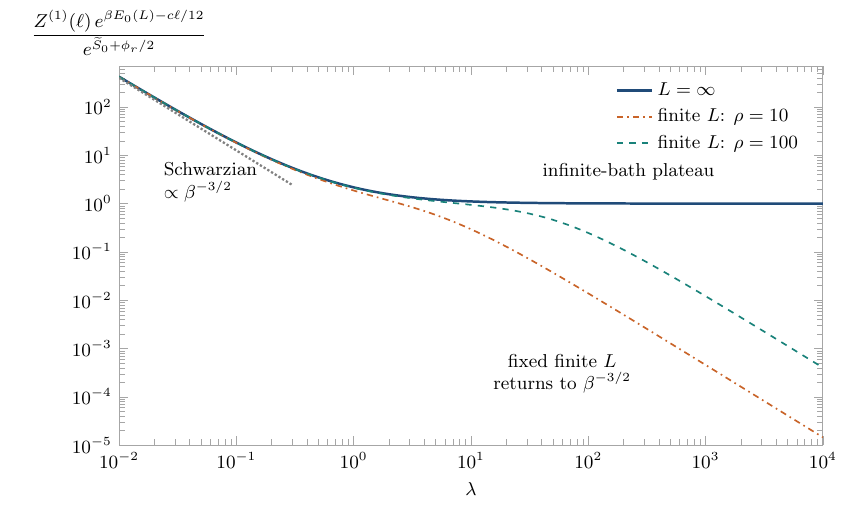}
    \caption{The $\beta$ dependence of the normalized one-loop factor.  The solid line is the infinite-bath result \nref{GFn}, with the classical Schwarzian contribution and ground state energy term $e^{-\beta E_0(L)}$ extracted. The other curves have fixed physical bath lengths such that $\rho =10$ and $\rho=100$.
    At low temperatures the finite-bath curves return to the Schwarzian-like $\beta^{-3/2}$ behavior.  }
    \label{fig:sec2_JTplusCFT}
\end{figure}

At high temperatures, or small $\lambda$,  the first two $\lambda$ dependent terms in~\eqref{GFn} dominate, reproducing the Schwarzian answer, with a subleading correction from the factor $\lambda^{-\lambda}$. As $\lambda $ becomes very large, we find that all the $ \lambda$ dependent terms in~\eqref{GFn}  go to a constant, and we find $e^{- \beta E_0 } e^{ \tilde S_0} $.  ${ \tilde S_0} $ can be viewed as the Affleck--Ludwig  \cite{Affleck:1990by}  boundary entropy of the BCFT.

We can now also write the finite $\ell $ answer as
\be \la{Fin1LExp}
  Z^{  (1)}(\ell)
  =  e^{ - \beta E_0}  e^{\tilde S_0}  \exp\left( \frac{\phi_r}{2 } + \frac{c\ell}{12}\right)   \lambda^{-3/2} \lambda^{ -\lambda } { \Gamma(2 + \lambda) \over \sqrt{ 2 \pi } } e^{\lambda} \prod_{n=2}^\infty { n + \lambda \over n + \lambda (1-q^n) }
\ee
and we have written the last factor so that it is manifestly convergent.  As we vary $\ell$ we see that we can neglect the $q^n$ correction with $\ell \gtrsim 1 $, or $ L \gtrsim \beta/(2\pi) $. As $\ell $ becomes smaller, then for modes satisfying $n \ell \ll 1$ we can expand the bracket in \eqref{eq:sec2_Itot} (or the denominator in \nref{Fin1LExp}) as
\be
 \phi_r \, n + { c \over 12 } (1 - e^{ - 2 n \ell } ) \sim \left( \phi_r + { c\over 6 } \ell \right)  n
 \ee
 which is essentially the same as the Schwarzian answer with a new Schwarzian coupling
\be \la{SchEffC}
 \phi_{r, {\rm eff}, {\rm phys}} = \phi_{r,{\rm phys}} + { c \over 6 }   L ~,~~~~~~~{\rm for } ~~~ L \lesssim \beta/(2\pi )
 \ee
 Note that the exponential in parentheses in \nref{Fin1LExp} can be interpreted as the classical contribution of the Schwarzian with this coupling \nref{SchEffC}, and it includes the thermal energy of the CFT in the bath.  In addition, there is also a shift in the ground state energy
 \be \la{ShiftGS}
 E_0(L) = E_0  - { c \over 24 \pi \phi_{r, {\rm phys}}} \int_0^\infty d x \log \left[{ x+1  \over x + 1 - e^{ -\rho x} }\right], \quad \rho =  \frac{\phi_c}{\phi_r} =  { c L \over 6 \phi_{r, {\rm phys}}}
 \ee
 where $E_0 = E_0(L=\infty)$. See figure \ref{fig:sec2_JTplusCFT} for the $\beta$ scaling behavior with different choices of the length.

\FloatBarrier
\subsection{Two point functions of operators outside}

\label{TwoOutsideSec}

We consider the situation where we insert two identical operators $V(v_i,\bar{v}_i)$ ($i=1,2$) of conformal dimension $(h,\bar{h})$ in the bath and we compute
\begin{figure}[t]
  \centering
  \includegraphics[width=0.95\linewidth]{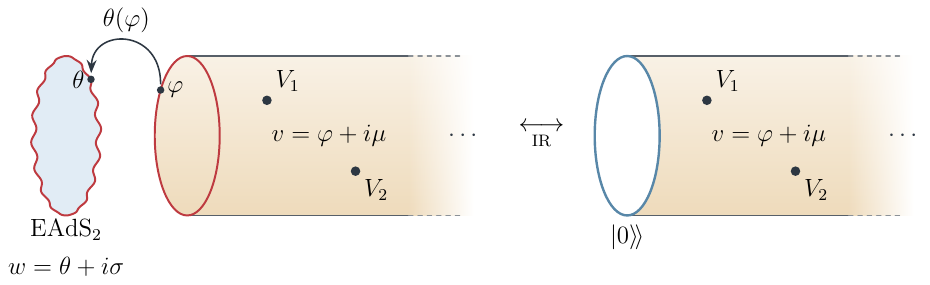}
  \caption{The infrared correspondence between the gravitational disk
  path integral with two bath insertions $V_i=V(v_i,\bar v_i)$ (left panel) and  the
  bath CFT with the vacuum Ishibashi state $|0\rangle\!\rangle$ at its
  boundary (right panel). Both bath cylinders extend to $\mu\to\infty$ on the right.
  }
  \label{fig:ishibashi-correspondence}
\end{figure}

\be  \la{TwoPt}
 \langle V(v_1, \bar v_1) V( v_2, \bar v_2) \rangle
\ee
where the expectation values include the sum over reparametrizations. We ignore the backreaction of these operator insertions and treat these operators as a probe of the system.

On the flat reference cylinder $z$~\eqref{CFP}, the correlation function of two identical operators is fixed by conformal symmetry
\longeq{
\langle V(z_1, &\bar z_1) V( z_2, \bar z_2) \rangle_z = \frac{1}{\left( 2 \sin \frac{z_{12}}{2}\right)^{2h} \left( 2 \sin \frac{\bar{z}_{12}}{2}\right)^{2 \bar{h}}}  \, ,
}
where $z_{ij} := z_i- z_j$,  and the conjugate is defined similarly. We can use the conformal map $z_i = F(v_i)$ to translate the expression back to the original coordinate. This introduces $\theta$ dependence implicitly, and what we actually compute is the functional integral over $\theta$
\longeq{
\label{eq:sec2_G}
\mathcal{G}(v_1,v_2) = \frac{\int d \mu [\theta] e^{-I[\theta]  }\mathcal{G}_F (v_1,v_2) }{\int d \mu [\theta] e^{-I[\theta]}},
}
where the action $I[\theta]$ is given by~\eqref{ActDisk} and
\lceq{
\label{eq:sec2_GF}
\mathcal{G}_F (v_1,v_2) := \left[\frac{F'(v_1) F'(v_2)}{4 \sin^2 \frac{F(v_1) - F(v_2)}{2}}\right]^{h} \left[\frac{\bar{F}'(\bar{v}_1) \bar{F}'(\bar{v}_2)}{4 \sin^2 \frac{\bar{F}(\bar{v}_1) - \bar{F}(\bar{v}_2)}{2}}\right]^{\bar{h}}  \, .
}
In general, it seems difficult to derive the full path integral~\eqref{eq:sec2_G}. Fortunately, for large $c$ we can perturbatively expand around the saddle and obtain corrections
\lceq{
\mathcal{G}(v_1,v_2) = \mathcal{G}^{(0)} (v_1,v_2) \left[1 +   \mathcal{G}^{(1)}(v_1,v_2) + O(c^{-2})\right] \, ,
}
The zeroth-order term $\mathcal G^{(0)}$ is evaluated at the saddle $\theta=\varphi$ and $F(v)=v$:
\lceq{
\label{eq:sec2_G0}
\mathcal{G}^{(0)}(v_1,v_2) = \frac{1}{\left( 2 \sin \frac{v_{12}}{2}\right)^{2h} \left( 2 \sin \frac{\bar{v}_{12}}{2}\right)^{2 \bar{h}}} \, .
}
To compute the correction at order $1/c$, we expand both the action and $\mathcal{G}_F$ to quadratic order in $\vare$. The resulting Gaussian integral is evaluated by Wick contracting the fluctuations using the propagator~\eqref{eq:sec2_varesq}. We expand functions $F$ and $\bar F$ in~\eqref{eq:sec2_GF} to linear order using ~\eqref{eq:sec2_Fourier}
\longeq{
\left[ 4 \sin^2\left(\frac{v_{12}}{2}\right) \right]^{ h}&\left[\frac{F'(v_1) F'(v_2)}{4 \sin^2 \frac{F(v_1) - F(v_2)}{2}}\right]^{h} \\
&= 1 + h \left[f'(v_1) + f'(v_2) - \left(f(v_1) - f(v_2) \right)\cot \frac{v_{12}}{2}\right] + O(f^2)
}
where we extracted the $\sin^{-2h}(v_{12}/2)$ factor in $\mathcal{G}^{(0)}$.  This particular combination of $f(v_i)$ and $f'(v_i)$ is actually fixed by the $\rm PSL(2,\R)$ symmetry. It is the unique linear combination that is invariant under the shift
\lceq{
f(v) \mapsto f(v) + a_0 + a_{-1} e^{-iv} + a_1 e^{i v} \, .
}
 We drop the $O(f^2)$ terms because they are made up of Fourier modes of a single sign and therefore will vanish when integrated around the circle.  Using the solution of the conformal welding problem~\eqref{gnvals} for $\ell = \infty $
we have
\lceq{
f'(v_1) + f'(v_2) - \left(f(v_1) - f(v_2) \right)\cot \frac{v_{12}}{2} = \sum_{n=2}^\infty \vare_n \mathcal{F}_n(v_1,v_2), \\
\mathcal{F}_n(v_1,v_2):= in \left(e^{in v_1} + e^{in v_2}\right) - \left(e^{in v_1} - e^{in v_2}\right) \cot \frac{v_{12}}{2} \, .
}
Combined with the antiholomorphic part, one can conclude the order $1/c$ correction is
\longeq{
\label{eq:sec2_G1}
\mathcal{G}^{(1)}(v_1, v_2) &= h \bar{h} \sum_{n=2}^\infty \langle \vare_n \vare^*_n \rangle  \mathcal{F}_n(v_1,v_2) \overline{\mathcal{F}_n(v_1,v_2)} = \frac{12h \bar{h}}{c} \sum_{n=2}^\infty \frac{  \mathcal{F}_n(v_1,v_2) \overline{\mathcal{F}_n(v_1,v_2)}}{n(n^2-1)} \\
&= \frac{12 h\bar{h}}{c} \left[-2 + \left(1 -\frac{2}{\chi}\right)\log (1-\chi)\right]= \frac{2 h \bar{h}}{c}\chi^2 {}_2 F_1 \left(2,2;4;\chi\right) \, ,
}
where ${}_2F_1(a,b;c;z)$ is the hypergeometric function and the whole expression is written only in terms of the conformal cross ratio $\chi$
\lceq{ \la{ChiDe}
\chi := { \sin({ v_1 - v_2 \over 2}) \sin ({ \bar v_1 - \bar v_2 \over 2 }) \over \sin({ v_1 - \bar v_2 \over 2} )\sin ({  \bar v_1 -  v_2 \over 2 }) }.
}
For the operators inserted in the bath, $\mu_1, \mu_2 > 0$, we have $0<\chi<1$ so $\log (1-\chi)$ is real without any branch ambiguity.

The same framework extends to multipoint correlators of bath operators. As shown in appendix~\ref{app:SDeq}, combining the CFT Ward identity with the Schwinger--Dyson equation for the reparametrization integral implies that the gluing condition~\eqref{IshCon} holds inside these correlators after integrating over $\theta$. This provides a further consistency check of the boundary-state interpretation.

\begin{figure}[t]
  \centering
  \includegraphics[width=0.8\linewidth]{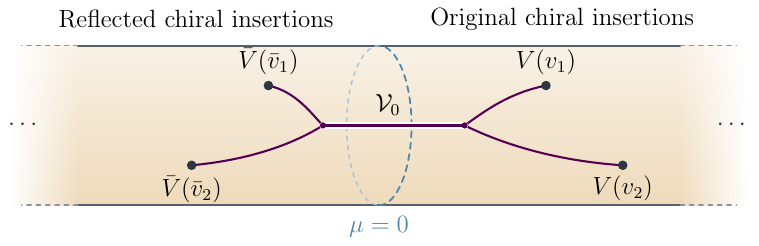}
  \caption{The chiral insertions at $v_i=\varphi_i+i\mu_i$ and their reflected images at $\bar v_i=\varphi_i-i\mu_i$. The graph depicts the vacuum block in the $V(v_1)V(v_2)$ channel. $\mathcal V_0$ includes the identity and its Virasoro descendants. The dashed circle marks the original boundary.}
  \label{fig:doubled-vacuum-block}
\end{figure}

\paragraph{The correlators in the presence of the Ishibashi state.} To test the vacuum Ishibashi state proposal, we compare the gravitational result \nref{eq:sec2_G1} with the correlator shown in the right panel of figure~\ref{fig:ishibashi-correspondence}
\lceq{
\mathcal G_{\rm Ishi}(v_1,v_2)
=  \langle 0|V(v_1,\bar v_1)V(v_2,\bar v_2)|0 \rangle\rangle
\, .
}
The infinite sum over descendants can be organized using the doubling trick, shown in  figure~\ref{fig:doubled-vacuum-block}.
In our cylinder coordinates, reflection across the interface sends $v_i$ to $\bar v_i$, allowing the antiholomorphic insertions to be represented as chiral insertions at the reflected points. Denoting the corresponding chiral fields by $V_h$ and $V_{\bar h}$, we obtain
\lceq{\la{IdChFa}
\frac{\mathcal G_{\rm Ishi}}{\mathcal G^{(0)}}
=
\frac{
\big\langle V_h(v_1)V_h(v_2)\,
\mathcal P_0\,
V_{\bar h}(\bar v_2)V_{\bar h}(\bar v_1)
\big\rangle_{\rm ch,cyl}
}{
\big\langle V_h(v_1)V_h(v_2)\big\rangle_{\rm ch,cyl}
\big\langle V_{\bar h}(\bar v_2)V_{\bar h}(\bar v_1)\big\rangle_{\rm ch,cyl}
}
=
\mathcal{K}_0(\chi)\, .
}
Here $\mathcal P_0$ projects onto the irreducible Virasoro vacuum module, it can be viewed simply as \eqref{Ishiexp} for the identity, with one of the kets replaced by a bra.  And $\mathcal{K}_0(\chi)$ is the normalized vacuum conformal block with external weights $(h,h,\bar h,\bar h)$. The projector specifies that the two   holomorphic insertions fuse into the identity and its descendants. The antiholomorphic insertions are similarly contracted into the antiholomorphic descendants of the identity. The vacuum Ishibashi state (the projector $\mathcal P_0$) acts as a mirror for energy, or a mirror within the descendants of the identity, and reflects them, turning holomorphic products of stress tensors into antiholomorphic ones.  Of course, we are getting just the identity because only the identity and its descendants can be absorbed by the gravity disk. Note that the operator $V$ does not have to factorize into well defined local operators $V_h$ and $V_{\bar h}$; the expression \nref{IdChFa} still makes sense.

The identity Virasoro conformal block has been studied in  ~\cite{Perlmutter:2015iya, Bombini:2018jrg}. It can be written as a sum over global conformal blocks
\lceq{
\mathcal{K}_0 (\chi) = 1 + \sum_{\substack{Q \, \, \text{quasiprimary} \\ Q \in \text{vacuum module}, Q\neq 1}
} \frac{C_{h h Q} C_{\bar{h}\bar{h} Q}}{\langle Q|Q \rangle} \chi^{h_Q} {}_2 F_1 \left(h_Q, h_Q ; 2 h_Q ; \chi\right),
}
For our purposes, it is enough to perturbatively expand to the order $1/c$ ~\footnote{One can also get this result directly by expanding the vacuum Ishibashi state via~\eqref{eq:sec2_BH_vac}.}
\lceq{ \la{LeadCo}
\mathcal{K}_0 (\chi) = 1 + \frac{2 h \bar{h}}{c} \chi^2 \, {}_2 F_1 (2,2;4;\chi) + O\left(c^{-2}\right).
}
The subleading contribution exactly matches the result~\eqref{eq:sec2_G1} derived from the gravitational path integral.

The leading order term in the small $\chi$ expansion of the full conformal block $\mathcal{K}_0 $ is the same as the one in \nref{LeadCo}, it goes as
\lceq{
\mathcal{K}_0(\chi) \sim 1 + \frac{2 h \bar{h}}{c} \chi^2 + \cdots , \quad ~~~{\rm as}~~~~\chi \rightarrow 0
}
which is the contribution from $T \bar T$.

We can also consider the OPE in the other channel,
as $v_1 \to \bar v_1$, which is when the operator approaches the boundary between the bath and the gravity region. In this case,
$\chi \to 1^-$. We  get a  term $\log (1-\chi) $   as \(\chi\to1^{-}\).
The logarithm can be interpreted as a first term in an expansion of an anomalous dimension. Indeed,  at finite $c> 1$, and for relatively small external dimensions $h$, $\bar h$, we find a power law behavior \cite{Collier:2018exn}. More precisely, introduce
\lceq{
\label{eq:sec2_cQ}
c=1+6Q^2,\qquad
h=\alpha(Q-\alpha),\qquad
\bar h=\bar\alpha(Q-\bar\alpha),
}
where the Liouville momenta $\alpha$ and $\bar{\alpha}$ are the smaller real roots. For sufficiently light external operators, $\alpha + \bar{\alpha}<Q/2$, the Virasoro conformal block gives the exact leading asymptotic behavior \cite{Collier:2018exn}
\lceq{
\label{eq:sec2_K0_div}
\mathcal K_0(\chi)
\sim  \mathcal{U}(c,h,\bar h)(1-\chi)^{-2\alpha\bar\alpha} = 1 - \frac{24h \bar{h}}{c} - \frac{12 h \bar{h}}{c} \log(1-\chi) + O (c^{-2}),
}
where we expanded for large $c $ to reproduce the logarithm we had in \nref{eq:sec2_G1}.
Note that $-2 \alpha \bar{\alpha} = - \frac{12 h \bar{h}}{c} + O(c^{-2})$ can be interpreted as the anomalous dimension of the boundary operator which is obtained by fusing $V_h(v)$ with its partner $\bar{V}_{\bar{h}}(\bar{v})$.
In general, bringing a bulk operator to the boundary we expect
\be
 V(\varphi, \mu ) \sim { \mu^{ -h - \bar h  +\Delta_{i} } } ( O_i(\varphi) + {\rm descendants} ) ~,~~~{\rm as } ~~\mu \to 0
\ee
where $O_i$ are the boundary operators. Here we see that there is a boundary operator with dimension
\be \la{AnomHHbar}
\Delta = h + \bar h - { 12 h \bar h \over c } + O(c^{-2})
\ee
The first term is the obvious total dimension of the bulk operator. The second term can be viewed as a gravitational correction to the dimension of the operator. This is present even if the operator factorizes into a product of two operators. For large $c$, we expect to trust this result.

Note that  \nref{eq:sec2_K0_div} is just the leading order behavior, there are  other subleading powers\footnote{The subleading powers correspond to dimensions $\Delta_m = h + \bar h + m - { 6 \over c} [ 2 h \bar h + 2 m (h + \bar h ) + m (m-1) ]$. The $h+\bar h+m$ part corresponds to the naive boundary expansion of $V(\varphi,\mu)$, while the term proportional to $1/c$ is a gravitational correction. } as well as a term with a logarithm that starts at dimension $h_{\rm int} = (c-1)/24$ of the form $ (1-\chi)^{ (c-1)/24 - h - \bar h} ( - \log (1-\chi) )^{-3/2} $ \cite{Collier:2018exn}.
For similar computations in $\AdS_3$ we can think of the discrete states as bound states of two particles, and the continuum as a black hole threshold. In our case, we expect that the discrete states are again a bound state of the two operators with the boundary graviton. We do not know whether the continuum has an interesting interpretation. Of course, one would assume that by summing over all gravitational topologies, the final answer should be interpreted as an average over local boundary conditions. For each particular boundary condition, there is only a discrete set of terms, with only powers, that appear in this crossed OPE channel. However, as a result of the average, one could, in principle, obtain logarithms or other functions.  On the other hand, if the full UV completion involves a string theory or gravity theory that corresponds to a single boundary condition, then by considering that exact theory we should recover the discrete powers.

Here, we will discuss aspects of the OPE limit as the operators approach the boundary.  This corresponds to operators acting at the boundary of the $\AdS_2$ region. It is particularly interesting to consider the situation where the operators approach the boundary and are separated in Lorentzian time. This is interesting because we expect that this correlator displays quasinormal mode behavior. For a fixed distance $\mu=\mu_1= \mu_2$ from the boundary and a time interval $t = t_{12} \gg 1 \gg \mu $
\be
(\chi -1 ) \propto  { \mu^2 e^{ - t }}~,~~~~~~~~~~ t = { 2 \pi \hat t_{\rm phys} \over \beta }
\ee
where $\hat t_{\rm phys}$  is the physical time.
And the correlator then behaves as
\be
\langle V_2 V_1 \rangle \sim e^{ - (h + \bar h ) t }  ( \chi-1)^{ -2 \alpha \bar \alpha } \sim \exp \left( - ( h + \bar h - { 12 h \bar h /c } +{\cal O}(1/c^2)  ) t \right)
\ee
where we used \nref{eq:sec2_K0_div}.\footnote{ In order to get a more careful result, including the overall phase,  it is necessary to keep track of the path that the cross ratio $\chi$ follows in the complex plane after the appropriate small displacements into Euclidean time are introduced.}
We see that to leading order in $c$ we have the expected quasinormal mode behavior for a conformal operator of dimension $\Delta = h + \bar h$, and the $1/c$ correction resembles a gravitational interaction mediated by the boundary graviton.

\paragraph{$T\bar T$ and a mirror for energy.}
The mixed stress-tensor correlator provides a direct diagnostic of energy reflection. At disk topology, the Ishibashi condition~\eqref{IshCon} identifies $T$ and $\bar T$ at the interface. The doubling trick therefore relates the mixed correlator to a holomorphic two-point function evaluated at the reflected point. In the bath cylinder coordinates, $v=\varphi+i\mu$, this gives
\be
\langle T(v_1)\bar T(\bar v_2)\rangle_{\rm conn}
=
\frac{c}{32\sin^4\!\left(\frac{v_1-\bar v_2}{2}\right)}.
\ee
Here the connected correlator subtracts the product of the
stress-tensor one-point functions. Continuing $\varphi=it$ and
defining $t_\pm=t\pm\mu$, we have $v=it_+$ and $\bar v=it_-$.
With our normalization $T(v)=-2\pi T_{vv}^{\rm E}$, the physical
Lorentzian components are
\be
T_{++}(t_+)=\frac{1}{2\pi}T(it_+),
\qquad
T_{--}(t_-)=\frac{1}{2\pi}\bar T(it_-).
\ee
Consequently, the Wightman correlator is
\be
\langle T_{++}(t_{1+})T_{--}(t_{2-})\rangle_{\rm conn}
=
\frac{c}{128\pi^2
\sinh^4\!\left(\frac{t_{1+}-t_{2-}-i\epsilon}{2}\right)},
\qquad \epsilon>0.
\ee
The singularity at $t_{1+}=t_{2-}$ follows the reflected null ray,
expressing the mirror behavior of energy.
\begin{figure}[!htbp]
 \centering
 \includegraphics[width=0.5\linewidth]{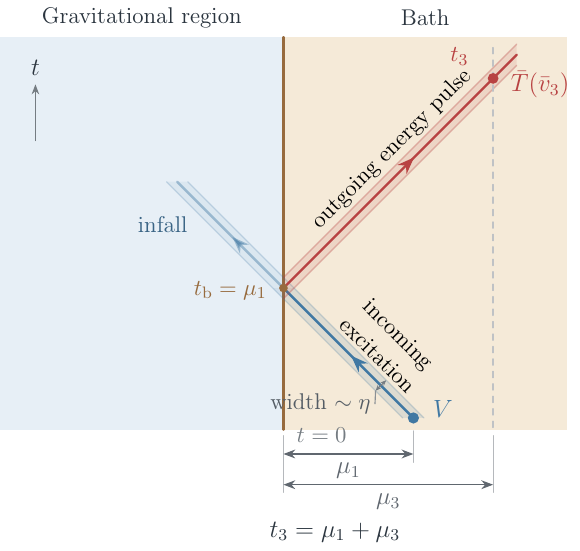}
 \caption{An incoming chiral excitation and the outgoing stress-tensor
 response. The excitation is centered at $t=0$, $\mu=\mu_1$.
 Its incoming null trajectory has $t+\mu=\mu_1$, while the outgoing
 response is centered on $t-\mu=\mu_1$. A probe at $\mu_3$ therefore
 sees the peak at $t_3=\mu_1+\mu_3$. The blue continuation into the
 gravitational region emphasizes that the diagram does not assert
 reflection of the matter operator itself. The ribbons schematically indicate the width
 controlled by the Euclidean regulator $\eta$.}
 \label{fig:chiral-stress-reflection}
\end{figure}

\paragraph{The outgoing pulse of energy.}

It is instructive to consider the special case of two holomorphic (or purely left moving) operators and a fluctuation of the right moving stress tensor over its background value $\delta \bar T = \bar T - { c \over 24 }  $. Before integrating over fluctuations, this vanishes. The integral over fluctuations leads to a reflective boundary condition for the stress tensor. Our aim here is to compute a correlation showing this. For this purpose, we imagine considering a state given by the action of a left-moving operator $V(v)$.

We will not do the computation explicitly, but we will simply use the $\SL(2,\R)$ symmetry to directly write the expected answer after integrating over reparametrizations
\be \la{CorrStre}
\langle V^\dagger(v_2) \delta \bar T(\bar v_3) V(v_1)  \rangle  = \frac{C_V}{4^{h+1}} { h \over  \sin^{ 2 h -2}( { v_2 - v_1 \over 2} ) \sin^{2} ( { v_1 - \bar v_3 \over 2 } ) \sin^2({v_2 - \bar v_3 \over 2 } ) }
\ee
where we subtracted the constant part, $c/24$, of the stress tensor and used that the remaining operator transforms as a primary under $\SL(2,\R)$ transformations that preserve the position of the boundary.   The overall coefficient is fixed separately by the stress tensor Ward identity, together with the normalization of the two point function, which fixes $C_V$.

An example is a free chiral Majorana fermion, $V =\psi $, with $h=1/2$. To create the incoming excitation, we consider fermionic insertions at $v_1 = - \eta + i \mu_1 $. Here we have set the Lorentzian time to zero and added a small  Euclidean time displacement $\eta$ in order to get a normalizable state. The same state in the bra leads us to set   $v_2 =  \eta + i \mu_1 $.  We consider the right moving stress tensor at a general Lorentzian position   $\bar v_3 = i t_3 - i \mu_3 $. Then~\eqref{CorrStre} is continued to
\be
 \la{CorrStrese}
\langle \psi(\eta + i \mu_1)  \delta \bar T(i(t_3-\mu_3))   \psi(-\eta + i \mu_1) \rangle  =   \frac{C_V}{16}{ \sin \eta \over  \left| \sinh (  { t_3 -\mu_3 - \mu_1 + i \eta \over 2 } ) \right|^{4}  }  \, .
\ee
Dividing by the two point function $\langle\psi_2\psi_1\rangle=C_V/(2\sin\eta)$ gives the outgoing pulse
\begin{equation}
 \delta T_{--}   = \frac{\sin^2\eta}{4 \pi (\cosh u-\cos\eta)^2
} \simeq \frac{\eta^2}{\pi (u^2 + \eta^2)^2} ~,~~~~~~~~u = t_3 - \mu_3 - \mu_1
\end{equation}
where we take the limit $|u|,\eta\ll1$. This is a positive outgoing energy pulse of width $O(\eta)$, centered at $t_3=\mu_1+\mu_3$. The incoming excitation reaches the interface after time $\mu_1$, and the outgoing response reaches the detector a further time
$\mu_3$ later, see figure~\ref{fig:chiral-stress-reflection}. We can interpret the outgoing response as the Hawking radiation coming from the region where the fermion crosses the boundary.
 The outgoing excess energy, integrated over the entire pulse,
equals the initial excess energy of the state above the thermal
background.

\paragraph{Lorentzian continuation of the two point function.}

\begin{figure}[t]
  \centering
  \includegraphics{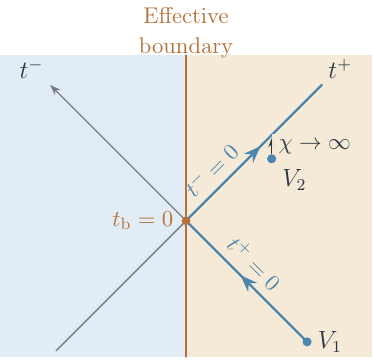}
  \caption{ We consider the two point function in Lorentzian signature. It has a singularity when the operator $V_2$ approaches the reflected light ray emanating from the first operator, denoted here by the solid blue line.
 }
  \label{fig:lorentzian-reflection}
\end{figure}

Here we discuss the Lorentzian continuation of the two point function of operators in the bath region \nref{TwoPt},
see figure~\ref{fig:lorentzian-reflection}.
Starting with the coordinate $v_j = \varphi_j + i\mu_j$, $j=1,2$,  we can analytically continue the Euclidean time to Lorentzian time $t$ via $\varphi_j \to \hat\epsilon_j +  i t_j $ where $\hat\epsilon_j$ is a small number to keep track of the order of the operator. Then $v_j$ and $\bar{v}_j$ can be written as
\lceq{
v_j = \hat \epsilon_j + i t_j^+, \quad \bar{v}_j = \hat \epsilon_j + i t_j^-, \quad t_j^\pm  = t_j \pm \mu_j \, ,
}
For a bath event $(t_j,\mu_j)$, $t_j^+$ is the time an incoming light ray reaches the interface, while $t_j^-$ is the time an outgoing light ray reaching that event left the interface.   Setting $\hat \epsilon = \hat \epsilon_2 - \hat \epsilon_1 >0$, the conformal cross ratio is given by
\lceq{
\chi = { \sinh \left({ t_{21}^+ -i \hat \epsilon \over 2} \right)\sinh \left({ t_{21}^- - i \hat \epsilon \over 2} \right) \over\sinh \left({ t_2^+ - t_1^- - i  \hat \epsilon \over 2} \right)\sinh \left({ t_2^- - t_1^+ - i  \hat \epsilon \over 2} \right)  } ~~~~~~~t_{21}^\pm = t_2^\pm - t_1^\pm
}
Take $t_2 > t_1$, one interesting point is that when these two operators are connected via a reflected ray (shown in figure~\ref{fig:lorentzian-reflection}). The operator $V_2$ approaches to the null light cone of the image operator $\tilde{V}_1$. It is described by the following condition
\lceq{
t_2^- = t_1^+ \qquad \Longleftrightarrow \qquad t_2 - t_1 = \mu_1 + \mu_2 ,
}
The conformal cross ratio $\chi$ approaches $\infty$. More precisely, denoting $\delta t  = t_1^+ - t_2^-$,
\lceq{
\chi  = - \frac{1}{M_{12} (\delta t + i  \hat  \epsilon)} + O(1) \rightarrow \infty, \quad M_{12} = \frac{\sinh (\mu_1+\mu_2)}{2\sinh \mu_1 \sinh \mu_2} \, ,
}
and the correction $\mathcal{G}^{(1)}$ of the two point correlators receives a logarithmic divergence
\lceq{
\label{eq:sec2_G1_log_div}
\mathcal{G}^{(1)} = - \frac{12 h \bar{h}}{c} \left[2  + \log \left(M_{12} (\delta t + i  \hat \epsilon)\right)\right] + \cdots
}
This singular behavior is the same as the one we obtained by moving one Euclidean insertion toward the interface as shown in~\eqref{eq:sec2_K0_div}.  It is explained by the symmetry
$\mathcal{K}_0(\chi) = \mathcal{K}_0( { \chi \over \chi - 1 } ) $ of the conformal block, which is due to the fact that the two operators have the same $(h,\bar h)$ dimensions.

Such a lightcone singularity is expected in general BCFTs, and it is closely related to the Regge limit
\cite{Mazac:2018biw}.

\subsection{Out of time order correlator }
\label{subsubsec:bdycorr}

Another interesting case is the out of time order correlator or OTOC. In this case, we think of the operators in Lorentzian time but with small $i\epsilon $ deformations that put the operators in an out of time order configuration. More precisely, it is convenient to consider the special case where the operator $V(v, \bar v) $  can be factorized as $V(v) W(\bar v)$. This can be the case if the CFT consists of free fermions, for example. In this case, the two point function becomes really a four point function and we use the result in \nref{eq:sec2_G1} to write the out of time order correlator of the form $\langle V(0) W(t) V(0) W(t) \rangle $.

 Introducing a small (but not too small)  Euclidean displacement, $\eta$, we can choose the points
\begin{equation}
 v_1=-\eta+i\mu,\qquad
 \bar v_1=it-i\mu,\qquad
 v_2=\eta+i\mu,\qquad
 \bar v_2=2\eta+it-i\mu,
 \label{eq:otoc_points}
\end{equation}
where $\mu>0$ and $0<\eta\ll1$. The Euclidean regulators
alternate between the barred and unbarred insertions, specifying
the out-of-time ordering. $\eta$ can be viewed as a small displacement in Euclidean time to make the states created by the operators normalizable. We continue $t$ from zero along the
positive real axis. Note that  $t=2\pi t_{\rm phys}/\beta$.

Using the cross ratio \nref{ChiDe}
we obtain
\begin{equation}
 \chi(t)=
 \frac{\sin^2\eta}
 {\sin[3\eta/2+i(t-2\mu)/2]\,
  \sin[\eta/2-i(t-2\mu)/2]}.
 \label{eq:otoc_chi}
\end{equation}
The normalized correlator at leading order in $1/c$ is
\begin{equation}
 \mathcal K_0(\chi)
 =1+\frac{12h\bar h}{c}
 \left[-2+\left(1-\frac2\chi\right)\log(1-\chi)\right]
 +O(c^{-2}).
 \label{eq:otoc_block}
\end{equation}
The logarithm must be followed continuously along the contour obtained by increasing $t$ continuously~\cite{Roberts:2014ifa}.
As we do this, we start from $\chi \sim 0$ and then  $\chi$ passes counterclockwise around $\chi =1$ and returns to $\chi \sim 0$, but after picking up a $  2 \pi i $ from the logarithm in \nref{eq:otoc_block}. By the way, notice that it crosses the real axis at $\chi> 1$ when $t= 2\mu$.  This is the reflected lightcone: the
propagation time to the interface and back is $2\mu$.

At late time the cross ratio scales as $\chi \propto e^{ - t } $ and the extra $2 \pi i$ from the logarithm gives a growing contribution to the OTOC
\begin{equation}
 \Delta\mathcal K_0
 =\frac{24\pi i h\bar h}{c}
       \left(1-\frac2\chi\right).
\end{equation}
 \begin{equation}
 \mathcal K_0^{\rm OTOC}(t)-1
 \simeq
 -\frac{12\pi i h\bar h}{c\sin^2\eta}\,
       e^{\,t-2\mu-i\eta} ~,~~~~~~~ {\rm for} ~~~t-2\mu\gg1
 \label{eq:otoc_growth}
\end{equation}
The magnitude of the connected correction grows
as $e^{2\pi t_{\rm phys}/\beta}$, giving the maximal
Lyapunov exponent $\lambda_L=2\pi/\beta$, even though the
boundary dynamics is induced entirely by the CFT.
This result applies in the perturbative window
\begin{equation}
\la{Scra}
 1\ll e^{t-2\mu}
 \ll \frac{c\sin^2\eta}{12\pi h\bar h},
\end{equation}
for fixed positive $h,\bar h$. Once the correction becomes
order one, higher orders in $1/c$ must be included. Of course, here $1/c$ is playing the role of $G_N$ and the upper bound in \nref{Scra} defines the scrambling time, up to the usual delay coming from the propagation in space which involves $\mu$.

This computation is very similar to the one discussed for a CFT in \cite{Roberts:2014ifa}, since it uses the identity block of the CFT, but the physical interpretation is different\footnote{   \cite{Roberts:2014ifa} considers a large $c$ CFT with a sparse spectrum, which justified their restriction to the identity block. Here, the large $S_0$ limit restricts us to the disk, which leads us to the identity block.}. In addition, this computation is essentially identical to the computation of the OTOC on the string worldsheet
\cite{Giombi:2022pas, Giombi:2023zte}, since the effective action for the reparametrization mode is the same.

\subsection{Deforming the CFT in the gravity region}

\label{sec:bulk_pri}

\begin{figure}[h]
  \centering
  \includegraphics[width=0.95\linewidth]{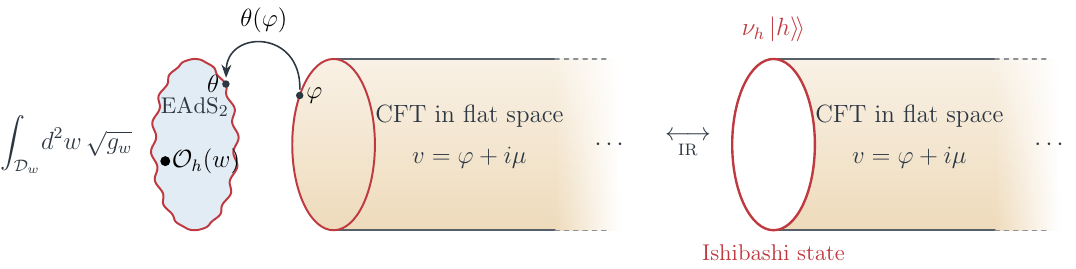}
  \caption{The proposal of the deformation~\eqref{eq:bulk_primary_deformation}. Integrating the positions of the operators over the gravitational region gives the Ishibashi states determined by the primary operators. }
  \label{fig:bulk-primary-ishibashi}
\end{figure}

It is natural to wonder whether we can modify the computation so that a different Ishibashi state, rather than the vacuum, would arise.
A natural idea is to add a deformation supported only in the gravitational region,
\begin{equation}
 \delta I=\lambda\int_{\mathcal D_w}d^2w\,\sqrt{g_w}\,
                 \mathcal O_h(w, \bar{w}),
 \label{eq:bulk_primary_deformation}
\end{equation}
where $\mathcal O_h$ is a scalar primary of conformal dimension $(h,h)$, regularized using the physical metric in $\AdS_2$.  In addition to neglecting the JT Schwarzian action, we work to first order in $\lambda$ and take the bath to be infinite. Figure~\ref{fig:bulk-primary-ishibashi} illustrates the proposal. For simplicity, choose a Hermitian primary with unit two-point normalization. We assume that  $h>1/2$, so the integral we perform below converges
near the $\AdS_2$ boundary.

An insertion at the disk center prepares the highest-weight state $|h, h \rangle$. Moving it away from the center generates the global descendant $L_{-1}^j \bar{L}_{-1}^k |h,h\rangle$. Boundary reparametrizations act through the Virasoro generators $L_n$ and $\bar L_n$ as described by the operator $U[\theta]$~\eqref{UOperator}, which generate further descendants while preserving the same Virasoro family. Moreover, integrating over its position covariantly preserves the $\SL(2)$ isometry of the disk and leaves the exterior sewing argument \eqref{eq:sec2_BH}-\eqref{IshCon} unchanged. The same arguments presented above imply that we now have the $h$-Ishibashi state, $|h\rangle \rangle $.
Thus, factoring out the common undeformed disk normalization, we have
\begin{equation}
 |{\rm BH}\rangle_{\rm disk,  \lambda}
 = e^{\tilde{S}_0}
 \left[
 |0\rangle\rangle-\lambda \nu_h|h\rangle\rangle
 +O(\lambda^2)
 \right].
 \label{eq:deformed_disk_state}
\end{equation}
 The minus sign follows from expanding $e^{-\delta I}$. In the notation $|{\rm BH}\rangle=\sum_i\alpha_i|h_i\rangle\rangle$ of section~\ref{sec:introduction}, the deformation thus gives $\alpha_i=-\lambda e^{\tilde S_0}\nu_h$ for $h_i=h$.

It is possible to perform a few explicit checks, whose details are in appendix~\ref{app:deform}. First, we can consider the computation of the one point function of the operator ${\cal O}_h$ in the bath. At leading order in the $1/c$ expansion, this is given by the two point function involving an insertion inside the disk and one outside the disk, integrated over the disk. We get the expected answer for the one point function in a BCFT as determined by the $\SL(2,\R)$ symmetry. This was done in \cite{Goto2016Causal}\footnote{For $h\leq 1/2$ the integral over the disk diverges near the boundary of the disk and we will need to include some operators at the boundary to cancel it. }. One can also consider a two point function of generic operators outside. In that case, we expect that the $|h\rangle \rangle $ Ishibashi state should give us an answer in terms of the conformal block, as in section \ref{TwoOutsideSec}, but now with the Virasoro state $h$ as intermediate state. It is possible to check this explicitly to order $1/c$ by performing the integral over the reparametrizations.

\paragraph{Deformation with descendants.} In addition to deforming the CFT by a Virasoro primary, as~\eqref{eq:bulk_primary_deformation}, we can also add a deformation by a Virasoro descendant and $\SL(2,\R)$ primary (appropriately including the Weyl factor of the $\AdS_2$ metric). Of course, the integral over $\AdS_2$ preserves the $\SL(2,\R)$ symmetry,
and the argument at the beginning of section \ref{subsec:Ishibashi} is still applicable. So we conclude that the state prepared by the descendant, after performing the integral over boundary reparametrizations, continues to give an $h$-Ishibashi state, up to an overall normalization.
In particular, we could insert $T\bar T $ in the disk, which is the $L_{-2} \bar L_{-2} $ descendant of the identity. This is a primary under $\SL(2,\R)^2$, and we can integrate over the gravitational region. It is certainly a valid irrelevant deformation of the theory. However, we claim that the addition of this operator does not give anything new. It simply gives the same identity Ishibashi state, but with a different overall coefficient. The reason is intuitively clear: the state propagating in the cylinder is still in the identity multiplet, so that when we integrate over the boundary wiggles, we get a projection again to the Ishibashi state.

As a simple illustration of this, we consider the insertion of a $T\bar{T}$ operator in the bulk action. Expanding to first order in the coupling deforms the state by
\begin{align}
    \int_{|z|<1} {d^2 z  (1 - |z|^2)^2}T(z) \bar{T}(\bar{z})|0 \rangle = 2 \pi \sum_{k \ge 2} {L_{-k} \bar{L}_{-k} \over k(k^2 - 1)}|0 \rangle
\end{align}
where the stress tensor is in the flat frame. The task is to show that
\begin{align} \la{OrigAc}
    \delta_\lambda | \mathrm{BH} \rangle = -\sum_{k \ge 2}  {2 \pi \lambda \over k(k^2 - 1)}\int d\mu[\theta] U^\dagger[\theta]  {L_{-k} \bar{L}_{-k} }|0 \rangle \propto \int d\mu[\theta] U^\dagger[\theta] |0 \rangle.
\end{align}
The trick is to write
\begin{align}
    \int d\mu[\theta] \left( U^\dagger[\theta]  L_{-k}U[\theta] \right) \left( U^\dagger[\theta] \bar{L}_{-k} U[\theta] \right)U^\dagger[\theta]|0 \rangle \label{eq:deformedstateU}
\end{align}
and show that the factors in the parentheses can be replaced by c-numbers after performing the $\theta$ integration and we consider them acting on the state on the right. This is simplest to see in the large $c$ limit. In this limit we can approximate the $U^\dagger$ operator as
\begin{align}
    U^{\dagger} = \exp\left[ {- i \sum_{n} \left( \varepsilon_n L_n - \varepsilon_{-n}\bar{L}_n\right)} \right]
\end{align}
dropping higher order terms of the variation in the exponent. This allows us to drop the path ordering symbol since the leading order term does not depend on the time parameter $s$. This approximation is justified because $\varepsilon_n \sim c^{-1/2}$ due to the overall $c$ in the quadratic action. Furthermore, we can think of the Virasoro generators as scaling with $c^{1/2}$ since $\langle 0 | L_{n}L_{-n} | 0 \rangle \sim c$. Simplifying the conjugations with the Baker--Campbell--Hausdorff formula and evaluating the Gaussian integral over $\varepsilon_n$, \eqref{eq:deformedstateU} becomes
\begin{align}
    (\ref{eq:deformedstateU}) \approx {c \over 12} {k (k^2 - 1)} \exp \left[{ \sum_{n \ge 2}{12 L_{-n}\bar{L}_{-n} \over c n (n^2 - 1)}} \right] |0 \rangle   =  {c \over 12} {k (k^2 - 1)}  \int d\mu[\theta]U^\dagger[\theta]|0 \rangle,
\end{align}
which, as shown on the right, can be written as a constant multiplied by the original state prepared by the bulk $|\mathrm{BH} \rangle$. A similar, but slightly more complicated, analysis applies to descendants of higher dimensional primaries $[L_{-k} \bar{L}_{-k}{\cal O}]$. Going back to \nref{OrigAc}  and summing over $k$ we obtain a divergent coefficient $\sum_k 1 $.
In order to study this divergence we reinstate the JT term in the action. This changes this sum to\footnote{ Adding a JT term also deforms the state at short distance by adding other operators at scales of order $\tau_c$, but such changes do not affect the IR region that involves longer times. }
\be
\sum_k 1  \to \sum_{k=2}^{k_{\max} } { 1 \over 1 + 12 \phi_r k/c } \propto  { c \beta \over \phi_{r, {\rm phys}} } \left( \log { \tau_c \over \epsilon_{\rm phys} }  + {\rm constant } \right) ~,~~~~~ k_{\max}  = { \beta \over  \epsilon_{\rm phys} }
\ee
with $\tau_c$ in \nref{DefTc}. This can be viewed as a shift in the ground state energy.

\subsection{Symmetry currents}
\label{sec:currents}

We can consider a CFT with continuous global symmetries, with associated left and right moving currents. For example, we can consider a $\U(1) $ current. We can first consider the case where there is no associated gauge field in the bulk. In this case, at the disk level, we see that charged objects can enter the black hole, whereas the Hawking radiation carries no net charge. Relatedly, the two point function $\langle J(v) \bar J(\bar v) \rangle$ will be zero, indicating that a current that goes into the black hole is not reflected and the charge is not conserved by the boundary. This is related to the general loss of information of the black hole at the disk level. In this case, we do not expect to recover it at higher orders.

The other possibility is that there is a gauge field in the gravitating region. In this case, we require $k = \bar k$ so that there is no $\U(1) $ anomaly for the vector symmetry. A gauge field in $\AdS_2$ with Dirichlet boundary conditions gives rise to a boundary mode, which at very low energies enforces that the total charge of the gravitating region (or the black hole) is zero \cite{Davison:2016ngz}. This is because the charged states have a higher ground state energy \cite{Davison:2016ngz}. This then leads to a reflection condition for the $\U(1) $ current at the interface, or a $J = \bar J$ boundary condition.

\section{The cylinder}
\label{sec:cylinder}
\subsection{Cylinder coupled to two baths}

\begin{figure}[htbp]
  \centering
  \includegraphics{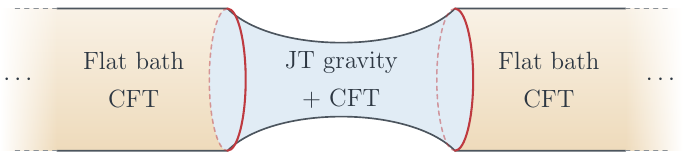}
  \caption{The double-trumpet gravitational region (blue)
  joined to two flat CFT baths (yellow). The red circles
  mark the interfaces, which are transparent to the CFT.}
  \label{fig:double-trumpet-setup}
\end{figure}

We now consider the next simplest gravitational geometry, the double trumpet with a metric
\begin{equation}
\label{eq:sec4_g_cyl}
 ds^2=\frac{\pi^2}{\kappa^2}
 \frac{d\theta^2+d\sigma^2}{\sin^2(\pi\sigma/\kappa)}
 =\frac{d\tau^2+d\widetilde\sigma^2}{\sin^2\widetilde\sigma},
 \qquad -\kappa<\sigma<0,
\end{equation}
where
\begin{equation}
 \tau=\frac{\pi}{\kappa}\theta,\qquad
 \widetilde\sigma=\frac{\pi}{\kappa}\sigma,\qquad
 \theta\sim\theta+2\pi,\qquad \tau\sim\tau+b,\qquad
 b=\frac{2\pi^2}{\kappa}.
 \label{eq:cyl_modulus}
\end{equation}
The modulus $b$ is the length of the minimal geodesic.
We glue it to two separate semi-infinite bath cylinders at the two ends, using appropriate cutoffs, etc.  The CFT lives on the whole space, see figure \ref{fig:double-trumpet-setup}.

\begin{figure}[t]
  \centering
  \includegraphics[width=0.95\linewidth]{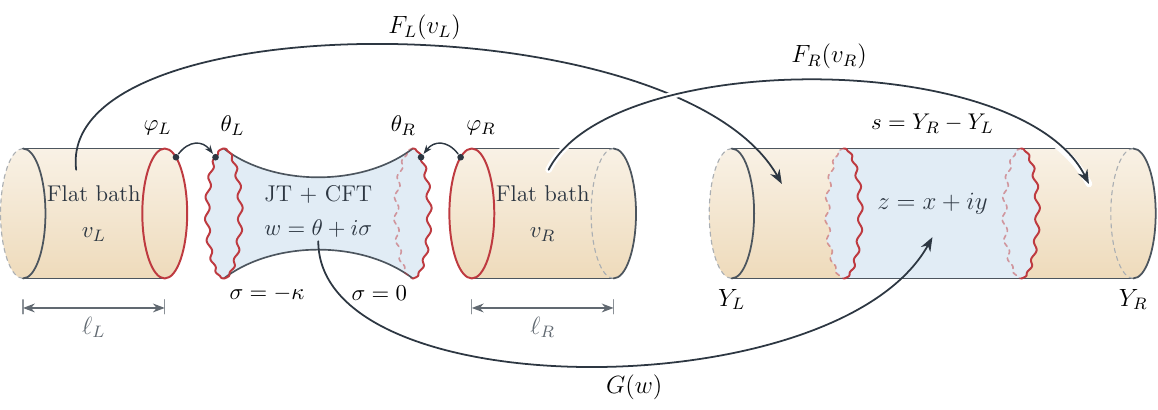}
  \caption{Conformal welding of the double trumpet coupled to two
  finite flat baths of lengths $\ell_L$ and $\ell_R$.
  The maps $F_L(v_L)$ and $F_R(v_R)$ map the baths to the reference
  cylinder, while $G(w)$ maps the gravitational region, shaded blue.
  The red curves denote the two interfaces and their images.
  The outer boundaries of the reference cylinder lie at
  $y=Y_L,Y_R$, with separation $s=Y_R-Y_L$. }
  \label{fig:double-trumpet-welding}
\end{figure}

\paragraph{JT gravity on the double trumpet.}     JT gravity~\eqref{JTAct} on the double trumpet was studied in \cite{Saad:2018bqo,Saad:2019lba}. We have two separate reparametrization modes living on the two boundaries $\theta_a(\varphi_a)$, $a=L,R$. The action \eqref{JTAct} leads to two decoupled actions
\begin{equation}
 I_{{\rm JT},a}
 =-\frac{\phi_{r,a}}{2\pi}\int_0^{2\pi}d\varphi_a\,
       \{e^{\gamma\theta_a},\varphi_a\},\quad \phi_{r,a}=\frac{2\pi \phi^{(a)}_{r,\text{phys}}}{\beta_a}, \quad \gamma = \frac{\pi}{\kappa} = { b \over 2 \pi } , \quad a = L,R
 \label{eq:cyl_JT}
\end{equation}
Note that we are free to consider two different temperatures at the two baths.
Again, due to the $\beta_a$ scaling, at low energies the Schwarzian action is usually negligible compared to the action induced by the conformal matter.
We will only need to retain this action to cure a certain divergence arising from some of the reparametrization modes, see section~\ref{subsubsec:cyl_identity} for more details.
It is important that each of the reparametrizations is integrated over ${\rm Diff}(S^1)/\U(1)$. In addition, we should integrate over $b$ and the twist mode, which leads to a factor of $b$ when we consider rotation invariant situations.
The measure was discussed in \cite{Stanford:2017thb,Saad:2019lba}.

The presence of the CFT matter leads to a divergence at small $b$,
$b\rightarrow 0$ \cite{Saad:2019lba}. We will regulate this degeneration by cutting off the integral at small $b$, $b\geq b_c$, leaving its microscopic resolution unspecified.
One concrete method would be to add a massive Pauli-Villars field that propagates only in the gravitating region. More physically, we can imagine that a more complete gravity theory takes over at shorter distances.

\paragraph{Two baths with lengths
$\ell_L$ and $\ell_R$.}
We attach a separate bath to each end of the gravitational region. This geometry gives a connected contribution for two black holes in contact with their respective baths. The metric for the bath region is simply
\lceq{
ds_a^2 = \frac{1}{\epsilon_a^2} \left(d \varphi_a^2 + d \mu_a^2\right), \quad a = L,R,
}
where we use the label $a$ to distinguish the quantities for the left and right baths. As in the disk, we have rescaled the coordinates so that all angular coordinates have period $2\pi$, namely we set $\varphi_a = 2 \pi u_a/\beta_a$ where $u_a$ is the physical Euclidean time of each bath region. We can consider the most generic case that both baths are of finite length, labeled by $L_a$ respectively. Setting $w=\theta+i\sigma$, $v_a=\varphi_a+i\mu_a$ and $\ell_a = 2\pi L_a/\beta_a$, the ranges of the baths are then
\begin{equation}
 -\kappa-\ell_L\leq\mu_L\leq-\kappa,\qquad
 0\leq\mu_R\leq\ell_R.
\end{equation}

\paragraph{Conformal welding condition.} To compute the conformal anomaly action~\eqref{ConfAn} in this geometry, we need two holomorphic maps $F_L,F_R$ on the baths and a single holomorphic map $G$ on the double trumpet, as illustrated in figure~\ref{fig:double-trumpet-welding}. The reference frame is still the flat cylinder $z$, with axial length restricted to $Y_L \le \Im z \le Y_R$. Similarly, we again choose the straight circle boundary condition on two ends of the $z$ coordinate, i.e.,
\lceq{
\label{eq:sec4_YLYR}
\Im F_L(\varphi_L-i(\kappa+\ell_L))=Y_L,\quad \Im F_R(\varphi_R+i\ell_R)=Y_R,
}
where $Y_L$ and $Y_R$ are both constants. The conformal welding condition now becomes
\begin{equation}
\label{CylBdy}
\begin{aligned}
 F_L(\varphi_L-i\kappa)
   =G(\theta_L(\varphi_L)-i\kappa),\quad F_R(\varphi_R)=G(\theta_R(\varphi_R)) \, .
\end{aligned}
\end{equation}
We have
$\theta_a(\varphi_a+2\pi)=\theta_a(\varphi_a)+2\pi$ with $\theta_a'>0$ and periodicity also requires
\lceq{
G(w+2\pi)  = G(w) + 2 \pi, \quad F_a (v_a + 2\pi) = F_a (v_a) + 2\pi \, .
}
Obviously if the gluing is at the round saddle, $\theta_a = \varphi_a$, then $F_L (v_L) = v_L, \, \, F_R (v_R) = v_R$ and $G(w) = w$ is clearly a solution of the conformal welding problem~\eqref{CylBdy}. We are mainly interested in the dynamics induced by the CFT action around this saddle.
\paragraph{The reference partition function.} The $Y_a$ are defined up to a common translation, while their separation is determined by the welding. As pointed out in~\eqref{eq:appA_dt_sep2} the change of separation with deformations of $\theta_a$ is
\longeq{
 \Delta y = s-s_0 &= \frac{1}{2\pi} \tpiint d \varphi_L \left(\theta_L' - 1\right) \Im \left[F_L(\varphi_L -i \kappa) - (\varphi_L-i \kappa)\right] \\
 & - \frac{1}{2 \pi } \tpiint d \varphi_R \left(\theta_R' - 1\right) \Im \left[F_R (\varphi_R) - \varphi_R\right] \, ,
 \label{eq:cyl_width}
}
where $s:=Y_R-Y_L
$ and $s_0=\kappa+\ell_L+\ell_R$. The change of the reference partition function will contribute to the effective action. When the state on the reference cylinder is uniform in space, we can write
the change in the reference partition function in terms of the energy
\lceq{
\label{eq:sec4_DeltaZFlat}
\Delta \log Z_{\rm flat} \simeq  - E \Delta y
}
For boundary states $|B_a\rangle=\sum_h \alpha_{a,h}|h\rangle\!\rangle$, the reference amplitude is
\begin{equation}
 Z_{\rm flat}(s)
 = \langle B_L|e^{-s(L_0+\overline L_0-c/12)}|B_R\rangle
 =\sum_h \alpha_{L,h}^*\alpha_{R,h}\,\chi_h(e^{-2s}).
 \label{eq:cyl_reference}
\end{equation}
Here $\chi_h(q)$ denotes the chiral Virasoro character of the
irreducible highest-weight module $\mathcal H_h$, generated by
a primary state of conformal weight $h$ and its Virasoro
descendants. For $c>1$ and $h>0$, this module contains no null
states, and hence
\begin{equation}
\label{eq:sec3_chih}
 \chi_h(q)
 :=\operatorname{Tr}_{\mathcal H_h}q^{L_0-c/24}
 =\frac{q^{h-c/24}}{\prod_{n=1}^{\infty}(1-q^n)},
 \quad q=e^{-2s} \, .
\end{equation}
In particular, the energy of the scalar primary propagating in the closed channel is
$E_h^{(0)}=2h-c/12$, and it is what we used in \nref{eq:sec4_DeltaZFlat}.
For a single energy eigenstate, the change of
$\log Z_{\rm flat}$ is exactly $-E_h^{(0)} \Delta y$. For one single conformal family labeled by $h$, one must keep the character in \eqref{eq:cyl_reference}. By~\eqref{eq:sec4_DeltaZFlat}, we have
\lceq{
\Delta \log Z_{\rm flat} \sim - \langle E  \rangle_h \Delta y, \quad \langle E\rangle_h
 =2h-\frac c{12}
  +2\sum_{n=1}^{\infty}\frac{n q_0^n}{1-q_0^n},
 \qquad q_0=e^{-2s_0}
 \label{eq:cyl_energy}
}
for small $\Delta y$.
The coefficients $\alpha_{a,h}$ will be suppressed when discussing a single
family. The infinite summation comes from the propagation of the descendants in the closed channel, which includes the dependence on the length of the cylinder.

\paragraph{Evaluation of the conformal anomaly.}
The derivation of the conformal anomaly is roughly the same as what we did in the disk case with a couple of modifications, as we show in appendix \ref{ConformalAnomalyCylinder}. Let us present first the final answer
\begin{equation}
    \begin{aligned}
        \label{ActCylM}
 -I_{\rm CFT}={}&-\frac{cb}{24}-  \Delta y E
 \\
 &\hspace{-0.5cm}+\frac{ic}{48\pi}\int_0^{2\pi}d\varphi_R\,\log\theta_R'
 \left(\frac{F_R''}{F_R'}-\frac{\overline F_R''}{\overline F_R'}\right)
 -\frac{ic}{48\pi}\int_0^{2\pi}d\varphi_L\,\log\theta_L'
 \left(\frac{F_L''}{F_L'}-\frac{\overline F_L''}{\overline F_L'}\right).
    \end{aligned}
\end{equation}
The first term is an important contribution that comes from the warp factor in the trumpet metric \nref{eq:sec4_g_cyl}. The rest are similar to what we had on the disk, with the different sign for the two boundaries accounted for by the different orientation.

\subsection{One loop partition function}

We consider the large $c$ approximation, and we expand the boundary  reparametrizations around the saddle $\theta_a = \varphi_a + \vare_a (\varphi_a)$ where $\vare_a$ is real and periodic in $2 \pi$, $a=L,R$.
Solving the conformal welding problem as explained in appendix \ref{WeldingCylinder} we get the quadratic action
\lceq{
\label{eq:sec4_I2}
- I^{(2)} = - \frac{c}{12}\sum_{n \ge 1} n \left(n^2 + \gamma_e^2\right) \left[a_{L,n} |\vare_n^L|^2 + a_{R,n} |\vare_n^R|^2 - b_n \left(\vare_{-n}^L \vare_n^R + \vare_{-n}^R \vare_n^L\right)\right],
}
with
\longeq{
\label{eq:sec4_cw_sol}
a_{L,n}&=\frac{2\sinh(n\ell_L)\sinh[n(\kappa+\ell_R)]}
                  {\sinh(ns_0)},
&a_{R,n}&=\frac{2\sinh(n\ell_R)\sinh[n(\kappa+\ell_L)]}
                  {\sinh(ns_0)},\\
b_n&=\frac{2\sinh(n\ell_L)\sinh(n\ell_R)}{\sinh(ns_0)}, &\gamma_e^2 &= 12 \langle E\rangle_h/c
\sim \gamma_h^2 = { 24 h \over c } -1
}
with   $\langle E \rangle_h$ in~\eqref{eq:cyl_energy} where we have assumed that $h/c$ is of order one so that we can neglect the rest of the terms from \eqref{eq:cyl_energy} because they would lead to terms of order $1/c$ in $\gamma_e$, assuming $q_0$ is order one. Also, $s_0$ is the length of the reference cylinder for trivial boundary reparametrizations.

Even though we start with two separate reparametrizations at the two ends, the final effective action \nref{eq:sec4_I2} is non-local and couples $\vare^L$ with $\vare^R$.

\paragraph{One loop partition function.}

We can now compute the integral over the reparametrizations in the Gaussian approximation. We use the measure~\eqref{eq:appG_dmuvarea} inherited from JT gravity on the trumpet geometry \cite{Saad:2019lba}. We first fix Ishibashi states $|h \rangle \rangle$ at the two external boundaries of the finite length baths, so that only the conformal towers associated with $h=\bar h$ propagate along the cylinder.  Let us present the result first for the fixed-$b$ amplitude $Z^{(1)}[b,\ell_L,\ell_R; h]$
\longeq{
\label{eq:sec4_Z1_gen}
Z^{(1)}[b,\ell_L,\ell_R; h] &=  Z_{\rm flat} e^{-\frac{c b}{24}} \int \prod_{n \ge 1} \left[n (n^2+ \gamma^2)\right]^2 d^2 \vare_n^L d^2 \vare_n^R e^{-I^{(2)}} \\
&=  e^{- \frac{cb}{24}} \chi_h (q_0) \prod_{n=1}^\infty \left(\frac{12 \pi}{c}\right)^2 \frac{(n^2 + \gamma^2)^2}{(n^2 + \gamma_h^2)^2 \left(a_{L,n} a_{R,n} - b_n^2\right)} \\
&=  \frac{c}{12 \pi} e^{-\frac{cb}{24} } \chi_h (q_L) \chi_h (q_R) \chi_h (q_\kappa)  \left[\frac{\gamma_h \sinh (\pi \gamma)}{\gamma \sinh (\pi \gamma_h)}\right]^2 \, .
}
where
\lceq{
q_a = e^{-2 \ell_a}, ~~a = L, R~~~~~~~~~~~ \quad q_\kappa  = e^{-2 \kappa} \, .
}
The denominator gives
\lceq{
\frac{1}{a_{L,n} a_{R,n} - b_n^2} = \frac{(1-q_0^n)}{(1-q_L^n) (1-q_R^n) (1-q_\kappa^n)}
}
and it leads to the cancellation of the factor $(1-q_0^n)^{-1}$ inside $\chi_h(q_0) = Z_{\rm flat}$.
The rest of the infinite product gives
\lceq{
\prod_{n=1}^\infty \frac{n^2 + \gamma^2}{n^2 + \gamma_h^2} = \frac{\gamma_h\sinh (\pi \gamma)}{\gamma \sinh (\pi \gamma_h)}, \quad ~~~~~~~~ \prod_{n=1}^\infty \left(\frac{12 \pi}{c}\right)^2 = \frac{c}{12 \pi} \, ,
}
where we used $\zeta$ function regularization.

Note that the factorization into three characters is consistent with the idea that the integral over reparametrizations gives Ishibashi states both for the external baths and for the internal gravitating side.
In other words, we expect that when a conformal family with sufficiently large $h$ is propagating through the interface, then the integral over reparametrizations at fixed $b$ results in the insertion of  a projection
\be   \la{IntProject}
\left.  \int { \cal D }_b \theta U[\theta] \right|_{h} = \mu(b) A(h,c) |h \rangle \rangle \langle \langle h |  ~,~~~~~~~{\rm for } ~~~ h> h_c
 \ee
where $h_c$ is a lower bound on $h$ that will be discussed below and
 \bea  \label{ExprForA}
 \mu(b ) &=& { \sinh( \pi \gamma ) \over \gamma } ~,~~~~~~~A(h,c) = \sqrt{ c \over 12 \pi } { \gamma_h \over \sinh (\pi \gamma_h)} \left[ 1 + O(1/c) \right]
 \\
 \gamma &=& { b \over 2 \pi } ~,~~~~~~~~~~~~~~\gamma_h^2 = { 24 h \over c } -1 \label{GammaH}
\eea
We have derived \nref{ExprForA} at large $c$. However, based on the results in \cite{Ang:2024ujz},  AI gives us a formula\footnote{Assuming that the regulated JT measure is the same as the welding measure defined purely in the IR theory. } for finite values of $c>25$. Using the following definitions
\lceq{
c = 1 + 6 Q^2 ~,~~~~~ Q = \mathfrak{b} + \mathfrak {b}^{-1}, \quad h = \frac{Q^2}{4} + P^2\, , ~~~~~ 0 < \mathfrak{b} < 1
}
we expect that
\lceq{
\label{eq:sec3_Ach}
A(h,c) = \mathcal{N}(c) \frac{2 \mathfrak{b} P}{\sinh (2 \pi \mathfrak{b} P)} \, , \quad  \mathcal{N}(c) = \sqrt{\frac{c}{12 \pi}} \left[1 + O(c^{-1}) \right], ~~~ h > h_c \equiv \half + {\mathfrak{b}^2 \over 4}
}
where $\mathcal{N}(c)$ is independent of $h$. The restriction on the range of $h$ is discussed in \cite{Ang:2024ujz}, and we discuss it further below.  When $h$ reaches the lower bound $h_c$ \nref{eq:sec3_Ach}, then $A(h,c)$  diverges. We are not proving \nref{eq:sec3_Ach}, but we will offer a check of the leading $1/c$ correction below.

The bound on $h$ in \nref{eq:sec3_Ach} is not visible in the large $c$ result in \nref{ExprForA},  which assumed $h/c$ stays finite. In fact, \nref{ExprForA} only diverges for $h/c = 0$. It is closely related to the divergence near the rim of the disk that we found when we discussed operator insertions on the disk in section \ref{sec:bulk_pri} and appendix \ref{app:deform}. This means that it is a divergence that comes from the region where the reparametrizations are large and cannot be seen in an expansion around the trivial reparametrization.

Note that the $\gamma_h$ dependent factor is smooth as $h$ decreases and $\gamma_h$ becomes imaginary, for $h< c/24$. However, as $h/c\to 0$ it does blow up. We will treat the $h=0$ case separately below.

Note that $\mu(b) $ takes into account the $b$ dependence of the measure. Here we only calculated it around the $\theta = \varphi $ point. However, since the measure is invariant under reparametrizations \cite{Stanford:2017thb}, this is the full $b$ dependence of the measure. The overall normalization of this factor is also related to the precise measure we use for the $b$ integral, which we here take to be
\be
\label{eq:sec3_bdb}
 \frac{1}{4 \pi}  \int b db
 \ee
We expect that  \nref{IntProject} holds because, as discussed in section \ref{sec:disk_stress_tensor}, the integral over reparametrizations imposes $T=\bar T$ both from the outside and inside points of view. Therefore, we should have an Ishibashi state both from the point of view of the bath and the gravitating region.

  The product of three characters in~\eqref{eq:sec4_Z1_gen} is consistent with applying the projection~\eqref{IntProject} at both boundaries. The two characters associated with the bath $\chi_h(q_a) $, $a=R,L$ are what we expect from general BCFT considerations. The fact that the middle gravitational region also gives rise to the character $\chi_h(q_\kappa)$ is a consequence of \nref{IntProject} applied to each of the two sides. If this description remains valid beyond the Gaussian approximation, higher-order corrections should modify
$A(h,c)$ without introducing additional dependence on
$q_L$, $q_R$, or $q_\kappa$ outside the characters.
From the point of view of the effective action derived using the conformal welding problem, this is highly non-trivial. For example, the expression for $E$ in \nref{eq:cyl_energy}  at order $1/c$ has subleading dependence on the $q_i$ through the last term in \nref{eq:cyl_energy}. A consistent calculation at this order must combine this correction with the cubic and quartic action and quadratic correction to the JT measure. We used GPT-6 Astra to numerically verify the computation at this order and found that the $q_i$ dependence canceled.

In fact, we have already computed the $h$ dependence of $A(h,c)$, in a different regime. The same reparametrization integral appeared in section~\ref{sec:bulk_pri}, where we studied a deformation of the CFT in the gravitating disk by an integrated primary operator, \nref{eq:bulk_primary_deformation}. To first order in $\lambda$, the disk then prepares the state \nref{eq:deformed_disk_state}, which contains $-\lambda\nu_h|h\rangle\rangle$ in addition to the vacuum Ishibashi state. We read off $\nu_h$ from the one point function of $\mO_h$ in the bath, and computed it in appendix~\ref{app:deform} for $h>1/2$ held fixed as $c\to\infty$, see \eqref{eq:sec2_1pt}:
\lceq{
\label{eq:sec3_alpha_h}
\nu_h=\frac{\pi\,2^{2-2h}}{2h-1}\left[1-\frac{6\,(6h^2-6h+1)}{c\,(2h-1)}+O(c^{-2})\right] .
}
Let us explain why $\nu_h$ is determined by $A(h,c)$. The insertion breaks the $\text{PSL}(2,\R)$ isometry of the disk down to the $\U(1)$ rotations around the insertion point. Since $\text{PSL}(2,\R)$ acts transitively on the disk $\mathbb{D}$, with stabilizer $\U(1)$,
\lceq{
\frac{\text{Diff}(S^1)\times\mathbb{D}}{\text{PSL}(2,\R)}\cong\frac{\text{Diff}(S^1)}{\U(1)}\, .
}
Therefore the integral over reparametrizations in $\text{Diff}(S^1)/\text{PSL}(2,\R)$, together with the integral over the insertion point, is equivalent to an integral over $\text{Diff}(S^1)/\U(1)$ with the operator fixed at the center of the disk. This is the same space that we integrate over at each boundary of the double trumpet. At the center, the $\AdS_2$ metric $4\,d\zeta d\bar\zeta/(1-|\zeta|^2)^2$ of appendix~\ref{app:deform} equals $4\,d\zeta d\bar\zeta$. So $\mO_h$, which is normalized with the physical metric, creates the state $2^{-2h}|\mO_h^{\text{flat}}\rangle$, where $|\mO_h^{\text{flat}}\rangle$ is the primary state created by the plane-normalized operator. The integral over $\text{Diff}(S^1)/\U(1)$ then projects this state onto $|h\rangle\rangle$ as in \nref{IntProject}, with $\mu(b)$ replaced by a disk measure factor $\mu_0$. This gives
\lceq{
\label{eq:sec3_alpha_A}
\nu_h=2^{-2h}\,\frac{\mu_0\,A(h,c)}{Z_{\rm vac}}\, ,
}
where $Z_{\rm vac}$ is the undeformed disk amplitude, which we factored out in \nref{eq:deformed_disk_state}. In the notation of \nref{DiskIshi}, $e^{\tilde S_0}=e^{S_0}Z_{\rm vac}$. For $h/c$ fixed, $2\mathfrak{b}P\to\gamma_h$, and \nref{eq:sec3_Ach} reduces to \nref{ExprForA} with $\mathcal{N}(c)\simeq\sqrt{c/12\pi}$. For $h$ fixed, $P$ is imaginary, $\mathfrak{b}^2=6/c+O(c^{-2})$, and
\lceq{
A(h,c)=\frac{\mathcal{N}(c)}{\pi\mathfrak{b}^2}\,\frac{1}{2h-1}\left[1-\frac{6\,(6h^2-6h+1)}{c\,(2h-1)}+O(c^{-2})\right] .
}
The prefactor $\mathcal{N}(c)/(\pi\mathfrak{b}^2)$ does not depend on $h$. So this is precisely the $h$ dependence of $2^{2h}\nu_h$ in \nref{eq:sec3_alpha_h}, including the $1/c$ correction. In this sense, the disk computation of section~\ref{sec:bulk_pri} is a nontrivial check of \nref{eq:sec3_Ach} for light operators.

In particular, note that there is a divergence at $h=1/2$ for $c=\infty$,  and this is shifted to $h= \half + {\mathfrak{b}^2\over 4} $ for finite $c$. This divergence can be removed by adding a counterterm at the interface; see appendix \ref{app:deform}. This issue arises for free fermions. However, the problem can be removed if we have some gauge symmetry in the gravitating region and the fermions are in non-anomalous chiral representations that do not allow a gauge invariant mass term for the fermions (as in the Standard Model without the Higgs).
Alternatively, we can consider the free orbifold CFTs in \cite{ji2026cftslargevirasoroprimary} which have an $h_{\rm gap} > 1$.

\subsubsection{The identity and its additional zero modes}
\label{subsubsec:cyl_identity}
For infinitely long baths $\ell_a \rightarrow \infty$, the reference cylinder projects onto the vacuum and $\gamma_h^2 = \gamma_0^2 = -1$. The quadratic action \nref{eq:sec4_I2} then vanishes for the modes  $\varepsilon_{\pm1}^{L,R}$. In fact, this is a consequence of a symmetry, so there are exact zero modes of the action. The symmetry in question is a pair of $\text{PSL}(2,\R)$ symmetries, acting on each of the external bath cylinders. If the cylinders are infinite, we can view them as disks, and we have the usual $\text{PSL}(2,\R)$ symmetries of the disk. The CFT partition function does not care whether we attach a bath disk or an $\text{SL}(2,\R)$ transformed bath disk at the end of a gravitating region.  This leads to two rather than three zero modes in the action because we are {\it not} integrating over the $\U(1) $ rotations of the cylinders, since the integral is over ${\rm Diff}(S^1)/\U(1)$. For a similar reason, we did not encounter any zero modes for the disk case, because the integral was over ${\rm Diff}(S^1)/\text{PSL}(2,\R)$.

Note that these are symmetries only in the low-energy limit, when we neglect the Schwarzian term and have only the induced action from the CFT. Therefore, once we remember that we also have the JT gravity coupling  in~\eqref{eq:cyl_JT}, this divergence is cured, at the cost of introducing some explicit dependence on $\beta$ that breaks the scaling symmetry. In order to do this properly, we write the integral over reparametrization as an integral over $\text{SL}(2,\R)/\U(1) $, which we treat non-linearly, times the remaining ones, which have a non-zero action at low energies, and we treat them in the Gaussian approximation. We discuss the details in appendix \ref{app:ZeroModes}.
The final result is that we replace the integral over $\vare^{L}_{\pm 1} $ by an integral involving the Schwarzian term in the action that gives the final answer
\be \la{JBetaEach}
{\cal J} \left(b, { \beta_L \over \phi_{r,{\rm phys}}^{(L)} } \right) = { 2 \over \pi (1 + \gamma^2) } { \beta_L \over 2 \phi^{(L)}_{r,\text{phys}}} \exp \left[ - { \phi^{(L)}_{r,\text{phys}} b^2 \over 4 \pi \beta_L } \right] ~,~~~~~~~~~ \gamma = { b \over 2 \pi}
\ee
And a similar expression for the right boundary in terms of ${ \beta_R /\phi_{r,{\rm phys}}^{(R)} }$. This depends both on the temperature $\beta$ and the Schwarzian coupling associated to each boundary. Our derivation of \nref{JBetaEach} was done to leading order in the $1/c$ expansion.

Interestingly, the rest of the integral involves the modes $\varepsilon^a_{n}$, with $n\geq 2$, which produce the identity character in the middle gravitating region. In other words, the final expression for the fixed $b$ partition function for infinite baths $\ell_{L,R} =\infty$ is
\be \la{FinReZer}
\left. Z^{(1)} [b;0] e^{ -( \ell_L + \ell_R) c/12 }\right|_{\ell_a=\infty} =  \left(\frac{c}{12 \pi} \right)^3 e^{-\frac{cb}{24}} \chi_0 \left(q_\kappa \right) \left[\frac{  \sinh (\pi \gamma)}{  \gamma  }\right]^2 {\cal J} \left(b, { \beta_L \over \phi_{r,{\rm phys}}^{(L)} } \right) {\cal J} \left(b, { \beta_R \over \phi_{r,{\rm phys}}^{(R)} } \right)
\ee
where $\chi_0$ is the vacuum Virasoro character~\eqref{eq:sec2_Z1chi}. Based on the above result, we conjecture that for $h=0$ we get
\be
 \left. \int { \cal D }_b \theta  e^{- I_{\rm JT}[\theta] }  U[\theta] \right|_{h=0} = \mu(b) \mathcal{J} \left(b,{\beta \over \phi_{r , {\rm phys}} }\right) A(0,c) |0\rangle \rangle \langle \langle 0 | , \quad A(0,c) := \left(\frac{c}{12 \pi}\right)^{3/2} \,.
\ee
We have only partially checked this by considering the contributions from finite-bath regions.
This means that the full integral is then, in the infrared limit,
\be \la{FullProjector}
\int { \cal D }_b \theta e^{- I_{\rm JT}[\theta] }  U[\theta] = \mu(b) \left[  \mathcal{J}\left(b,{\beta \over \phi_{r , {\rm phys}} }\right)  A(0,c) |0\rangle \rangle \langle \langle 0 |  + \sum_{h_i>h_c} A(h_i ,c) |h_i \rangle \rangle \langle \langle h_i |  \right] \, .
\ee
where we assumed that the theory does not have any operator with dimension $h\leq h_c$ other than the identity, and we have kept the $\beta/\phi_{r,{\rm phys}} \gg 1 $ dependence only for the identity term.

\subsubsection{The \texorpdfstring{$b$}{b}-integral and its divergences}
\label{subsec:cyl_modulus_integral}
In the previous discussion, we derived the fixed $b$ partition function. After factoring out the two bath characters, we denote the connected cylinder amplitude by
\begin{equation}
 A_{h\bar h}^{(2,0)}
 = Z_{\text{CFT}}\left[
     \,
     \vcenter{\hbox{\figTopoCylinderH}}
   \, \right]
 = A(h,c)^2 F(h,c),
 \qquad h=\bar h.
 \label{eq:proposal_cylinder_match}
\end{equation}
The label $h$ specifies the conformal family of the primary propagating through the cylinder. The path integral includes the integration over the neck length $b$, with
\begin{equation}
 F(h,c)
 =\frac{1}{4\pi}\int_0^\infty b\,db\,
 e^{-cb/24}\chi_h\!\left(e^{-4\pi^2/b}\right)
 \left[\frac{\sinh(b/2)}{b/(2\pi)}\right]^2.
 \label{FhcDef}
\end{equation}
where we have dropped the descendant contributions by using $\gamma_h$ defined in~\eqref{GammaH}, and $\chi_h$ was defined in \nref{eq:sec3_chih}.  When $b \rightarrow \infty$, the integrand behaves as
\lceq{ \label{c25}
  b    e^{-\frac{cb}{24} } \chi_h (e^{- \frac{4\pi^2}{b}})  \left[\frac{ \sinh (b/2)}{b/(2\pi) }\right]^2 \quad \rightarrow \quad b^{-3/2} e^{- \frac{c-25}{24}b} \, , ~~~~~~{\rm as } ~~~~~~~b \to \infty
}
which leads to a convergent integral (at infinity)
as long as $c\geq25$, which is satisfied by the large $c$ assumption.
For $c< 25$ we can make the integral converge at large $b$ by retaining a non-zero JT coupling. This, however, would lead to additional $\beta$ dependence. It would be interesting to know whether this means that there is some RG flow of boundary conditions. From now on, let us assume that $c> 25$ to avoid this issue.

The problematic regime is when $b \rightarrow 0$ $(\kappa \rightarrow \infty)$, corresponding to the long thin wormhole configuration. Notice that
\begin{equation}
 \chi_h(e^{-2\kappa})
 \sim\exp\!\left[\left(\frac c{12}-2h\right)\kappa\right]
 =\exp\!\left[\frac{4\pi^2}{b}
                    \left(\frac c{24}-h\right)\right].
 \label{eq:cyl_pinch}
\end{equation}
When the conformal dimension $h$ lies below the threshold value $\frac{c}{24}$, we get an exponential small-$b$ divergence. This is a long standing problem for two dimensional gravity \cite{Saad:2019lba}.

For $h>c/24$, the integral is perfectly finite and it gives a function of $h$ and $c$.
We might be tempted to define the integral via analytic continuation for $h< c/24$. There is a logarithmic branch cut that starts at $h=c/24 $ (and other ones at lower values) which implies that the analytic continuation gets an imaginary part whose sign depends on whether we continue via the upper or lower half plane. In either case, we cannot interpret it as giving us $\overline{|\alpha_i|^2}$ (with $h_i=h$), which is positive.
More precisely, near $ y = h - c/24 \sim 0 $, the first nonanalytic term in  $F(h,c)$ behaves as  $y^2 \log y $. In particular, it is finite at $h= c/24$.

\paragraph{Periodic fermions.}

A useful way to avoid the negative closed-channel energy is to take $N_f$ free Majorana fermions and impose periodicity around the thermal circle. This is computing $\Tr[(-1)^F e^{-\beta H}]$ on each bath. The closed-channel states are
in the Ramond sector, with
\begin{equation}
 c=\frac{N_f}{2},\qquad
 h_{\rm R}=\overline h_{\rm R}=\frac{N_f}{16}=\frac c8,\qquad
 E_{\rm R}=2h_{\rm R}-\frac c{12}=\frac c6>0 .
 \label{eq:cyl_ramond}
\end{equation}
where $h_R=\bar h_R$ denote the Ramond sector ground-state weights. Thus the integral over $b$ is well defined. Of course, in the same theory we could compute the ordinary thermal partition function, and we recover the divergence.

\paragraph{Short baths and the induced Schwarzian.}
We first take $\ell_L,\ell_R\ll\kappa$ at fixed modulus $\kappa$. For modes satisfying $n\ell_a\ll1$, the coefficients~\eqref{eq:sec4_cw_sol} of the quadratic action become
\begin{equation}
 a_{a,n}=2n\ell_a+O(\ell^2),\qquad
 b_n=\frac{2n^2\ell_L\ell_R}{\sinh(n\kappa)}+O(\ell^3).
 \label{eq:cyl_sb_coefficients}
\end{equation}
Defining $E(s)=-\partial_s\log Z_{\rm flat}(s)$, so that $\gamma_e^2=12E(s_0)/c$ with $s_0=\kappa+\ell_L+\ell_R$, we obtain
\begin{equation}
 I_{\rm CFT}^{(2)}
 =\frac c6\sum_{a=L,R}\ell_a\sum_{n\ge1}
 n^2\left(n^2+\frac{12E(\kappa)}c\right)|\varepsilon_n^a|^2
 +O(\ell^2).
 \label{eq:cyl_sb_quadratic_general}
\end{equation}
Thus the two boundaries decouple at this order. Though it looks like the quadratic expansion of the double trumpet Schwarzian~\eqref{eq:cyl_JT}, in general $E(\kappa)$ is not simply proportional to $1/\kappa^2$. To obtain the trumpet Schwarzian, we must impose another assumption.

For ordinary conformal boundary conditions whose open spectrum contains the identity, the hierarchy $\ell_a\ll\kappa\ll1$ gives open-channel vacuum dominance, i.e.,
\begin{equation}
 Z_{\rm flat}(s)\simeq n_0 e^{c\pi^2/(12s)},\qquad
 E(\kappa)\simeq\frac{c\pi^2}{12\kappa^2}
 =\frac{c\gamma^2}{12},\qquad \gamma=\frac\pi\kappa.
 \label{eq:cyl_open_vacuum}
\end{equation}
Here $n_0>0$ is the vacuum multiplicity. The quadratic action becomes
\begin{equation}
 I_{\rm CFT}^{(2)}\simeq
 \sum_{a=L,R}\phi_{c,a}\sum_{n\ge1}n^2(n^2+\gamma^2)
           |\varepsilon_n^a|^2,
 \qquad \phi_{c,a}=\frac{c\ell_a}{6}.
 \label{eq:cyl_sb_quadratic_schwarzian}
\end{equation}
This is precisely the quadratic expansion of Schwarzian theory on double trumpet. We prove that this is true in appendix~\ref{app:short_bath}  without expanding $\theta_a$ around the round configuration. With the assumption $\ell_a \ll \kappa \ll 1$, we conclude
\begin{equation}
 -I_{\rm CFT}\simeq
 \sum_{a=L,R}\frac{\phi_{c,a}}{2\pi}
 \int_0^{2\pi}d\varphi_a\,
       \{e^{\gamma\theta_a},\varphi_a\},\qquad
 \phi^{(a)}_{c, \text{phys}}:=\frac{\beta_a\phi_{c,a}}{2\pi}
 =\frac{cL_a}{6}.
 \label{eq:cyl_induced_schwarzians}
\end{equation}
We used $\ell_a=2\pi L_a/\beta_a$. The effective coupling $\phi_{c,a}$ matches with the nonlinear derivation, see appendix~\ref{app:short_bath}. In physical time $\varphi_{a , \text{phys}}=\beta_a\varphi_a/(2\pi)$, the action coefficient is $\phi^{(a)}_{c, \text{phys}}/(2\pi)=cL_a/(12\pi)$, the same as for the short-bath
disk~\eqref{eq:sec2_phic}.

\paragraph{Relation to the no-bath limit.}
As an unrelated curiosity, we can consider the situation of a CFT living in the gravitational region, with some boundary conditions at the $\AdS_2$ boundary, without any bath. In that case, for a finite UV cutoff $\epsilon$ there is also an induced action which takes the Schwarzian form at leading order in $\epsilon$. We show the derivation in appendix~\ref{app:CFT reduced to the boundary}. For the case of the cylinder we have two $\epsilon_a$ and an effective action of the form
\begin{equation}
 -I_{\rm CFT}^{\rm no\ bath}\simeq
 -\sum_{a=L,R}\frac{c\epsilon_a}{8\pi}
    \int_0^{2\pi}d\varphi_a\,
      \{e^{\gamma_{\rm eff} \theta_a},\varphi_a\},
\label{eq:cyl_sb_no_bath_comparison}
\end{equation}
with $\gamma_{\rm eff }$ defined in~\eqref{eq:appB_gammaeff}. When we work on the relatively short cylinder with $b \gg 1$, the energy is approximated by the Casimir energy of the vacuum propagating in the open channel, $E(\kappa) \simeq c\gamma^2/12$, which leads to $\gamma^2_{\rm eff} \simeq \gamma^2$, reproducing the usual Schwarzian theory on double trumpet.

A  curiosity is that this is roughly similar to taking \eqref{eq:cyl_induced_schwarzians} and setting $\ell_a \to -{ 3 \over 2} \epsilon_a $.
However, the bath length $\ell_a$
and the gravitational cutoff $\epsilon_a$ describe different
deformations and should not be identified.  It is perhaps not surprising that in many situations we obtain a Schwarzian, since it is the leading local correction available.

\subsection{Torus geometry and open--closed duality}
\label{subsec:cylinder_torus}
We now consider the case in which we have a single finite-length bath with a gravitating region on each side. One reason to focus on this geometry is that the BCFT interpretation leads to a non-trivial consistency condition: open-closed duality. If we have an interval with two conformal boundaries at the two ends at finite temperature, then we can think of the problem in two ways: the open and the closed string channels; see figure \ref{fig:open-closed-Intro}.

For these purposes, the interesting regime is when the bath is very short, $\ell= 2\pi L/\beta \to 0$, where we expect to obtain the contribution of the identity in the open channel.
As we mentioned in the introduction, we will get such a contribution by considering the following connected geometry
\begin{equation}
  \langle \BH |e^{-\ell H}| \BH\rangle_{\rm con} =  \vcenter{\hbox{%
    \includegraphics{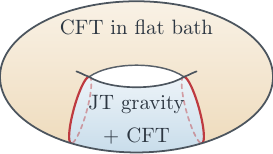}%
  }}
  \label{eq:torus-bath}
\end{equation}
Here the conformal matter lives on a torus and this computes the connected contribution to the boundary-state overlap. In addition to this contribution, we have a more obvious contribution which consists of two disks attached to the two ends of the interval; see figure \ref{fig:open-closed-geometries}(a). However, as analyzed around~\eqref{IdCo}, the disconnected contribution (figure~\ref{fig:open-closed-geometries}(a)) is smaller than the connected contribution \nref{eq:torus-bath}  for $\ell \ll c/S_0 $. We continue to neglect the bare JT action.

\paragraph{Finding the leading saddle.} Let us discuss first the saddle point. We will assume that the saddle point for the reparametrizations is for the trivial ones, $\theta_a(\varphi_a) = \varphi_a$, $a=L,R$. Then we should search for the saddle point of the $ b$-integral. We assume that $\ell$ is small, and we aim to find a saddle point where $\kappa = 2\pi^2/b$ is also small. The CFT is on a torus, and the modular parameter of the torus is such that, if we set the length of the Euclidean time to be $2\pi$, then the length of the spatial circle is $\ell + \kappa$, which by assumption is very small. This means that the vacuum is propagating along the Euclidean time direction and the partition function of the CFT  given by the identity contribution has the leading form
\be
 Z_{\rm flat } \simeq \exp\left[  { 4 \pi^2 \over \kappa + \ell } { c \over 12 }  \right] \, .
\ee
The conformal anomaly gets a contribution from the warp factor of the double trumpet geometry, which is \nref{ActCylM}
\be
- I_{\rm CFT}  = - {  c b \over 24 } = - { 2\pi^2 \over \kappa } { c \over 24} \, .
\ee
 This means that we need to extremize
 \be
 \log Z = { c \pi^2 \over 12} \left[ -{ 1 \over \kappa } + {  4 \over \kappa + \ell } \right] + {\rm const}
\ee
over $\kappa$. The saddle is at
\be \la{saddlpk}
\kappa_* = { 2 \pi^2 \over b_*} =  \ell  ~,~~~~~~~\log Z_* = { 2 \pi^2 \over \ell } { c \over 24} ~~~~~{\rm for }~~~~~~ \ell \ll 1   ~,~~~~~~\phi^L_{r,{\rm phys}} + \phi^R_{r,{\rm phys}} \ll c L
\ee
and we can a posteriori check that indeed $\kappa \ll 1$, so that our approximations are valid as long as we are in the regime $\ell \ll 1$, and the last condition in \nref{saddlpk} comes from demanding that we can still neglect the Schwarzian coupling.  We also see that the saddle-point value of the full action reproduces the expected contribution from the identity in the open channel.

\begin{figure}[t]
  \centering
  \includegraphics{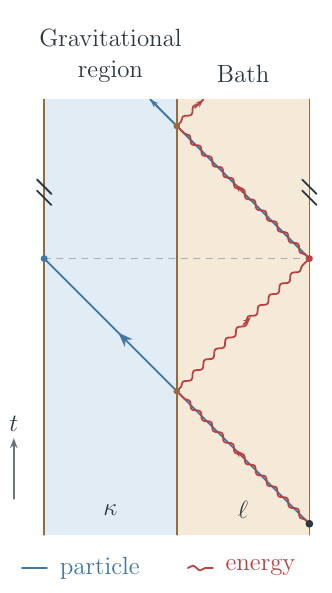}
  \caption{Lorentzian torus saddle at $\kappa=\ell$.
  Matching double slashes identify the outer edges. Blue particle rays
  cross the gravitational strip, while red energy pulses reflect within
  the bath. All propagation centerlines are null. The dashed guide marks
  equal arrival times through gravity and the bath. }
  \label{fig:lorentzian-torus}
\end{figure}

The saddle point geometry can be viewed as an eternally traversable wormhole, similar in spirit to the one discussed in \cite{Maldacena:2018lmt}. The interaction between the two sides is provided by the CFT on the interval and the negative energy necessary for traversable wormholes \cite{Gao:2016bin} is provided by the Casimir energy of the CFT on the cylinder. This geometry can also be continued to Lorentzian signature, as shown in figure~\ref{fig:lorentzian-torus}.
A curious feature is that $\kappa=\ell$ means that this wormhole is barely consistent with boundary causality, which requires $\kappa \geq \ell$. A signal that goes through the wormhole comes out from the other side in the same time it takes the signal to go from one side of the bath to the other.

This motivates us to rewrite the general saddle point equation as
\be \la{neweq}
{ c \over 24 } +   { 2 \kappa^2 \over (\kappa + \ell)^2 }\tilde E_{\text{CFT}}(\tilde \beta) =0  ~,~~~~~~\tilde \beta = { 4 \pi^2 \over \kappa + \ell }
\ee
where $\tilde E_{\text{CFT}}$ is the energy for the finite temperature CFT on a spatial circle of length $2\pi$ and an inverse temperature  $\tilde \beta = { 4 \pi^2 \over \kappa + \ell } $.  When $\kappa + \ell \ll 1 $ this is
$-c/12$ so that \nref{neweq} reproduces \nref{saddlpk}. As $\ell$ increases, $\tilde E_{\text{CFT}}$ starts to receive a contribution from non-vacuum states and \nref{neweq} is hard to solve in general.  However, we see that, since $\tilde E_{\text{CFT}} > - c/12$, the solution will be such that $\kappa_* > \ell$, again consistent with causality, but not at the verge of inconsistency.  We also expect that as $\ell $ becomes larger, the solution will cease to exist.

\begin{figure}[t]
  \centering
  \includegraphics{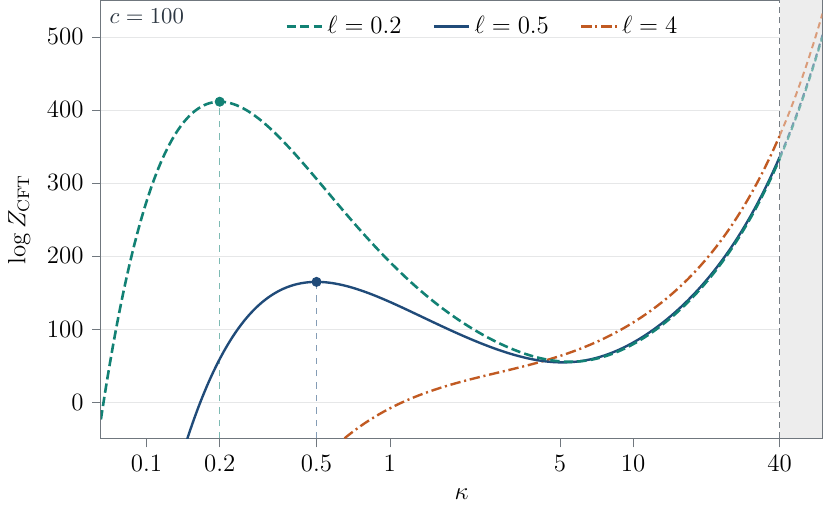}
  \caption{Numerical check of the saddle point $\kappa_*\simeq\ell$. In the above figure, we place $n = 100$ copies of self-dual compact bosons, each with $c=1$ on the torus with $\ell=0.2$, $\ell=0.5$ and $\ell =4$. The partition function of the $n$ copies of self-dual compact boson is $Z_{\rm flat} (i \mathfrak{t}) = \left[\frac{\vartheta_3 (0|i \mathfrak{t})^2}{\eta (i \mathfrak{t})^2}\right]^n$ for $\mathfrak{t} = \frac{2 \pi}{\kappa + \ell}$. We plot the logarithm of the reference partition function plus the conformal anomaly. One can see that $\kappa_*\simeq \ell$ is a local maximum for relatively short $\ell$, though we have a UV divergence when $\kappa \rightarrow \infty$.
  In this figure, the value of the local maximum for $\ell =0.5$ is lower than the value at the cutoff $\kappa_{\rm cutoff} =40$. As $\ell \to 0$, the value at the local maximum becomes larger than the value at the fixed cutoff (at $\kappa =40$);  see the plot for $\ell = 0.2$. We also see that when $\ell $ is not small, then we would lose the maximum; see the plot for $\ell =4$.     }
  \label{fig:torus-boson-saddle}
\end{figure}

More generally, we expect that as a function of $b$ the whole partition function should have the general qualitative form shown in figure~\ref{fig:torus-boson-saddle}.

In principle, we also need to consider the expansion around the saddle point to compute the overall coefficient of the identity contribution. However, since the Euclidean time circle is not contractible in this geometry we can conclude that it should also be manifest from the gravity + CFT point of view that we are performing a trace. We expect that the vacuum state of the wormhole is unique and it should contribute with a unit normalization to the trace. This leads us to conclude that the saddle point contribution to the partition function on the torus geometry has the form
\be
 Z  = \exp\left( { 2 \pi^2 \over \ell } { c \over 24 } \right) \left[ 1 + O( e^{ - \Delta_{\rm gap} { 2 \pi^2 \over \ell } } ) \right]   ~,~~~~~~~{\rm for } ~~~ \ell \ll 1
 \ee
where $\Delta_{\rm gap}$ is the dimension of the lightest operator in the BCFT. Here we imagine taking the small $\ell $ limit keeping the small $b$ (or large $\kappa$) cutoff fixed, see figure \ref{fig:torus-boson-saddle}.

\paragraph{The one loop result.}
We have checked this expectation by performing the gravitational path integral around the saddle. We leave the lengthy calculation and discussion to appendix~\ref{app:OpenClosed}. Around the saddle point, the one loop partition function is\footnote{In the sum over $h_i$ we assumed that both $h_i$ and $\bar h_i$ are non-zero. If one of them is zero then there should only be one  factor of $(1-q_o)$ in the denominator. }
\bea
\label{eq:sec4_Ztor}
Z_{\text{one loop}} (\ell) &=& q_o^{-c/24}  \mathcal{V} (q_o)^2  \left[1 + \sum_{h_i, \bar h_i }  { q_o^{h_i + \bar h_i } \over (1 - q_o)^2 }  +  O(c^{-1})\right], \\
\quad q_o &=& e^{-2 \pi^2/\ell}, \quad \mathcal{V}(q) = \prod_{n=2}^\infty \frac{1}{1 - q^n} \, .
\eea
The factor $q_o^{-c/24}$ is the saddle point value~\eqref{saddlpk}. Apart from $\mathcal{V}(q_o)^2$, all one-loop factors combine into a unit coefficient. Appendix~\ref{app:OpenClosed} derives~\eqref{eq:sec4_Ztor} both directly and in the closed channel, verifying the open-closed duality in the small-$\ell$ limit.

\paragraph{The open-channel states.}
One interesting question is whether we can read off any other open string channel operator besides the identity.
In general, these can be found by looking at subleading terms in the small $\ell$ expansion. One subtle issue is that our computation is just the leading order term in the $1/c$ expansion; therefore, we should be careful not to trust terms that could be overwhelmed by $1/c $ corrections. There are two simple cases we can consider in order to get oriented: a large $c$ theory with a sparse spectrum \cite{Hartman:2014oaa} or a system of $2c $ free Majorana fermions.

Let us analyze the first case. In this case, the vacuum channel is a good approximation to the partition function for fixed $\ell < \pi $ \cite{Hartman:2014oaa}, which would make $\kappa + \ell = 2\ell $.
In this case, we can write the partition function of the CFT as
\be \la{ClosedCh}
Z_{\rm flat} = \chi_0^2(q_c) + \sum_{h_i,\bar h_i} \chi_{h_i}(q_c) \chi_{\bar h_i}(q_c) + \cdots  ~,~~~~~~~~~~q_c = e^{ - { 4 \pi^2 \over \kappa + \ell }}
\ee
where $\chi_h$ are the characters of Virasoro representations and $(h_i,\bar h_i) $ are the left and right conformal dimensions of the various fields which are of order one in the large $c$ expansion. Note that the theory can have fields with spin, with $h_i - \bar h_i \not =0$.
When we expand around the saddle point for a value of $\ell$ that is fixed in the large $c$ limit, but with $\ell < \pi$, then we can trust the various terms in \nref{ClosedCh}. We are supposed to rewrite this expression in terms of the open string characters $\chi_{h}(q_o)$. First note that
\be
q_o = e^{ - { 2 \pi^2 \over \ell }} = q_c|_{\kappa = \ell}
\ee
The first observation is that the identity contribution in \nref{ClosedCh} contains the identity plus some extra terms
\begin{equation}
 q_o^{-c/24}\mathcal V(q_o)^2
 =\chi_0(q_o)+\sum_{r=2}^{\infty}N_r\chi_r(q_o),\qquad
 \sum_{r=2}^{\infty}N_rq^r=(1-q)[\mathcal V(q)-1].
 \la{NewPrimPar}
\end{equation}
The first term is the full identity contribution in the full open channel, and it comes from the operator $T=\bar T$ at the boundary. All the terms in the sum represent new Virasoro primaries.  All of these should be boundary operators of the BCFT. What are these new operators? Note that the stress tensor is already taken care of by the first term.
To leading order in $1/c$, the idea is that there is a new operator $\hat T $ of dimension two which is a linear combination of the Schwarzian
$\{ \tan{ \theta \over 2 } , \varphi \} $ and the stress tensor,
\be \la{NewTOp}
\hat T \sim T - {c\over 12 }  \{ \tan{ \theta \over 2 } , \varphi \}
\ee
See appendix \ref{app:boundary_operator} for more discussion.
This field can have $\SL(2,\R)$ descendants, which are just $\partial_\varphi^n \hat T $. Counting all the multiparticle combinations which are $\SL(2)$ primaries we get the terms in brackets in \nref{NewPrimPar}. See appendix \ref{app:boundary_operator}.
From the point of view of the wormhole, we interpret these as related to the fluctuations of the length of the wormhole, see appendix \ref{app:wormhole_length}.

The operator $\hat T $ has dimension 2 at leading order in the $1/c$ expansion, but using AI we computed the first correction to its dimension and found that it is
\be \la{FiOCo}
h= 2 - 24/c + O(1/c^2)
\ee

There is a somewhat similar story with the operators
$V_{h \bar h} $ in \nref{ClosedCh}, only one of the factors $\chi_h$ corresponds to the descendants under the single Virasoro of the strip. All other factors denote further open primaries. The first operator is simply the bulk operator evaluated at the boundary; the next is the normal derivative of the bulk operator at the boundary; the rest probably result from multiparticle operators by the addition of the operator $\hat T $.

The results derived here for sparse theories are likely valid for more general theories if we consider operators whose dimensions remain of order one in the large-$c$ limit.

The second observation concerns the second term in \nref{ClosedCh}, which appears to arise from the states of the full CFT on the cylinder. These contributions can be viewed as boundary operators of the BCFT  with boundary dimensions $\Delta_{\rm bdy} = h+ \bar h + n  + O(1/c)$. These are the same as the operators that we encountered when we computed the boundary OPE of bulk operators with a gravitational disk around \nref{AnomHHbar}, where the order $1/c$ correction was also noted.

One curious feature is that the boundary OPE of a bulk operator does not produce the operator $\hat T $, at any $c$ and for the disk gravity topology; see appendix \ref{app:boundary_operator}.  One way we might insert the operator $\hat T $ is by locally changing the value of the Schwarzian coupling.

\section{ The black hole as a boundary state }
\label{sec:bh_as_bd_state}

\paragraph{The black hole boundary state.} Our previous discussion motivates the conjecture that the low energy limit of a black hole interacting with a CFT should be a proper boundary condition, or a BCFT.

At finite temperature, the idea is that we can view the gravitational path integral as preparing a proper boundary state, a so called Cardy state
\cite{Cardy:1989ir}, which is a particular linear combination of Ishibashi states
\be
|\mathrm{BH} \rangle = \sum_i \alpha_i |h_i\rangle\rangle
\ee
where $h_i$ is summed over the spin zero primary operators of the CFT. These coefficients are constrained by consistency conditions \cite{Cardy:1989ir}. In addition, we should have a spectrum of boundary operators and appropriate structure constants \cite{Cardy:1989ir,Kusuki:2022boundaries}.   Of course, the idea is that this should be the situation in a full theory of quantum gravity, such as string theory. We have not given a concrete example, so this remains a speculative conjecture, as an exact statement.

\paragraph{Higher order topologies in our toy model.}
  A natural conjecture, inspired by \cite{Coleman:1988cy,Giddings:1988cx,Saad:2019lba}, is that higher topologies come from averaging over possible boundary conditions for a fixed two dimensional CFT. In fact, this was already proposed in \cite{Kusuki:2022boundaries}, in a very closely related context.

\begin{figure}[t]
  \centering
  \includegraphics{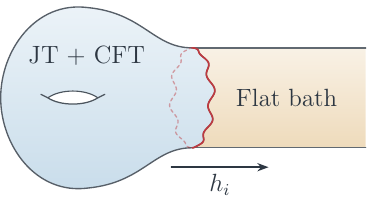}
  \caption{In the gravitational region, we can have a higher order contribution corresponding to a disk with a handle. This can lead to states $h_i$ propagating across the interface into the bath, which may differ from the identity.   }
    \label{fig:DiskHandle}
\end{figure}

\paragraph{Single bath.} As an example, we can analyze the computations that we did in sections \ref{sec:disk}, \ref{sec:cylinder}  from this point of view.
For the disk, and the unperturbed CFT, we would conclude that
\be
\overline{ \alpha_i} = e^{\tilde S_0} \delta_{i 0} + O ( e^{ - \tilde S_0 } )
\ee
This is saying that the averages of the coefficients are all zero, except for the coefficient of the identity.
Adding a bulk deformation to the CFT as in section \ref{sec:bulk_pri} would shift these averages away from zero.
Similarly, adding a handle to the disk, see figure \ref{fig:DiskHandle}, can also lead to a contribution with a non-trivial $h_i$ flowing through the interface. Such a contribution is down by a factor of $e^{-2 S_0} $ compared to the disk, and can also shift the averages of $\alpha_{i}$ away from zero, for $i\not =0$.

\paragraph{The cylinder.} The computation of the gravitating cylinder with two independent baths attached at the two ends, see figure \ref{fig:double-trumpet-setup},  can be viewed as providing the average values for
\be
\overline{ \alpha^*_i \alpha_j }_{\rm con} = \overline{ \alpha_{i}^* \alpha_j } - \overline{\alpha_i^* } \, \, \overline{\alpha_j} = \delta_{ij}  A(h_i,c)^2 F(h_i,c)
+ O ( e^{ - 2 \tilde S_0 } ) \ee
where $A(h,c)$ is the constant that arises from the integral over reparametrizations, see \nref{eq:sec3_Ach}, and $F(h_i,c)$ is the result of the $b$ integral \nref{FhcDef}. Notice that, since the cylinder is of order one in the $e^{-S_0}$ expansion, this average is also of order one.

Of course, for $ h < c/24$ we find that the integral over $b$ diverges at small $b$ and $F$ does not have a value that is calculable purely in the low energy theory. For these representations, the low-energy theory does not determine the variance of the boundary-state coefficients; its value depends on the resolution of the degeneration region. It is natural to suspect that this variance is larger than naively expected (larger than order one in the $e^{ - {\tilde S}_0}$ expansion).
  Of course, this regularization could completely alter the discussion about the sum over topologies, since it could involve a higher dimensional string theory, for example. In this way, it is unclear whether even the average interpretation remains appropriate.

In a string theory example, we expect that the bulk deformations of the CFT for small $h$ arise from integrating out massive fields at the string scale and would lead to corrections that would go like $  ( l_s/R)^{2h-2} $ with $R$ the AdS radius. Assuming that $S_0$ goes as some power of the radius of the space in string units, $S_0 \sim { 1 \over g_s^2} ( R/l_s)^n $, we conclude that these corrections are expected with coefficients that go like inverse powers of $S_0$ (for $h>1$). These are much larger than  $e^{ - S_0} $ as long as $h$ is not too high. Furthermore, these values might differ as we change from one string-theory realization to another.
Therefore, we can conclude that the fact that for relatively small $h$ we cannot compute the average from the low-energy theory is somewhat consistent with expectations from string theory. However, we do not understand why the proposed UV explanation yields the precise limit $h < c/24$ found in our calculation.

For $h=0$ and infinite baths, we had an additional complication due to residual
$\beta$-dependence arising from the $\SL(2,\R)$ zero modes, \nref{JBetaEach} and \nref{FinReZer}.
The leading $\beta$ dependence is just a factor of $\beta$ for each bath region. We propose to interpret this as follows. The idea is that each boundary condition we are averaging over could have a slightly different ground state energy $E_0$. So we write
$E_0 = \bar E_0 + \delta E_0$ where $\bar E_0$ is what we see at the disk level. The small fluctuations are then computed by the cylinder
\be
{ \rm{cylinder } \over {\rm Disk}_L { \rm Disk}_R } \sim \beta_L \beta_R \overline{    (\delta E_0)^2  }
 \ee
This implies that these fluctuations are of order $ \delta E_0 \propto e^{-S_0}$.
 However, there is the additional complication that the $b$ integral over \nref{FinReZer} is divergent at small $b$, so the overall coefficient is not calculable, and its final scaling with $e^{ - \tilde S_0} $ will depend on the proper resolution of this divergence. Even then, we are supposed to trust this only while the right hand side is smaller than one, otherwise we should include other diagrams beyond the cylinder one, so we cannot make $\beta_{L,R}$ arbitrarily large.
We can similarly think about subleading terms in the large $\beta$ expansion of \nref{FinReZer} as averages of $ \overline{   \delta E_0 \delta S_0  }  $, $\overline{(\delta S_0)^2 } $, as well as terms involving $\beta_L/\beta_R$ which can be viewed as correlations of the specific heat and $\delta E_0$.

\begin{figure}[t]
  \centering
  \includegraphics{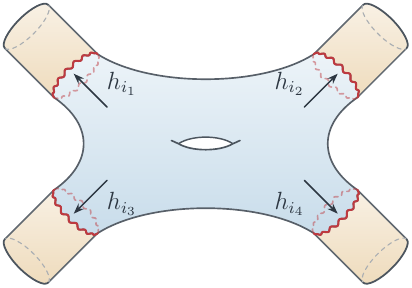}
  \caption{We see a higher order connected contribution that has many bath regions attached. Decomposing the diagram as a sum in terms of operators $h_{i_k } $ flowing across the interfaces, we can then view each term, with definite $(h_{i_1}, h_{i_2}, h_{i_3}, h_{i_4}) $, as a contribution to  the fully connected correlator $\overline{\alpha_{i_1} \alpha_{i_2}\alpha_{i_3} \alpha_{i_4}}_{\rm con} $. We get other contributions by considering other topologies with the same boundaries.  }
    \label{fig:MultiBdy}
\end{figure}

\paragraph{Many baths connected by the bulk.} One could also consider higher order diagrams like the one in figure \ref{fig:MultiBdy}.  This can be viewed as computing statistical correlations for the coefficients of the boundary state.
In principle, we can compute the diagram as follows. The integral over reparametrizations yields the projector \nref{FullProjector} on each interface. We can then think of the diagram as the sum of the primaries with $ h_i = \bar h_i$ that cross each interface. This leads to expressions for both the bath and interior contributions. If we integrate over the moduli of the interior,
we find divergences from the regions where the minimal geodesics become very small, and we need to cut off these regions. Due to these divergences, this does not appear to be a well defined problem.   If we were to ignore these divergences, then we can view the answer as providing some expectation values for averages of the form
\be
 \overline{ \alpha_{i_1} \alpha_{i_2} \cdots \alpha_{i_n}}_{\rm con}
 \ee

These diagrams can then be viewed as computing expectation values of correlators of the boundary state coefficients $\alpha_{i} $.

  \paragraph{Large $S_0$ for fixed $c$.}
One curious aspect of our discussion is that $S_0$
could be arbitrarily large, even for a fixed $c$. Bounds on the boundary entropy were discussed in \cite{Collier:2021ngi}. Those bounds depend on the value of the gap in the spectrum of boundary dimensions. Their method only gives a bound in the case that $h_{\rm gap} > { c - 1 \over 24}$.
In our case, there is a primary of dimension two \nref{NewTOp} for large $c$. For $h_{\rm gap}=2$, we would have a bound on $\tilde S_0$ for $c<49$. However, AI found the correction \nref{FiOCo}. If we only took this first correction into account, there would be a bound on $\tilde S_0$ only below the reduced value $c= 29.4$. So we see that the correction shrinks the region where we expect a bound to hold. However, for this value the first correction to the anomalous dimension is already significant, and we would need to go to higher order in $1/c$.  It is possible that as we consider higher corrections to \nref{FiOCo}, the bound on $S_0$ is completely eliminated (at least in the region $c\geq 25$ where the extra large $b$ divergences vanish).

\paragraph{Summary.} We expect that in a full theory of quantum gravity, a black hole interacting with a CFT of the form that we discussed should become a boundary condition in the CFT. We might expect that in a full theory of quantum gravity there will be no divergences in the bulk. This full theory might yield a factorizing answer or one that we will have to interpret in terms of an average.
Another separate issue is that the additional $\beta$-dependence we found for trumpets might lead to further RG flow, so that the true BCFT is only found after this flow and could have a smaller boundary entropy.

It is important to note that the discussion in section \ref{subsec:cylinder_torus} involves a saddle point whose contribution is independent of the UV resolution of the small-handle divergences.

\section{Magnetic black holes and tabletop quantum gravity}
\label{sec:Magnetic}

Magnetically charged black holes in the Standard Model are interesting because they can approach extremality and remain as extremal black holes for a long time, if the monopole creation process is sufficiently suppressed. In particular, we can have relatively light black holes, say of order a kilogram, which we could hold in a laboratory\footnote{In order to hold a magnetic black hole in the gravitational field of the earth you need a magnetic field of a few hundred Tesla, so you might need an orbiting laboratory.}. They would not evaporate immediately as a Schwarzschild black hole of the same mass would. Instead, they decay to extremality and, if fundamental monopoles are heavy enough, they would stay as extremal black holes for longer than the age of the universe. In addition, for this mass range, the energy scale at which quantum gravity effects associated with reparametrizations begin to dominate could be comparable to atomic-physics scales. Therefore, for such a black hole we could test some non-perturbative aspects of quantum gravity in the laboratory!
Of course, this presupposes that there is some process that produces these black holes in the first place, perhaps in the very early universe.  We do not have a very compelling scenario that produces them, but we will still analyze the physics of these solutions. In particular, we would like to explain how the toy model that we analyzed in this paper is relevant for those black holes.

\begin{figure}[t]
    \begin{center}
\includegraphics[width=\linewidth]{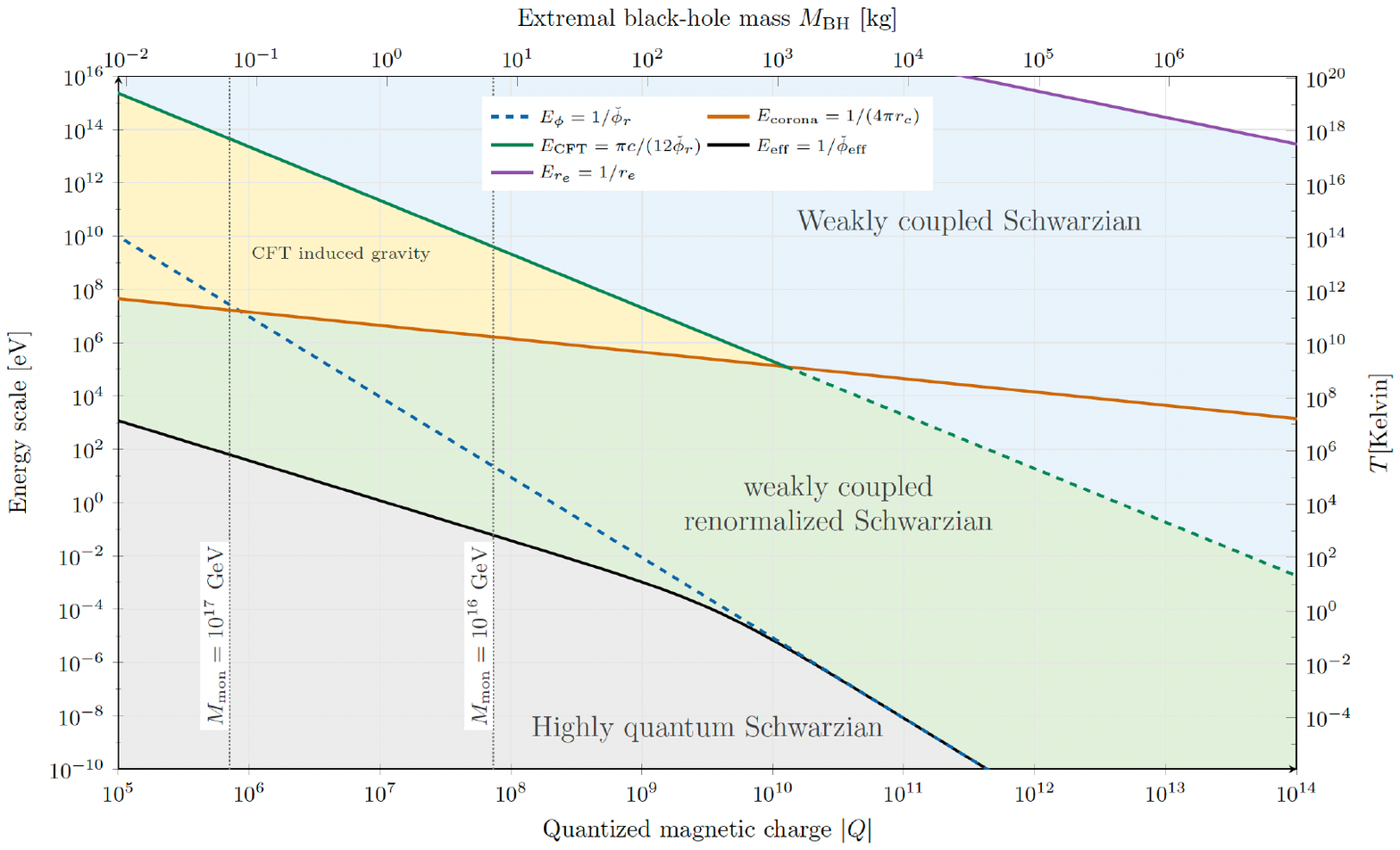}
  \end{center}
  \caption{ Various energy scales that separate different behaviors. Below the purple line, we have the effective description in terms of the reparametrization mode.  Below the solid green line, the induced gravity action becomes more important than the original Schwarzian, and we go into the regime where we can approximate the bath as being infinite. As we lower the energy, we hit the scale corresponding to the bath's size. Below this energy, we again obtain a Schwarzian but with a different coefficient. In the region where the green line is dotted, we first encounter the finite size bath scale when all the CFT matter can be integrated out, and the physics is dominated by the usual Schwarzian term. In other words, a separate CFT-dominated window exists only to the left of the intersection. The vertical lines indicate that black holes with larger charge have a lifetime longer than the age of the universe for that value of the unit charge monopole mass. Notice that for charges of order $Q\sim 10^7$ the quantum behavior of the Schwarzian starts at energies which are of the order of atomic energy scales or around an eV.     }
    \label{fig:Scales}
\end{figure}

A standard magnetically charged black hole has the following relevant parameters
\be \la{ParamVaM}
M = { r_e \over G_N} ~,~~~~~~~
S_e = { \pi r_e^2 \over G_N} ~,~~~~~~
\check \phi_r = \pi \phi_{r,\rm{\phys} } = { 2 \pi^2 r_e^3 \over G_N } ~,~~~r_e = { \sqrt{ \pi G_N} \over g  } Q
\ee
where $Q = { 1 \over 2 \pi } \int_{S^2} F $ is an integer representing the quantized magnetic charge, which we take to be positive. $g $ is the coupling, defined from the action $I = - { 1 \over 4 g^2 } \int F^2$.
We have computed the Schwarzian coupling by looking at the near extremal entropy of such black holes and comparing with the Schwarzian result. We defined $\check \phi_r$ so that the action is given by $-I = \check \phi_r T$, with $T$ the temperature, with no extra numerical constant.
The energy scale $E_s= { 1 \over \check \phi_r}$ indicates when the reparametrization mode becomes strongly coupled, in the absence of other relevant fields; see figure \ref{fig:Scales}.

  Due to the weak gravity conjecture, we expect that such a black hole should decay by nucleating magnetic monopoles. The unit charge monopole creation rate goes as $\exp\left( - {g^2 \over 2} m^2/{\cal F} \right) = \exp\left( - \pi G_N m^2 Q\right) $ \cite{Pioline:2005pf}, where ${\cal F}$ is the size of the magnetic field in the throat, ${\cal F}^2 = \half F_{\mu \nu} F^{\mu \nu}$. ${\cal F} $ scales like $ M_{pl}^2/Q$.  We see that for large $Q$, monopole creation will be exponentially suppressed.
  Monopoles in grand unified theories have a mass that is somewhat dependent on the precise setup, but they range around $10^{17}$ GeV, with lighter masses also possible\footnote{We assumed a susy GUT value here.}. For the sake of the argument, let us assume this value. If we demand that the decay time be greater than the age of the universe,  we obtain a charge roughly $Q \sim 10^6$;  see figure \ref{fig:Scales}. We get correspondingly larger values if we lower the mass.

  This means that black holes with masses on the order of kilograms or higher could have survived from the beginning of the universe to today as very, very near extremal black holes. Furthermore, their  Schwarzian couplings $\check \phi_r$ could be small enough that we could enter the strongly coupled quantum regime for the reparametrization mode \cite{Brown:2024evaporation}.

  For the relevant values of the charges,  it is important to take into account the effects of the magnetic field on the matter of the Standard Model. In fact, in this regime the magnetic field is so large that the electroweak symmetry is restored in the near horizon region. In that region, the $\mathrm{SU}(2)$ component of the magnetic field is screened, and we need only consider the hypercharge component. So the coupling $g$ appearing in \nref{ParamVaM} should be the $\mathrm{U}(1)_Y$ coupling evaluated at the scale set by the magnetic field,  $\sqrt{\cal F} $.
  The region where the electroweak symmetry is restored is separated from the usual vacuum by an electroweak ``corona''. The inner radius of this corona is at
  \be \la{RCorona}
  r_c = \sqrt{ Q \over 2 } { 1 \over m_H }
  \ee
  where $m_H$ is the value of the Higgs mass; see figure \ref{fig:Scales}. This corona is a rather complicated object \cite{Gervalle:2024yxj,Gervalle:2025awa} and we are going to ignore its dynamics here.

  For values of the radius smaller than $r_c$, all the Standard Model fermions are massless, and their lowest Landau level on the $S^2$ gives rise to order $Q$ massless fermions in the radial and time directions. These can be viewed as a CFT with central charge $c_L =c_R = 9 Q$, including all three families  \cite{Maldacena:2020skw}. Of course, there could be extra fields such as higgsinos which would contribute more to the central charge. The important part is that $c$ is linear in $Q$ and therefore rather large for the charges in question. These fields are almost free fermions, but they are subject to gauge and Yukawa interactions which induce four fermion couplings. These are classically marginal in two dimensions and could give rise to small masses via the magnetic catalysis phenomenon \cite{MiranskyShovkovy2015}. A naive estimate of these effects, using the formulas for the mass in  \cite{MiranskyShovkovy2015}, suggests that they are not relevant for our discussion; they give a very tiny mass in units of the radius (a full analysis is beyond the scope of this paper). So for our purposes, we will treat the fermions as  massless within the radius \nref{RCorona}.
 This implies that we have roughly the model we discussed in this paper.\footnote{Depending on the temperature, some charged fermions can continue beyond the corona. The neutrinos can also continue beyond, but would lose the large degeneracy factor.}  We have an $\AdS_2 \times S^2$ near horizon region described by JT gravity and coupled to a CFT, this CFT has transparent boundary conditions at the end of the throat where we transition into almost flat space, which we view as the bath in our model. This region has length $L = r_c$, because beyond that region the fermions become massive (or, in the case of the neutrino, the large $Q $ degeneracy will be lost). The corona is a complicated object, and approximating it as a simple boundary condition for all the fermions is, of course, an approximation.

  There are a few relevant length scales (or time scales) as follows. $\check \phi_r$ sets the scale of the original Schwarzian coupling. $ 12 \check \phi_r /( \pi c) $ is the evaporation time scale of the black hole due to the CFT. $L=r_c$ is the size of the bath.
  Finally we also have the effective Schwarzian coupling at scales longer than $L$, which is
  \be \la{NewCoupL}
  \check \phi_{\rm eff} = \check \phi_r + { c \over 6 } \pi r_c
  \ee
  We plot the inverse of all these scales as a function of $Q$ or the mass $M(Q)$ of the black hole in figure \ref{fig:Scales}.
  From this plot we can see which approximation governs the physics at each length scale. We start at very high temperatures or high energies where the basic Schwarzian dominates, then at the energy scale  $c/\check \phi_r $ we could enter the regime where the CFT dominates. When we consider $\beta > 4 \pi r_c$ we can now go back to a new Schwarzian description with the new coupling \nref{NewCoupL}.

  As a fun example, consider a black hole of a mass of 10 kg. If we start at very high energy scales, the original Schwarzian term dominates. As we lower the energy,  we then cross the solid green line of figure \ref{fig:Scales} into the yellow region, which is described by the CFT induced gravity. As we continue to lower energy, we cross the orange line, which marks the size scale of the corona; below it, we again have a Schwarzian term with a renormalized coefficient that is initially weakly coupled. However, as we cross the solid black line, we enter the highly quantum Schwarzian regime at an energy scale of $1/\check \phi_{\rm eff}$, which is roughly comparable to reasonable lab energies and temperatures. This is a toy model estimate. For example, for this case, the electron Compton wavelength is roughly 36 times the corona radius. This implies that the induced coupling could be a factor of 4 larger\footnote{ A factor 36 times a factor 1/9 from the fact that the central charge is now reduced to only the electron contribution. } and therefore the quantum gravity energy scale a factor of 4 smaller. So we could test quantum gravity on a tabletop, {\it if we could find one of these black holes.}

  Our objective here was to emphasize that black hole physics need not involve astronomically large black holes; the laws of the Standard Model allow small, long-lived ones where the quantum effects arise at accessible energies. Of course, these black holes are very hard to produce, even in the early universe.

\section*{Acknowledgments}

We would like to thank Andreas Blommaert, Thomas Dumitrescu, Luca Iliesiu,  David Kolchmeyer, Thomas Mertens, Brandon Rayhaun, Douglas Stanford, Mykhaylo Usatyuk, Yifan Wang, Edward Witten, and Zhenbin Yang for discussions.

The work of JM was supported in part by U.S. Department of Energy grant DE-SC0009988.

Most of the work here was done by its human authors. However, we have used AI for some computations and to help with writing the paper.

\appendix

\section{Shift in the size of the reference cylinder after welding}

\label{app:relative_radial_scale}

The CFT partition function $Z_{\text{CFT}}$ contains both the conformal anomaly and the reference partition function $Z_{\text{flat}} (s)$, where we emphasize the dependence on the total length of the cylinder $s$, measured in the flat reference metric with the circumference fixed to $2 \pi$. A fluctuating reparametrization along the interface between the gravitational region and the bath, in general, changes the length $s$ of the reference cylinder $s \rightarrow s+\Delta y$, which leads to the variation
\lceq{
\Delta \log Z_{\rm flat} (s) = \log Z_{\rm flat} (s+\Delta y) - \log Z_{\rm flat} (s) \, .
}
In most of the discussions below we consider infinitesimal variation so it is enough to keep the linear order of $\Delta y$, then \lceq{
\Delta \log Z_{\text{flat}} =  \Delta y \partial_{s} \log Z_{\text{flat}}  \,  \, .
}
In this appendix we explain how to determine the shift $\Delta y$ for the cases we studied in the main text.

Of course, if the contribution to  $Z_{\rm flat}$  involves an energy eigenstate, such as the vacuum or a conformal primary state flowing along the closed channel, then the relation holds for any $\Delta y$, not necessarily small:
\be
\Delta \log Z_{\rm flat} = - \Delta y E
\ee
where $E$ is the energy ($E= - c/12$ for the vacuum, for example).

\subsection{Geometries with the bath}
\paragraph{Double trumpet geometry.}
It is more intuitive to start with the case where the double trumpet is coupled to two baths of finite size. We use $w = \theta+ i \sigma$ in the gravitational region, $v_a = \varphi_a + i \mu_a$ ($a=L,R$) in the baths, and the conformal maps $z = x+iy = G(w) = F_a (v_a)$. For a fixed waist length $b$, $\kappa = 2\pi^2/b$ with $-\kappa< \sigma<0$. The bath coordinate runs over $-\kappa -\ell_L  \le \mu_L \le -\kappa $ and $0 \le \mu_R \le \ell_R$. The conformal welding condition is shown in~\eqref{CylBdy}. The conformal maps send the two boundaries of the baths  to circles of constant $y$ in the reference cylinder, denoted by $Y_L$ and $Y_R$:
\lceq{
\Im F_L (\varphi_L - i (\kappa + \ell_L)) = Y_L, \quad \Im F_R (\varphi_R + i \ell_R ) = Y_R \, .
}
The variation of the length of the reference cylinder is
\lceq{
\Delta y = Y_R - Y_L - \left(\kappa + \ell_L + \ell_R \right) \, .
}
Since $Y_L$ is independent of $\varphi_L$, one can write the above equations using the angular mean, for example,
\lceq{
Y_L - (-\kappa - \ell_L) = \frac{1}{2 \pi}\int  d \varphi_L \Im \left[ F_L (\varphi_L+i \mu_L) - (\varphi_L+i \mu_L)\right] \, .
}
$F_L(v_L) - v_L$ is holomorphic, so we can move the integral to the interface. The same conclusion holds for the holomorphic functions $F_R(v_R) - v_R$ and $G(w) - w$. We can write the $\Delta y$ in terms of the integral along the interface
\lceq{
\label{eq:appA_dt_sep}
\Delta y = \frac{1}{2\pi}\int d \varphi_R \Im \left[F_R(\varphi_R) - \varphi_R\right] - \frac{1}{2\pi}\int d \varphi_L \Im \left[F_L(\varphi_L - i \kappa ) - (\varphi_L - i \kappa )\right] \, .
}
One will find that if we use this version, we need to solve the conformal welding problem to the quadratic order, since $f_0^a$ (see Fourier expansion in~\eqref{eq:cyl_map_modes}) contains a quadratic term of the fluctuation $\varepsilon$. To avoid introducing extra complexities, notice that we have the following identity
\longeq{
0 &= \tpiint d \theta_R \Im \left[ G (\theta_R)  -\theta_R\right] - \tpiint d \theta_L \Im \left[ G(\theta_L - i \kappa) - (\theta_L-i\kappa) \right]   \\
&= \tpiint d \varphi_R \theta_R' \Im \left[F_R(\varphi_R) - \varphi_R\right]  - \tpiint d \varphi_L \theta_L' \Im \left[F_L (\varphi_L -  i \kappa) - (\varphi_L -  i \kappa)\right] \, ,
}
where we use the conformal welding condition~\eqref{CylBdy} and $\theta_a - \varphi_a$ is real. Subtracting this identity from~\eqref{eq:appA_dt_sep} gives
\longeq{
\label{eq:appA_dt_sep2}
\Delta y = \frac{1}{2 \pi } \tpiint d \varphi_L (\theta_L'-1) &\Im \left[F_L(\varphi_L - i \kappa) - (\varphi_L-i \kappa)\right] \\
&- \frac{1}{2 \pi} \tpiint d \varphi_R (\theta_R' - 1) \Im \left[F_R(\varphi_R) - \varphi_R\right] \, .
}
Now, since $\theta'-1$ is of first order in $\varepsilon$ we only need to solve the conformal welding problem to first order in $\varepsilon$ in order to get the quadratic terms in the action. Note that \nref{eq:appA_dt_sep2} is valid for any reparametrization and is not restricted to quadratic order; it holds for all orders.

\paragraph{Disk geometry.} For the disk geometry, $w = \theta+i \sigma$ describes the disk region with $-\infty < \sigma < \sigma_b$ and the bath $v = \varphi + i \mu$ has $0 \le \mu \le \ell$, if the bath is finite. The outer bath boundary is straight in the reference cylinder; we have the boundary conditions~\eqref{eq:sec2_bdy}. Define two offsets by
\lceq{
\Delta y_G = \lim_{\sigma \rightarrow -\infty} \left[\Im G(w) - \sigma \right], \quad \Delta y_F = \Im F(\varphi+ i \ell) - \ell \, ,
}
the shift of the length is then
\lceq{
\Delta y =  \Delta y_F- \Delta y_G  = \frac{1}{2\pi}\int d \varphi \Im F(\varphi) - \frac{1}{2\pi}\int d \theta \Im G(\theta) \, ,
}
where we have used the holomorphic property of $F(v) -v$ and $G(w) - w$ and keep the terms at the leading order of the cutoff. For the infinite bath case, due to the holomorphicity, the expression is exactly the same by taking $\ell \rightarrow \infty$. This is exactly the expression we write in~\eqref{eq:sec2_delta}.

\subsection{Geometries without the bath}
In appendix~\ref{app:CFT reduced to the boundary}, we also investigate the geometries without coupling to the bath. Without the bath, the CFT ends on the gravitational cutoff curves. Those computations also include the effect of the change in the length of the reference cylinder after we turn on the gravitational fluctuations and apply the conformal mapping. In preparation, we determine the changes here.
\paragraph{Double trumpet without the bath.} For the double trumpet region $w = \theta + i \sigma$ with $-\kappa < \sigma<0$, the CFT ends on the cutoff curves located at $\sigma = -\kappa+ \sigma_L$ and $\sigma = -\sigma_R$, with $\epsilon_a$ expansion shown in~\eqref{eq:app_sigmaa}. At the gravitational saddle $\theta_a = \varphi_a$, $\sigma_a = \epsilon_a+O(\epsilon_a^3)$. The forward map $z = H(w)$ sends these curves to the straight ends at the flat cylinder and the change of the length is exactly
\lceq{
\Delta y = \Im H(\theta_R - i \sigma_R) - \Im H (\theta_L + i (-\kappa+\sigma_L)) - (\kappa - \epsilon_L - \epsilon_R)\, .
}
Decompose the map $H(w) = w + \eta(w)$ with $\eta = O(\epsilon)$. Due to the fact that $\eta(w)$ is holomorphic and $\Im H$ at the boundaries are constants,
\longeq{
\Im H (\theta_R - i \sigma_R) &= - \frac{1}{2\pi} \int d \theta_R \sigma_R + \frac{1}{2\pi} \int d \theta_R \Im \eta(\theta_R - i \sigma_R) \\
&\hspace{-0.5cm}= -  \frac{1}{2\pi} \int d \theta_R \sigma_R + \frac{1}{2\pi} \int d \theta_L \Im \eta(\theta_L + i (-\kappa+\sigma_L))\\
&\hspace{-0.5cm}= -\frac{1}{2\pi} \int d \theta_R \sigma_R + \Im H(\theta_L +i (-\kappa + \sigma_L)) - \frac{1}{2\pi} \int d \theta_L(-\kappa + \sigma_L) \, .
}
Therefore
\lceq{
\label{eq:appA_dt_w/o_bath}
\Delta y = - \frac{1}{2 \pi} \sum_{a = L,R} \int_0^{2 \pi} d \varphi_a \theta_a' (\sigma_a - \epsilon_a) = - \sum_{a = L,R} \epsilon_a \left[\frac{1}{2 \pi}\int_0^{2 \pi} d \varphi_a \theta_a'^2 -1 \right] + O(\epsilon^2)\, .
}

\paragraph{Disk without the bath.}
For the disk, the discussion is similar; one can find the corresponding shift in the disk case by freezing one side of the gravitational fluctuation, say $\theta_L \equiv \varphi_L$, and taking $\kappa \rightarrow \infty$. The final expression is simply
\lceq{
\label{eq:appA_disk_w/o_bath}
\Delta y = - \frac{\epsilon}{2 \pi} \int^{2 \pi}_0 d \varphi (\theta'^2 - 1) + O(\epsilon^2) \, .
}

\subsection{ Form of the welding action in the mathematics literature}
\la{TTAction}

For convenience, we rewrite the action here in the form of \cite{TakhtajanTeo2004}, with minor changes to the notation to make it more similar to ours.

They consider a reparametrization $\theta(\varphi)$ and a disk $\mathbb{D}$ with coordinate
$\tilde w$, $|\tilde w |\leq 1$, and such that for $|\tilde w|=1$ we have $\tilde w = e^{ i\theta } $. We use $\tilde v$ to describe the exterior of the
disk $\mathbb{D}^*$ with $|\tilde v|\ge 1 $, such that for $|\tilde v|=1$ we have $\tilde v = e^{ i \varphi} $. We also have the coordinate $z$ on the plane and the holomorphic maps
 $g:\mathbb{D}\to\Omega$
and $f:\mathbb{D}^*\to\Omega^*$ satisfying $g \circ \theta =f $ on
$S^1$. Here  $\Omega$ and $\Omega^*$ are the images of $\mathbb{D}$ and $\mathbb{D}^*$  in the $z$ plane. In general, they are some wiggly deformations of disks.
 With
$g(0)=0$ and $f(\infty)=\infty$, the Takhtajan--Teo action
(or ``universal Liouville action'') is
\begin{equation} \label{TTAct}
S_{\mathrm{TT}}[\theta]
=
\int_{\mathbb{D}}
\left|\frac{g''(z)}{g'(z)}\right|^2\,d^2z
+
\int_{\mathbb{D}^*}
\left|\frac{f''(z)}{f'(z)}\right|^2\,d^2z
+
4\pi\log\left|\frac{g'(0)}{f'(\infty)}\right|,
\end{equation}
where $d^2z=dx\,dy$ and
$f'(\infty):=\lim_{z\to\infty}f(z)/z$. The last term in \nref{TTAct} is related to the $E \Delta y$ term in \eqref{eq:sec2_sumI}, with $E = - c/12$.
The first two terms in \nref{TTAct} can be integrated by parts to give something purely at the boundary, producing the rest of \nref{eq:sec2_sumI}.

\section{ JT gravity with a CFT and no bath at finite cutoff}
\label{app:CFT reduced to the boundary}

In the usual JT gravity discussion, the action reduces to Schwarzian dynamics near the cutoff. In this appendix we derive the first correction in the cutoff for a CFT confined to the gravitational region with no external bath. We revisit the disk calculation of~\cite{Yang:2018gdb} and extend it to the double trumpet. The summary is that we induce a Schwarzian term in the action with a coefficient that is proportional to the cutoff $\epsilon$.

A more general discussion of gravity with an induced action given by the conformal anomaly on the disk (with no bath) was given in \cite{Chaudhuri:2024yau}. Their semiclassical limit is $c$ very negative. What we discuss here can also be viewed as a simple limit of theirs.

\paragraph{Disk geometry.}
For the metric~\eqref{Diskme} and cutoff curve $w_b=\theta(\varphi)+i\sigma_b(\varphi)$, the induced metric condition gives
\lceq{
\label{eq:nb_disk_cutoff}
\frac{\theta'^2+\sigma_b'^2}{\sinh^2\sigma_b}=\frac1{\epsilon^2},
\qquad
\sigma_b=-\epsilon\theta'
-\epsilon^3\left(\frac{\theta''^2}{2\theta'}-\frac{\theta'^3}{6}\right)
+O(\epsilon^5).
}
In contrast with ~\eqref{eq:sec2_displacement}, here the cubic term is needed because the radial integral contains
$1/\sigma_b$ and we are interested in the action up to the linear order of the cutoff. Introduce the forward conformal map and the boundary condition
\lceq{
\label{eq:nb_disk_map}
z=H(w)=w+\eta(w),\qquad \eta(w+2\pi)=\eta(w),\qquad
\Im H(w_b)=0,\qquad \eta=O(\epsilon).
}
It maps the regulated disk to the half-cylinder $z=x+iy$ ($y\le 0$) with reference metric $d\hat s^2=|dz|^2$ and straight circle at the end. The Weyl factor is simply
\lceq{
\label{eq:nb_disk_rho}
\rho=\rho_0+\rho_H,\qquad
\rho_0=-\log|\sinh\sigma|,\qquad
\rho_H=-\log|H'(w)|.
}
Since the extrinsic curvature of the straight circle is zero, we only need to evaluate the first term in~\eqref{ConfAn}. We can split the integral into three terms as we did in~\eqref{eq:sec2_split}. Direct computation gives
\longeq{
\label{eq:nb_disk_background}
-\mL_{\rm b}
&=\int_0^{2\pi}d\varphi\,\theta'
  [\sigma-\coth\sigma]_{\sigma_{\min}}^{\sigma_b}\\
&=\frac{2\pi}{\epsilon}
 -2\pi(\sigma_{\min}-\coth\sigma_{\min})
 -\frac\epsilon2\int_0^{2\pi}d\varphi
 \left[\theta'^2+\left(\frac{\theta''}{\theta'}\right)^2\right]
 +O(\epsilon^3).
}
To evaluate the mixed term $-\mL_\times$, we define
$\psi(\alpha)=\theta'(\theta^{-1}(\alpha))$. The boundary condition
in~\eqref{eq:nb_disk_map} implies
\lceq{
\label{eq:nb_disk_map_data}
\Im\eta(\alpha)=\epsilon\psi(\alpha)+O(\epsilon^2),\quad
\eta(w)=\frac\epsilon{2\pi}\int_0^{2\pi}d\alpha\,
 \cot\frac{w-\alpha}{2}\,\psi(\alpha)+O(\epsilon^2),
\quad \Im w<0.
}
Since
$\rho_H=-\Re\eta'+O(\epsilon^2)$, the Cauchy--Riemann equations give
\lceq{
\label{eq:nb_disk_normal}
\partial_n\rho_H\,ds_w
=\partial_\sigma \rho_H \,d\alpha+O(\epsilon^2)
=\epsilon\psi''(\alpha)\,d\alpha+O(\epsilon^2).
}
The field $\rho_H$ is harmonic and its derivatives vanish
at the center. Integration by parts therefore gives
\longeq{
\label{eq:nb_disk_mixed}
-\mL_\times
&=2\oint\rho_0\partial_n\rho_H\,ds_w
=-2\epsilon\int_0^{2\pi}d\alpha\,
       \log\psi\,\psi''+O(\epsilon^2)\\
&=2\epsilon\int_0^{2\pi}d\alpha\,\frac{\psi'^2}{\psi}
  +O(\epsilon^2)
=2\epsilon\int_0^{2\pi}d\varphi
       \left(\frac{\theta''}{\theta'}\right)^2+O(\epsilon^2).
}
We used $\rho_0|_\partial=-\log(\epsilon\psi)+O(\epsilon^2)$ and
$d\alpha=\theta'd\varphi$. The constant $\log\epsilon$ multiplies
total derivative and hence drops out exactly. Also
$-\mL_H=O(\epsilon^2)$, since $\rho_H=O(\epsilon)$ is regular. The reference partition function contributes $\Delta\log Z_{\rm flat}=-E\Delta y$, with $E=-c/12$ and length shift $\Delta y$ derived in~\eqref{eq:appA_disk_w/o_bath}. As in the main-text disk calculation, the dependence on the auxiliary regulator $\sigma_{\min}$ cancels once the smooth filling at the disk center is restored. Combining all these contributions, and subtracting the $1/\epsilon$ divergence, we get
\begin{equation}
\label{eq:nb_disk_result}
-I_{\rm CFT}
=-\frac{c\epsilon}{8\pi}
\int_0^{2\pi}d\varphi\,
\left\{\tan\frac{\theta}{2},\varphi\right\}
+O(\epsilon^2).
\end{equation}
Notice that the Schwarzian coefficient has the opposite sign from the one in JT gravity.  Under the formal continuation $\epsilon\to-\frac{2}{3}\ell$, this expression reproduces the short-bath Schwarzian action derived in~\eqref{ESchCou}. In appendix~\ref{app:short_bath}, we explicitly show that this match holds at leading order in $\ell$, without expanding the boundary reparametrization.

\paragraph{Double trumpet geometry.}
We now generalize the discussion to the double trumpet, equipped with the metric~\eqref{eq:sec4_g_cyl}. Write the two cutoff curves as $w_L=\theta_L+i(-\kappa+\sigma_L)$ and $w_R=\theta_R-i\sigma_R$, where $\sigma_a>0$. Requiring flat induced metric along these curves we have
\lceq{
\label{eq:app_sigmaa}
\frac{\gamma^2}{\sin^2(\gamma\sigma_a)}
 (\theta_a'^2+\sigma_a'^2)=\frac1{\epsilon_a^2},\qquad
\sigma_a=\epsilon_a\theta_a'
+\epsilon_a^3\left[\frac{\theta_a''^2}{2\theta_a'}
                  +\frac{\gamma^2}{6}\theta_a'^3\right]
+O(\epsilon_a^5).
}
Similarly, to evaluate~\eqref{ConfAn}, we introduce the conformal map $H(w)=w+\eta(w)$ that maps the double trumpet to a flat cylinder with two straight circles as boundaries, labeled as $y=Y_L$ and $y=Y_R$ respectively. Its boundary conditions are therefore
\begin{equation}
\label{eq:nb_dt_map_data}
\Im\eta(w_L)=Y_L+\kappa-\sigma_L,\qquad
\Im\eta(w_R)=Y_R+\sigma_R,\qquad
\eta(w+2\pi)=\eta(w).
\end{equation}
Their separation is $s=Y_R-Y_L=s_0+\Delta y$, where
$s_0=\kappa-\epsilon_L-\epsilon_R+O(\epsilon^3)$ and $\Delta y$ is derived in~\eqref{eq:appA_dt_w/o_bath}. The Weyl factor now is
\lceq{
\rho_0=\log\gamma-\log|\sin(\gamma\sigma)|,\qquad
\rho_H=-\log|H'|.
}
The background term is
\longeq{
\label{eq:nb_dt_background}
-\mL_{\rm b}
={}&2\pi\left(\frac1{\epsilon_L}+\frac1{\epsilon_R}\right)
-2\pi\gamma^2\kappa+\frac12\sum_{a=L,R}\epsilon_a\int_0^{2\pi}d\varphi_a
 \left[\gamma^2\theta_a'^2
          -\left(\frac{\theta_a''}{\theta_a'}\right)^2\right]
+O(\epsilon^3).
}
Let
$\psi_a(\alpha)=\theta_a'(\theta_a^{-1}(\alpha))$, by similar consideration as in~\eqref{eq:nb_disk_mixed}, the cross term yields
\lceq{
\label{eq:nb_dt_mixed}
-\mL_\times=2\sum_{a=L,R}\epsilon_a\int_0^{2\pi}d\varphi_a
 \left(\frac{\theta_a''}{\theta_a'}\right)^2+O(\epsilon^2)
}
and the term $-\mL_H$ is negligible to the first order of $\epsilon_a$. For fixed conformal boundary conditions $B_L,B_R$, the reference partition function is~\eqref{eq:cyl_reference}. Similarly, define
\lceq{
\label{eq:nb_dt_energy}
E(s):=-\partial_s\log Z_{\rm flat}(s),\quad
\log Z_{\rm flat}(s)=\log Z_{\rm flat}(\kappa)
+\frac{E(\kappa)}{2\pi}\sum_a\epsilon_a
  \int_0^{2\pi}d\varphi_a\,\theta_a'^2+O(\epsilon^2).
}
After subtracting the divergences, the full answer through first order is
\longeq{
\label{eq:nb_dt_expanded}
-I_{\rm CFT}
={}&-\frac{c\pi^2}{12\kappa}+\log Z_{\rm flat}(\kappa)\\
&+\frac{c}{16\pi}\sum_a\epsilon_a\int_0^{2\pi}d\varphi_a
 \left[\left(\frac{\theta_a''}{\theta_a'}\right)^2   + \left(\frac{\gamma^2}{3} + \frac{8 E(\kappa)}{c}\right) \theta_a'^2\right]
+O(\epsilon^2).
}
Equivalently, using $\int\{e^{\gamma\theta_a},\varphi_a\}=-\frac12\int[(\theta_a''/\theta_a')^2+\gamma^2\theta_a'^2]$, we can write
\lceq{
\label{eq:nb_dt_result}
-I_{\rm CFT}
=- \frac{c\pi^2}{12 \kappa} + \log Z_{\rm flat}(\kappa) - \frac c{8\pi}\sum_{a=L,R}\epsilon_a
 \int_0^{2\pi}d\varphi_a\,\{e^{\gamma_{\rm eff}\theta_a},\varphi_a\}
+O(\epsilon^2),
}
with the effective $\gamma_{\rm eff}$
\lceq{
\label{eq:appB_gammaeff}
\gamma_{\rm eff}^2 := \frac{\gamma^2}{3} + \frac{8 E(\kappa)}{c} \, .
}
Notice the effective coupling is again of the opposite sign and consistent with the formal continuation $\epsilon_a \rightarrow - \frac{2}{3} \ell_a$ to match the coefficients in~\eqref{eq:cyl_sb_quadratic_schwarzian}.

\section{Nonlinear short-bath limit}
\label{app:short_bath}
In the main text, we argued that the CFT action reduces to a Schwarzian action in the short-bath limit, both for the disk and, under the stated assumptions on the reference partition function, for the double trumpet. We checked this at quadratic order in the reparametrization fluctuation $\varepsilon$. Here we derive the result without expanding around the round configuration, instead expanding in the bath lengths at fixed smooth boundary reparametrizations.

\paragraph{Disk coupled to a short bath.}
\label{app:short_bath_disk}
Use $w=\theta+i\sigma$ with $\sigma<0$ in the disk and $v=\varphi+i\mu$ with $0\leq\mu\leq\ell$ in the bath. The maps satisfy~\eqref{eq:sec2_welding} and~\eqref{eq:sec2_bdy}. Taylor expansion across the bath then yields
\begin{equation}
 \operatorname{Im}F(\varphi)
+\ell\,\operatorname{Re}F'(\varphi)=\ell+O(\ell^2),
 \qquad
 \operatorname{Im}F(\varphi)=\ell(1-\theta')+O(\ell^2).
 \label{eq:disk_loop_short_boundary}
\end{equation}
Differentiating along the interface and expanding the ratio gives
\lceq{
 \Im \frac{F''}{F'} =\frac{\Im F''}{\theta'}
   -\frac{\theta''\Im F'}{\theta'^2}+O(\ell^2)=-\ell\,\partial_\varphi\!\left(\frac{\theta''}{\theta'}\right)
   +O(\ell^2).
}
We can then easily derive the effective action~\eqref{ActDisk} as
\longeq{
 - I_{\rm CFT}
 &=\frac{c\ell}{12}
   -\frac{c\ell}{24\pi}\int_0^{2\pi}d\varphi\,
       \left[\left(\frac{\theta''}{\theta'}\right)^2
                         -(1-\theta')^2\right]+O(\ell^2) \\
 &=\frac{c\ell}{12\pi}\int_0^{2\pi}d\varphi\,
       \left\{\tan\frac{\theta}{2},\varphi\right\}+O(\ell^2) \, .
}
where we have chosen $E = -c/12$. It reproduces the Schwarzian action explicitly.

\paragraph{Double trumpet coupled to two short baths.}
\label{app:short_bath_double} For the double trumpet coupled to two short baths, we have the conformal welding condition~\eqref{CylBdy} and the boundary conditions~\eqref{eq:sec4_YLYR}.  Since $G'=1+O(\ell_L+\ell_R)$, we have $F_a'=\theta_a'+O(\ell_L+\ell_R)$ at the interfaces. Taylor expansion across the baths gives
\begin{equation}
\begin{aligned}
\Im F_L(\varphi_L-i\kappa)
&=Y_L+\ell_L\theta_L'+O((\ell_L+\ell_R)^2),\\
\Im F_R(\varphi_R)
&=Y_R-\ell_R\theta_R'+O((\ell_L+\ell_R)^2).
\end{aligned}
\label{eq:sb_double_interface_data}
\end{equation}
As in the disk case, this implies
\begin{equation}
\begin{aligned}
\Im\frac{F_L''}{F_L'}
&=\ell_L\partial_{\varphi_L}
  \left(\frac{\theta_L''}{\theta_L'}\right)
  +O((\ell_L+\ell_R)^2),\\
\Im\frac{F_R''}{F_R'}
&=-\ell_R\partial_{\varphi_R}
  \left(\frac{\theta_R''}{\theta_R'}\right)
  +O((\ell_L+\ell_R)^2),
\end{aligned}
\label{eq:sb_double_log_derivatives}
\end{equation}
where the ratios are evaluated at their respective interfaces.
The length-shift formula~\eqref{eq:cyl_width} then gives
\begin{equation}
s=Y_R-Y_L
=\kappa+\sum_{a=L,R}\frac{\ell_a}{2\pi}
\int_0^{2\pi}d\varphi_a\,\theta_a'^2
+O((\ell_L+\ell_R)^2).
\label{eq:sb_double_length}
\end{equation}
Substituting into~\eqref{ActCylM}, restoring the round reference
contribution $\log Z_{\rm flat}(s_0)$, and integrating by parts gives
\begin{equation}
\begin{aligned}
-I_{\rm CFT}
={}&-\frac{cb}{24}+\log Z_{\rm flat}(\kappa)\\
&-\frac{c}{24\pi}\sum_{a=L,R}\ell_a
\int_0^{2\pi}d\varphi_a
\left[
\left(\frac{\theta_a''}{\theta_a'}\right)^2
+\frac{12E(\kappa)}{c}\theta_a'^2
\right]
+O((\ell_L+\ell_R)^2).
\end{aligned}
\label{eq:sb_double_general}
\end{equation}
Here $E(\kappa)$ is the energy used in~\eqref{eq:cyl_sb_quadratic_general}. Under the open-channel vacuum-dominance
assumption~\eqref{eq:cyl_open_vacuum},
this reproduces the Schwarzian action~\eqref{eq:cyl_induced_schwarzians}, up to the constant $\log n_0$, without expanding around the round
configuration.

\section{Deforming the CFT and non vacuum Ishibashi states}
\label{app:deform}
We derive the order-$1/c$ corrections to the one- and two-point functions discussed in the main text. The scalar insertion $\mathcal O_h$ has weights $(h,h)$ and unit two-point normalization, while the identical bath operators $V$ have weights $(h_V,h_V)$. We work in the infinite-bath limit, neglecting the Schwarzian action, up to the first order of the coupling $\lambda$, while the perturbative expansion is based on order $1/c$. We set $h>1/2$ and $h,h_V$ fixed as $c\to\infty$.

\paragraph{Virasoro conformal block at large $c$.}
\label{app:insertion_block_review} As a warm-up, we recall the necessary knowledge of Virasoro conformal blocks in the CFT literature. For convenience in this appendix, we introduce the plane coordinates
\lceq{
 \zeta=e^{-iw},\quad \xi=e^{-iv},\quad \mathfrak{z}=e^{-iz} \, .
}
where $w=\theta+i\sigma$ is the coordinate of the disk with $\sigma<0$, $v=\varphi+i\mu$ with $\mu>0$ is the coordinate of the bath and $z$ is the coordinate of the reference flat cylinder. The doubling construction in the main text turns the two bath insertions into four chiral insertions at $(\xi_1,\xi_2,1/\bar\xi_2,1/\bar\xi_1)$, each of weight $h_V$. The presence of the Ishibashi state $|h\rangle \rangle$ is equivalent to inserting a projector $\mathcal{P}_h$ that selects the Virasoro family of $\mathcal O_h$ to propagate in the intermediate channel. Namely, the two point correlator in the presence of the Ishibashi state $|h\rangle \rangle$ can be written as
\lceq{
\frac{\langle0|V_1V_2|h\rangle\!\rangle}{\mathcal G^{(0)}}
 =C_{VV\mathcal O_h}\,\chi^h\mathcal F_h(\chi),\quad \mathcal F_h(\chi)={}_2F_1(h,h;2h;\chi)+\mathcal F_h^{(1)}(\chi)+O(c^{-2}).
}
where $\chi$ is the conformal cross ratio~\eqref{ChiDe} written in the plane coordinate
\lceq{
\chi=\frac{|\xi_1-\xi_2|^2}{|1-\xi_1\bar\xi_2|^2} \, .
}
Here $0<\chi<1$, $\mathcal F_h(0)=1$, and $\mathcal F_h^{(1)}$ includes its factor of $1/c$. The full nonchiral coefficient $C_{VV\mathcal O_h}$ appears once. We also normalize the expression by the two point function $\mathcal{G}^{(0)}$~\eqref{eq:sec2_G0}.  In~\cite{Bombini:2018jrg} the authors derive $\mathcal{F}_h^{(1)}$ explicitly. Choose the basis $|h;q,A\rangle$ at level $q$, with Gram matrix $(M_q)_{AB}=\langle h;q,A|h;q,B\rangle$. The projector $\mathcal{P}_h$ is
\begin{equation}
 \mathcal P_h=\sum_{q\ge0}\sum_{A,B}
 |h;q,A\rangle(M_q^{-1})_{AB}\langle h;q,B|,
 \label{eq:insapp_projector}
\end{equation}
with null states removed before inversion. The global descendants $L_{-1}^q|h\rangle$ give the leading contribution
\begin{equation}
 \mathcal F_h^{(0)}(\chi)
 =\sum_{q\ge0}\frac{(h)_q^2}{q!(2h)_q}\chi^q
 ={}_2F_1(h,h;2h;\chi),\qquad
 (a)_q=\frac{\Gamma(a+q)}{\Gamma(a)}.
 \label{eq:insapp_global_block}
\end{equation}
At the next order, one needs states with one additional generator $L_{-s}$, $s\ge2$. However, those states are not orthogonal to the global descendants, so it is essential to remove the overlap. We can define those states as
\lceq{
 |q,s\rangle_\perp
 =L_{-s}L_{-1}^{q-s}|h\rangle-a_{q,s}L_{-1}^q|h\rangle, \quad  a_{q,s}=[h(s+1)+q-s]\frac{(2h)_{q-s}}{(2h)_q} \, ,
}
with the norm
\begin{equation}
 \begin{aligned}
 {}_\perp\langle q,s|q,s\rangle_\perp
 &=\frac c{12}s(s^2-1)(q-s)!(2h)_{q-s}+O(1).
 \end{aligned}
 \label{eq:insapp_orthogonalization}
\end{equation}
The remaining overlaps between distinct $s$ are $O(1)$ and affect the inverse Gram matrix only at order $c^{-2}$. Those states contribute to the Virasoro conformal block at the subleading order
\begin{equation}
 \mathcal F_h^{(1)}(\chi)
 =\frac{12}{c}\sum_{q=2}^{\infty}\chi^q\sum_{s=2}^q
 \frac{\left[X_{q,s}-a_{q,s}(h)_q\right]^2}
 {s(s^2-1)(q-s)!(2h)_{q-s}}, \quad  X_{q,s}=[h_V(s-1)+h+q-s](h)_{q-s}
 \label{eq:insapp_target}
\end{equation}
We will show explicitly that the gravitational computation reproduces this result through order $1/c$, providing a perturbative check of the proposal.

\paragraph{Conformal welding problem on the plane.} To prepare for the following computation, we solve the conformal welding problem in the plane coordinate first. Exponentiating the cylinder welding problem $z=G(w)=F(v)$ gives
\begin{equation}
 \begin{gathered}
 \widehat G(\zeta)=e^{-iG(w)},\qquad
 \widehat F(\xi)=e^{-iF(v)},\qquad
 \widehat G(e^{-i\theta(\varphi)})
       =\widehat F(e^{-i\varphi}).
 \end{gathered}
 \label{eq:insapp_plane_welding}
\end{equation}
The maps $\widehat G$ and $\widehat F$ are holomorphic inside and outside the unit circle, respectively. We keep the exterior normalization $F(v)-v\to0$ as $\mu\to\infty$, or equivalently $\widehat F(\xi)/\xi\to1$. We can expand the maps based on the order of $\vare$, which we denote using the superscript, and solve the linear order conformal welding~\eqref{eq:insapp_plane_welding}
\begin{equation}
 \begin{aligned}
 \widehat F(\xi)&=\xi+\widehat f^{(1)}(\xi)
                         +\widehat f^{(2)}(\xi)+\cdots,&
 \widehat f^{(1)}(\xi)&=-i\sum_{n\ge2}\varepsilon_n\xi^{1-n},\\
 \widehat G(\zeta)&=\zeta+\widehat g^{(1)}(\zeta)
                         +\widehat g^{(2)}(\zeta)+\cdots,&
 \widehat g^{(1)}(\zeta)&=i\sum_{n\ge2}\varepsilon_{-n}\zeta^{n+1}.
 \end{aligned}
 \label{eq:insapp_linear_plane}
\end{equation}

\paragraph{One point function.}
Due to the presence of the boundary, the one point function of the primary operator $\mathcal O_h$ is no longer zero. In the plane coordinates, the disk metric and the bath metric become
\begin{equation}
 ds_w^2=\frac{4\,d\zeta\,d\bar\zeta}{(1-|\zeta|^2)^2}, \quad ds_v^2 = \frac{d \xi d \bar{\xi}}{\epsilon^2|\xi|^2} \, .
\end{equation}
The Weyl transformation of the primary gives
\begin{equation}
 \int_{\mathcal D_w} d^2w \sqrt{g_w} \mathcal O_h(w, \bar{w})
 =
 2^{2-2h}\int_{|\zeta|<1}d^2\zeta\,
 (1-|\zeta|^2)^{2h-2}\,
 \mathcal O_h^{\rm flat}(\zeta,\bar\zeta).
\label{AppOflat}\end{equation}
Here the ``flat'' denotes the plane normalization of the primary\footnote{
For the plane-normalized operator $\mO_h^{\rm flat} (\xi, \bar{\xi})$, the two point function is simply $\langle \mO_h^{\rm flat} (\xi_1, \bar{\xi}_1) \mO_h^{\rm flat} (\xi_2, \bar{\xi}_2)\rangle = |\xi_1 - \xi_2|^{-4h}$. Based on the transformation law of the primary operator, it is related to the cylinder operator $\mO_h(v,\bar{v})$ via
\lceq{
\mO_h(v,\bar{v}) = \left|\frac{d \xi}{d v}\right|^{2 h} \mO^{\rm flat}_h(\xi, \bar{\xi}) = |\xi|^{2h} \mO^{\rm flat}_h (\xi, \bar{\xi}),
}
with the normalization $\langle \mO_h(v_1,\bar{v}_1)\mO_h(v_2,\bar{v}_2)\rangle =|2 \sin \frac{v_{12}}{2}|^{-4h}$ as shown in~\eqref{eq:sec2_G0}.
}.
Using
\lceq{
\mathcal O_h^{\rm flat}(\zeta,\bar\zeta)|0\rangle
=e^{\zeta L_{-1}+\bar\zeta\bar L_{-1}}|h,h\rangle = \sum_{j,k=0}^\infty \frac{\zeta^j \bar{\zeta}^k}{j! k!} L_{-1}^j \bar{L}_{-1}^k \ket{h,h},
}
the angular integral pairs equal left and right levels and yields \cite{Goto2016Causal}
\begin{equation}
 \left[\int_{\mathcal D_w} d^2w \sqrt{g_w} \mathcal O_h(w, \bar{w}) \right] \ket{0}
 =
 \nu_h^{(0)}
 \sum_{n=0}^{\infty}
 \frac{L_{-1}^{\,n}\bar L_{-1}^{\,n}}
      {n!(2h)_n}|h,h\rangle,
 \qquad
 \nu_h^{(0)}=\frac{\pi\,2^{2-2h}}{2h-1},
 \label{eq:primary_global_ishibashi}
\end{equation}
For the plane-normalized bath operator $\mO^{\rm flat}_h (\xi, \bar{\xi})$, to the leading order of $\lambda$, the correction to the one point function is simply
\longeq{
\delta_\lambda \langle \mathcal{O}_h^{\rm flat} (\xi, \bar{\xi})\rangle  &= - \lambda \int_{\mathcal{D}_w} d^2 w \sqrt{g_w} \langle  \mO_h^{\rm flat} (\xi, \bar{\xi}) \mO_h (w, \bar{w})\rangle_0 \\
&= -\lambda 2^{2-2h} \int_{|\zeta|<1} d^2 \zeta \left(1- |\zeta|^2\right)^{2h-2} \left\langle \frac{|\widehat{F}'(\xi) \widehat{G}' (\zeta)|^{2h}}{|\widehat{F}(\xi) - \widehat{G}(\zeta)|^{4h}} \right\rangle_0 \, ,
}
where the subscript $0$ denotes the undeformed path integral governed by the action~\eqref{ActDisk}. Practically, to the order $1/c$, we can expand the above expression to quadratic order of $\vare$ and perform the Gaussian integral weighted by~\eqref{eq:sec2_inf_I2}. Direct computation gives
\lceq{
\label{eq:sec2_1pt}
\delta_\lambda \langle \mO^{\rm flat}_h (\xi, \bar{\xi}) \rangle  = - \frac{\lambda \nu_h}{\left(|\xi|^2 - 1\right)^{2h}}, \quad \nu_h = \nu_h^{(0)} \left[1 - \frac{6 (6h^2 - 6h+1)}{c(2h-1)} + O(c^{-2})\right] \, .
}
From the BCFT perspective, in the presence of the $h$ Ishibashi state as the boundary state, the one point function is
\lceq{
\langle 0 |\mO_h^{\rm flat} (\xi, \bar{\xi} ) |h\rangle \rangle  = \frac{1}{\left(|\xi|^2 -1\right)^{2h}} ,
}
which exactly matches with~\eqref{eq:sec2_1pt} up to the normalization. Higher orders in $1/c$ will only change  the overall normalization, while leaving $h$ and the position dependence unchanged.

Notice that the integral in \nref{AppOflat} diverges near $|\zeta| =1$ for $h\leq \half $.
This can be seen very clearly just from the leading order expression
\be
 \int_{|\zeta|<1} d^2 \zeta \left(1- |\zeta|^2\right)^{2h-2}   \frac{ 1 }{| \xi -  \zeta|^{4h}}   \, , ~~~{\rm with} ~~~~~~ |\xi| > 1,
\ee
which diverges near $|\zeta|=1$ if $h\leq \half$.

For $h \leq \half $, we would need to add a counterterm located at the interface, which is essentially given by the limit of the bulk operator to the boundary. Its final renormalized coupling is additional data.  This operator is relevant (or marginal for $2h=1$) and there will be an RG flow due to such a boundary operator and we would need to understand what effect it has on the boundary conditions. It might remove the transparent boundary conditions.  A free fermion mass term is an example with $h=1/2$. One way to avoid this issue is to consider a CFT where all scalar primary operators (other than the identity) have $h> 1/2$. The statements in this paragraph were to leading order in $c$, for general $c$ we will need a counterterm starting at $h \leq h_c$ \nref{eq:sec3_Ach}.

\paragraph{The two point function.}
More generally, we can place two identical primary scalar operators of conformal dimension $h_V$ in the bath and we are interested in how the gravitational path integral corrects the two point correlation function. We normalize the correction $\delta_\lambda \langle V_1 V_2 \rangle$ by the cylinder two point function $\mathcal{G}^{(0)}$ in~\eqref{eq:sec2_G0}. The complete expression is
\longeq{
\frac{\delta_\lambda \langle V_1 V_2\rangle}{\mathcal{G}^{(0)}} &= - \frac{\lambda}{\mathcal{G}^{(0)}} \int_{\mathcal{D}_w} d^2 w \sqrt{g_w} \langle V_1 V_2 \mO_h (w, \bar{w}) \rangle_0 \\
&= -\lambda 2^{2-2h} C_{VV\mathcal O_h}   |\xi_{12}|^{4 h_V} \int_{|\zeta|<1} d^2 \zeta \left(1 - |\zeta|^2\right)^{2h-2} \\
&\hspace{4cm}\times \left\langle \frac{|\widehat{F}_1'|^{2h_V}|\widehat{F}_2'|^{2 h_V}|\widehat{G}'|^{2h}}{|\widehat{F}_1 -\widehat{F}_2|^{4h_V - 2h} |\widehat{F}_1 - \widehat{G}|^{2h} |\widehat{F}_2 - \widehat{G}|^{2h}}  \right\rangle_0 \, ,
}
where $\widehat{F}_i = \widehat{F}(\xi_i)$ and $\xi_{12} = \xi_1 - \xi_2$. For simplicity, we can use the global conformal transformation to fix $\xi_1 \rightarrow \infty$ and $\xi_2 = 1/\sqrt{\chi}$, after taking the limit we obtain
\begin{equation}
\begin{aligned}
\frac{\delta_\lambda\langle V_1V_2\rangle}{\mathcal G^{(0)}}
=-\lambda 2^{2-2h} C_{VV\mathcal O_h}
\int_{|\zeta|<1}d^2\zeta\,
(1-|\zeta|^2)^{2h-2}\left\langle
\frac{|\widehat F'(1/\sqrt{\chi})|^{2h_V}
      |\widehat G'(\zeta)|^{2h}}
     {|\widehat F(1/\sqrt{\chi})-\widehat G(\zeta)|^{2h}}
\right\rangle_0.
\end{aligned}
\label{eq:insapp_two_point_limit}
\end{equation}
Expanding the kernel up to the order $c^{-1}$ gives
\lceq{
\left\langle
\frac{|\widehat F'(1/\sqrt{\chi})|^{2h_V}
      |\widehat G'(\zeta)|^{2h}}
     {|\widehat F(1/\sqrt{\chi})-\widehat G(\zeta)|^{2h}} \right\rangle_0= \frac{1}{\left|\xi_2 - \zeta\right|^{2h}} \left[1 + \langle |\mathfrak{B}_1|^2\rangle  + 2 \Re \langle \mathfrak{B}_2 \rangle + 2 \Re \langle \mathfrak{B}_3 \rangle   \right] \, ,
}
where
\begin{equation}
\begin{aligned}
\mathfrak B_1
&=h_V\widehat f^{(1)\prime}
  +h\widehat g^{(1)\prime}
  -h\frac{\widehat f^{(1)}-\widehat g^{(1)}}{\xi_2-\zeta},
\\
\mathfrak B_2
&=\frac12\mathfrak B_1^2
  -\frac{h_V}{2}(\widehat f^{(1)\prime})^2
  -\frac h2(\widehat g^{(1)\prime})^2
  +\frac h2
   \frac{(\widehat f^{(1)}-\widehat g^{(1)})^2}
        {(\xi_2-\zeta)^2},
\\
\mathfrak B_3
&=h_V\widehat f^{(2)\prime}
  +h\widehat g^{(2)\prime}
  -h\frac{\widehat f^{(2)}-\widehat g^{(2)}}{\xi_2-\zeta}.
\end{aligned}
\end{equation}
Here $\mathfrak B_1$ is linear in $\varepsilon$, while $\mathfrak B_2$ and $\mathfrak B_3$ are quadratic. Note that $\mathfrak B_3$ requires solving the conformal welding problem~\eqref{eq:insapp_plane_welding} to the second order
\lceq{
\langle \widehat{f}^{(2)}\rangle_0=0,
\qquad
\langle \widehat{g}^{(2)}\rangle_0=-\frac9c\,\zeta \, .
}
In summary, to the subleading order $1/c$, ~\eqref{eq:insapp_two_point_limit} produces the exact same result as what we expect in~\eqref{eq:insapp_target}
\lceq{
\label{eq:sec2_VVO}
\frac{\delta_\lambda \langle V_1 V_2\rangle}{\mathcal{G}^{(0)}} = - \lambda \left\{\nu_h C_{VV\mathcal O_h} \chi^h \left[{}_2F_1(h,h;2h;\chi)+ \mathcal{F}_h^{(1)} (\chi) \right]+ O (c^{-2})\right\},
}
where $\nu_h$ is also of the same form as what we obtain in the one point function~\eqref{eq:sec2_1pt}.

\section{ Explicit check of the Schwinger--Dyson equation for the disk}
\label{app:SDeq}
The Ward identity provides a direct check for general multipoint exterior correlators in a semi-infinite bath. Let $\mathcal G_F$ now denote the normalized correlator of primary operators $\prod_i V_i(v_i,\bar v_i)$ at fixed $\theta$, with dimensions $(h_i,\bar h_i)$ and distinct insertion points in the bath, $\mu_i>0$. As in the two-point case, it is obtained from the reference cylinder correlator by setting $z_i=F(v_i)$, $\bar z_i=\bar F(\bar v_i)$ and multiplying by the factors $F'(v_i)^{h_i}\bar F'(\bar v_i)^{\bar h_i}$.
\lceq{
\label{eq:sec2_GFn}
\mathcal{G}_F (v_1, \ldots, v_n) = \prod_i F'(v_i)^{h_i} \bar{F}' (\bar{v}_i)^{\bar{h}_i} \left \langle \prod_i V_i(z_i, \bar{z}_i) \right \rangle_z \, ,
}
where the $\langle\ldots \rangle_z$ denote the correlators evaluated in the $z$ coordinate. Varying the welding condition $G(\theta(\varphi))=F(\varphi)$ and using $G'(\theta)\theta'=F'(\varphi)$ gives
\begin{equation}
 \delta F(\varphi)-\delta G(\theta(\varphi))
   =F'(\varphi)\frac{\delta\theta}{\theta'}.
 \label{eq:multi_sd_welding_variation}
\end{equation}
Since $\delta F$ and $\delta G$ are holomorphic on the bath and gravity
sides respectively, introducing the plane coordinate $\mathfrak{z} = e^{-iz}$ and defining the contour $\mathcal{C}_{\mathfrak{z}} = \{\mathfrak{z} = e^{-iF(\varphi)}|0\le \varphi< 2\pi\}$, the Cauchy integral formula gives
\longeq{
 \frac{1}{2 \pi i} \oint_{\mathcal{C}_{\mathfrak{z}}} \frac{\delta G}{\mathfrak{z} - \mathfrak{z}_i}\,d\mathfrak{z} = 0 \, ,
}
and
\lceq{
\delta F(v_i) =\frac{1}{2\pi i}\int_0^{2\pi}d\varphi F'(\varphi)^2\frac{\delta\theta}{\theta'} \left[\frac{1}{2} \cot \frac{F(\varphi) - z_i}{2} - \frac{i}{2}\right] \, .
}
Differentiating the above formula gives
\lceq{
\pl_{v_i} \left[\frac{1}{2} \cot \frac{z - z_i}{2} - \frac{i}{2}\right] = \frac{F'(v_i)}{4 \sin^2 \frac{z -z_i}{2}},
}
which leads to
\begin{equation}
 \begin{aligned}
 \frac{\delta F'(v_i)}{F'(v_i)}
 &=\frac{1}{2\pi i}\int_0^{2\pi}d\varphi\,
      F'(\varphi)^2\frac{\delta\theta}{\theta'}
      \frac{1}{4\sin^2\!\frac{F(\varphi)-z_i}{2}}.
 \end{aligned}
 \label{eq:multi_sd_cauchy}
\end{equation}
Now varying the correlator~\eqref{eq:sec2_GFn} gives
\longeq{
\delta \mathcal{G}_F = &\left(\prod_j F'(v_j)^{h_j} \bar{F}'(\bar{v}_j)^{\bar{h}_j} \right) \sum_i \left[h_i \frac{\delta F'(v_i)}{F'(v_i)} + \delta F(v_i) \pl_{z_i}\right] \left \langle \prod_j V_j (z_j, \bar{z}_j) \right \rangle_z \\
& + \text{antiholomorphic part}.
}
Combined with the integral representation~\eqref{eq:multi_sd_cauchy} we can write
\longeq{
\label{eq:sec2_deltaGF1}
\delta \mathcal{G}_F = &\frac{1}{2 \pi i} \tpiint d \varphi F'(\varphi)^2 \frac{\delta \theta}{\theta'} \prod_j F'(v_j)^{h_j} \bar{F}'(\bar{v}_j)^{\bar{h}_j} \\
& \times \sum_i \left[\frac{h_i}{4 \sin^2 \frac{F(\varphi) -z_i}{2}} + \left(\frac{1}{2} \cot \frac{F(\varphi) - z_i}{2} - \frac{i}{2}\right) \pl_{z_i}\right] \left\langle \prod_j V_j (z_j, \bar{z}_j ) \right\rangle_z \\
&+ \text{antiholomorphic part} \, .
}
Since $\sum_i \pl_{z_i} \left\langle \prod_j V_j \right \rangle_z = 0$, we can drop the constant $-i/2$ term. Using the Ward identity one can write the above expression in terms of the stress tensor, i.e.,
\longeq{
\sum_i &\left[\frac{h_i}{4 \sin^2 \frac{F(\varphi) -z_i}{2}} + \left(\frac{1}{2} \cot \frac{F(\varphi) - z_i}{2}\right) \pl_{z_i}\right] \left\langle \prod_j V_j (z_j, \bar{z}_j ) \right\rangle_z =  \\
& \hspace{5.5cm}\left\langle \left(T(z) - \frac{c}{24}\right)\prod_j V_j (z_j, \bar{z}_j) \right\rangle_z  \, .
}
Using the transformation law of the stress tensor, together with the antiholomorphic part, the equation~\eqref{eq:sec2_deltaGF1} is transformed to
\begin{equation}
 \begin{aligned}
 \delta\mathcal G_F
 =\frac{1}{2\pi i}\int_0^{2\pi}d\varphi\,
    \frac{\delta\theta}{\theta'}
    \left\{\left\langle
        [T(\varphi)-\bar T(\varphi)]\prod_iV_i(v_i,\bar v_i)
       \right\rangle
       -\left\langle T(\varphi)-\bar T(\varphi) \right\rangle
                    \mathcal G_F\right\}.
 \end{aligned}
 \label{eq:multi_sd_normalized_variation}
\end{equation}
Here the expectation values are taken at fixed $\theta$, and the stress tensors are evaluated at the interface. Combining this result with the first-variation formula $\delta I$~\eqref{eq:disk_stress_metric_response} derived earlier we can compute the variation
\longeq{
 \delta \left(e^{-I[\theta]}\mathcal G_F\right) &= e^{-I} \left(\delta \mathcal{G}_F - \mathcal{G}_F \delta I\right)\\
 &=\frac{e^{-I[\theta]}}{2\pi i}\int_0^{2\pi}d\varphi\,
     \frac{\delta\theta}{\theta'}
     \left\langle[T(\varphi)-\bar T(\varphi)]
                  \prod_i V_i(v_i,\bar v_i)\right\rangle .
 \label{eq:multi_sd_full_variation}
}
We can view the above computation from the perspective of the Schwinger--Dyson equation. As we mentioned before, the measure $d\mu [\theta]$ is invariant under the diffeomorphism of $\theta$. Consequently, the path integral gives
\lceq{
0 = \int d \mu[\theta] \tpiint d \varphi \xi(\varphi) \theta'(\varphi) \frac{\delta}{\delta \theta(\varphi)} \left(e^{-I[\theta]}\mathcal G_F\right)
}
with $\delta \theta  = \xi(\varphi) \theta'(\varphi)$ and $\xi(\varphi)$ is an arbitrary infinitesimal reparametrization
of the boundary coordinate. The equation is then obeyed if and only if  \begin{equation}
 \left\langle[T(\varphi)-\bar T(\varphi)]
              \prod_i V_i(v_i,\bar v_i)\right\rangle=0,
 \label{eq:multi_sd_gluing}
\end{equation}
where the brackets now include the integral over reparametrizations. This verifies that the no-flux condition holds inside multipoint exterior correlators, in agreement with the Ishibashi condition
argued above.

We should emphasize that we have only shown this equation at the disk level. We expect that for higher genus surfaces we would have trouble with this equation if $h\leq h_c$ \nref{eq:sec3_Ach},  due to divergences from the boundary of the integration region.

\section{Conformal anomaly in the cylinder case}
\subsection{Evaluation of the action of the cylinder}
\label{ConformalAnomalyCylinder}
Here we discuss the derivation of the action~\eqref{ActCylM}. For the gravitational region, the Weyl factor becomes
\lceq{
\rho  = \rho_0 + \rho_G, \quad \rho_0 = \log \gamma - \log \left|\sin\frac{\pi \sigma}{\kappa}\right|, \quad \rho_G = -\frac{1}{2} \log |G'|^2 \, ,
}
giving the bulk integral involving $\rho_0$
\lceq{
-\mL_b = \int_{\mathcal{D}_w} d^2 w \sqrt{g_w} \left(\nabla \rho_0\right)^2 = 2\pi \left(\frac{1}{\epsilon_L} + \frac{1}{\epsilon_R}\right) - \pi b + O(\epsilon_a) \, .
}
$\rho_G$ is of the same form as in the disk case so the cross term $-\mL_\times$ and the term $-\mL_G$ are roughly the same, with two separate terms due to the extra gluing boundary
\longeq{
-\mL_\times &=  -i \tpiint d \theta_L  \log (\epsilon_L\theta_L') \left[\frac{G'' (\theta_L - i \kappa)}{G' (\theta_L - i \kappa)} - \frac{\bar{G}''(\theta_L +i \kappa)}{\bar{G}' (\theta_L +i \kappa)}\right] \\
& \hspace{3cm} + i \tpiint d\theta_R \log (\epsilon_R \theta_R') \left[\frac{G''(\theta_R)}{G'(\theta_R)} - \frac{\bar{G}''(\theta_R)}{\bar{G}'(\theta_R)}\right] \,  ,\\
- \mL_G &= -  \frac{i}{4} \tpiint d \theta_L \log |G'(\theta_L - i\kappa )|^2 \left( {G''_L \over G'_L} - { \bar G''_L \over \bar G'_L } \right)
\\ & \hspace{3cm} +\frac{i}{4} \tpiint d \theta_R \log |G'(\theta_R)|^2 \left( {G''_R \over G'_R} - { \bar G''_R \over \bar G'_R } \right) \, .
}
The minus sign before the integral at the left boundary comes from the opposite boundary orientation. In the disk case we argued that the logarithmic divergence $\log \epsilon_a$ disappears. The same conclusion holds in the cylinder case while the proof is different. Differentiating the welding conditions gives
\lceq{
\label{eq:sec4_dF_dG}
 F_a'=\theta_a' G', \quad \frac{F_a''}{F_a'} = \theta_a' \frac{G''}{G'} + \frac{\theta_a''}{\theta_a'} \, .
}
For notational convenience we suppress the arguments with $F_a$ and its derivatives evaluated at $v_a = \varphi_a+i \mu_a$ and $G$ and its derivatives at $w_a = \theta_a +i \sigma_a$ on the corresponding interface. The last term is real so $\Im F_a''/F_a' = \theta_a' \Im G''/G'$. Therefore
\lceq{
\tpiint d \theta_a \Im \frac{G''}{G'} =  \tpiint d \varphi_a \Im \frac{F_a''}{F_a'}  = 0,
}
where, in the last equality, we use the holomorphic property of $F''_a/F_a'$ to move the integral contour to the end of the bath. Notice that $\Im F_a$ is constant at the outer end, so its derivatives vanish there.

For the bath region, the only non vanishing term is
\longeq{
-(\mL_{F,L}+\mL_{F,R}) &= \frac{i}{4} \int d \varphi_L \log |F_L'|^2 \left(\frac{F''_L}{F'_L} - \frac{\bar{F}_L''}{\bar{F}_L'}\right)
\\ &\quad -  \frac{i}{4} \int d \varphi_R \log |F_R'|^2 \left(\frac{F''_R}{F'_R} - \frac{\bar{F}_R''}{\bar{F}_R'}\right)   \,.
}
Combined with the change of the reference partition function $\Delta \log Z_{\rm flat}(s)$ we obtain the action shown in~\eqref{ActCylM}.

\subsection{Solving the conformal welding problem on the cylinder}
\label{WeldingCylinder}
We perturb the boundary reparametrization around the saddle $\theta_a = \varphi_a + \vare_a (\varphi_a)$ where $\vare_a$ is real and periodic in $2 \pi$. It admits the Fourier expansion
\lceq{
\label{eq:sec4_vare_a}
\vare_a (\varphi_a) = \sum_{n \in \Z} \vare_n^a e^{i n \varphi_a}, \quad \left(\vare_n^a\right)^*  = \vare_{-n}^a \, .
}
The conformal maps are deformed due to the fluctuation $\vare_a$. Write the conformal maps as
\begin{equation}
 \begin{aligned}
 F_a (v_a) & = v_a + f_a (v_a), &f_a (v_a) &= \sum_{n \in \Z} f_n^a e^{in v_a} \, ,  \\
 G(w)&=w+g(w),&
 g(w)&=\sum_{n\in\mathbb Z}g_n e^{inw}.
 \end{aligned}
 \label{eq:cyl_map_modes}
\end{equation}
Here $f_a(v_a)$ and $g(w)$, as well as their Fourier modes, are all of order $O(\vare_a)$. We can then expand the action~\eqref{ActCylM} to the quadratic order in terms of these functions. By direct expansion of~\eqref{eq:cyl_width} and~\eqref{ActCylM} we have
\longeq{
\label{eq:cyl_length_quadratic_maps}
 \Delta y^{(2)}
 & =\frac{1}{2\pi}\int_0^{2\pi}d\varphi_L\,
       \varepsilon_L'\operatorname{Im}f_L(\varphi_L-i\kappa)
 -\frac{1}{2\pi}\int_0^{2\pi}d\varphi_R\,
       \varepsilon_R'\operatorname{Im}f_R(\varphi_R).
 \\
 \frac{ic}{48\pi}&\int_0^{2\pi}d\varphi_a\,\log\theta_a'
 \left(\frac{F_a''}{F_a'}-\frac{\overline F_a''}{\overline F_a'}\right) = \frac{ic }{4 8 \pi} \tpiint d \varphi_a \vare_a' \left(f_a'' - \bar{f}_a''\right) + O(\vare^3) \, .
}
As previously found, it is enough to solve the conformal welding problem to the linear order.
Put the Fourier expansion~\eqref{eq:sec4_vare_a} and~\eqref{eq:cyl_map_modes} into the conformal welding condition~\eqref{CylBdy} and straight circle boundary condition~\eqref{eq:sec4_YLYR}, up to the linear order of $\vare_a$, we have the solution
\begin{equation}
 \label{eq:cyl_linear_solution}
 \begin{aligned}
 f_n^L&=\frac{e^{-n(\kappa+\ell_L)}}{\sinh(ns_0)}
   \left\{\sinh[n(\kappa+\ell_R)]\,\varepsilon_n^L
                  -\sinh(n\ell_R)\,\varepsilon_n^R\right\},\\
 f_n^R&=\frac{e^{n\ell_R}}{\sinh(ns_0)}
   \left\{\sinh[n(\kappa+\ell_L)]\,\varepsilon_n^R
                  -\sinh(n\ell_L)\,\varepsilon_n^L\right\},\\
 g_n&=f_n^R-\varepsilon_n^R
     =f_n^L-e^{-n\kappa}\varepsilon_n^L.
 \end{aligned}
\end{equation}
Substituting these solutions into the expansion yields the quadratic action~\eqref{eq:sec4_I2}.

\section{The identity case and the treatment of the extra zero modes}
\label{app:ZeroModes}
\paragraph{The measure on trumpet geometry.}
We first specify the JT measure on the trumpet. For one boundary $a$ $(a=L,R)$, the symplectic form from the JT gravity on trumpet geometry \cite{Stanford:2017thb,Saad:2019lba} is
\lceq{
\label{eq:appG_sympletic}
\Omega_a  =\frac{1}{8 \pi} \tpiint d \varphi_a \left[\frac{\delta \theta_a' \wedge \delta \theta_a''}{\theta_a'^2} + \gamma^2 \delta \theta_a \wedge \delta \theta_a'\right] \, ,
}
where $\delta$ is the differential operator in the field space, and primes denote the derivative with respect to $\varphi_a$. For convenience, we fix the $1/8 \pi$ as the overall normalization. At the round saddle, the measure for each $\vare_a$ is
\lceq{
\label{eq:appG_dmuvarea}
d \mu [\vare_a] = \prod_{n \ge 1}  n (n^2+ \gamma^2)  d^2 \vare_n^a\, , \quad d^2 \vare_n^a = d \Re \vare_n^a d \Im \vare_n^a \, ,
}
which is the measure we used in the Gaussian integral~\eqref{eq:sec4_Z1_gen}. If a non vacuum state is propagating along the cylinder, $\gamma_e^{2} \neq -1$ in equation~\eqref{eq:sec4_cw_sol}, the zero mode is $n=0$. However, we cannot arbitrarily choose $\vare_0^L$ and $\vare_0^R$ since the finite cylinder only has one global $\U(1)$ rotation, which rotates the two boundaries by the same angle. The relative twist $\tau_{\text{tw}}$ needs to be retained and the measure for $\tau_{\text{tw}}$, together with the neck length $b$, is
\lceq{
\label{eq:appG_dbdtau}
\frac{d bd \tau_{\text{tw}}}{4 \pi} \, .
}
Integrating $\tau_{\rm tw}$ over $0\le\tau_{\rm tw}<b$ gives the measure in~\eqref{eq:sec3_bdb}.

\paragraph{Infinite bath case.}
If we consider infinite bath $\ell_a \rightarrow \infty$, $\langle E \rangle_h = 2h-\frac{c}{12}$. Moreover, if there are no insertions at the infinite end of the whole geometry, only the vacuum state propagates along the closed channel and $\gamma_e^2 = -1$. We then have extra zero modes $\vare_{\pm 1}^a$, as we saw in the disk case. Due to the vanishing of the quadratic terms in $|\vare_1^a|^2$, we have a divergence when we try to perform the integral of this particular mode.

To resolve the divergence, we can retain a small positive JT coupling as written in~\eqref{eq:cyl_JT}. The large-$c$ expansion suppresses fluctuations transverse to the M\"{o}bius transformation but leaves the zero modes $\vare_{\pm 1}^a$ unsuppressed, so these directions cannot generally be treated as small fluctuations around the identity. We therefore parametrize the finite M\"{o}bius transformation explicitly and separate their measure from the transverse modes $n \ge 2$. This permits an exact treatment of the JT-regulated integral over the M\"{o}bius family, while the remaining fluctuations are evaluated in the Gaussian approximation.

\paragraph{Finite M\"{o}bius transformation and the measure.}
We split general reparametrizations in $\text{Diff}(S^1)/\text{U}(1)$ as the composition of a finite $\rm SL(2,\R)/U(1)$ and an infinitesimal reparametrization in $\text{Diff}(S^1)/\rm SL(2,\R)$. After separating the rotations, each finite M\"{o}bius transformation is parametrized by a complex number in a disk $p_a \in \mathbb{D}$, where $\mathbb{D} = \{p:|p| <1\}$, the corresponding transformation $ M_{p_a}$ acts on the $\varphi_a$ in the following way
\lceq{
\label{eq:sec4_Mobius}
e^{i M_{p_a} (\varphi_a)} = \frac{e^{i \varphi_a} - p_a}{1 - \bar{p}_a e^{i \varphi_a}} \, .
}
To include the remaining fluctuations, we write
\lceq{
\theta_a  = f_a \circ M_{p_a}, \quad f_a (\varphi_a) = \varphi_a + \vare_a (\varphi_a), \quad \vare_a(\varphi_a) = \sum_{n \neq 0,\pm 1} \vare_n^a e^{i n \varphi_a} \, .
}
We then try to evaluate the symplectic volume around $\vare_a = 0$ with $p_a$ kept finite. At $\varepsilon_a=0$, the chain rule gives the variation of $\theta_a$ as
\lceq{
\delta \theta_a = \delta \vare_a \circ M_{p_a} + \delta M_{p_a}.
}
Introduce another variable $\tilde{\varphi}_a := M_{p_a} (\varphi_a)$ and $\eta_a (\tilde{\varphi}_a) := \delta \theta_a \left( M_{p_a}^{-1} (\tilde{\varphi}_a) \right)$, the symplectic form~\eqref{eq:appG_sympletic} is written in terms of integrating over $\tilde{\varphi}_a$
\lceq{
\Omega_a = \frac{1}{8 \pi} \tpiint d \tilde{\varphi}_a \left[\pl_{\tilde{\varphi}_a} \eta_a \wedge \pl_{\tilde{\varphi}_a}^2 \eta_a + \gamma^2 \eta_a \wedge \pl_{\tilde{\varphi}_a} \eta_a\right] \, .
}
Direct variation with respect to~\eqref{eq:sec4_Mobius} gives
\lceq{
\eta_a = \delta \vare_a + \frac{i}{1 - |p_a|^2} \left[\left(e^{-i \tilde{\varphi}_a} + \bar{p}_a \right) \delta p_a - \left(e^{i \tilde{\varphi}_a} + p_a\right) \delta \bar{p}_a\right] \, .
}
Straightforward differentiation of the above expression gives the symplectic form
\lceq{
\Omega_a^{(0)} = \frac{1 + \gamma^2}{\left(1 - |p_a|^2\right)^2} d \Re p_a \wedge d \Im p_a - \sum_{n \ge 2} n(n^2+\gamma^2) d \Re \vare_n^a \wedge d \Im \vare_n^a \, ,
}
where the superscript $(0)$ emphasizes we evaluate the symplectic form around $\vare_a = 0$ and we use $\delta \vare_n^a \wedge  \delta \bar{\vare}_n^a = -2 i d \Re \vare_n^a \wedge d \Im \vare_n^a$. One can see that there is no mixed term between $p_a$ and higher modes $\vare_n^a$. Therefore the integral measure that we are going to use is
\lceq{
\label{eq:sec4_dmu_p}
d \mu_a [p_a, \vare_n^a] = (1+\gamma^2) \frac{d^2 p_a}{(1 - |p_a|^2)^2} \prod_{n \ge 2} n (n^2 + \gamma^2) d^2 \vare_n^a \, .
}

\paragraph{The JT regulator.}
For each boundary we retain a small JT action~\eqref{eq:cyl_JT}. For M\"{o}bius transformation parametrized by $p_a$, the weight coming from the JT action is simply
\begin{equation}
 I_{{\rm JT},a} [p_a]
=\frac{\phi_{r,a}}2\left[(1+\gamma^2)\frac{1 + |p_a|^2}{1 -|p_a|^2}-1\right] \, .
 \label{eq:cyl_collective_action}
\end{equation}
The exact integral over the family of M\"{o}bius transformation is then
\begin{equation}
 I = \int_{\mathbb D} (1 + \gamma^2) \frac{d^2 p_a}{(1 - |p_a|^2)^2}   e^{-I_{{\rm JT},a} [p_a] }
 =\frac{\beta_a}{2\phi^{(a)}_{r,\text{phys}}}\,
   \exp\!\left[-\frac{\phi^{(a)}_{r,\text{phys}} b^2}{4\pi\beta_a}\right] ,
 \label{eq:cyl_collective_integral}
\end{equation}
where we restore the $\beta_a$ dependence explicitly. Including the overall factor with $\mathcal{J} = 2 I/[\pi (1+\gamma^2)]$, this is the
additional term shown in~\eqref{JBetaEach} for each boundary.
The factor is present because the infinite product in \nref{eq:sec4_dmu_p} is missing the $n=1$ term but it is present in our definition of $\mu$ \nref{ExprForA}.   Schwarzian corrections to the nonzero-mode integral can still be neglected in the infrared limit.

\section{Open-closed consistency}
\label{app:OpenClosed}
\begin{figure}[htbp]
  \centering
  \includegraphics[width=0.6\linewidth]{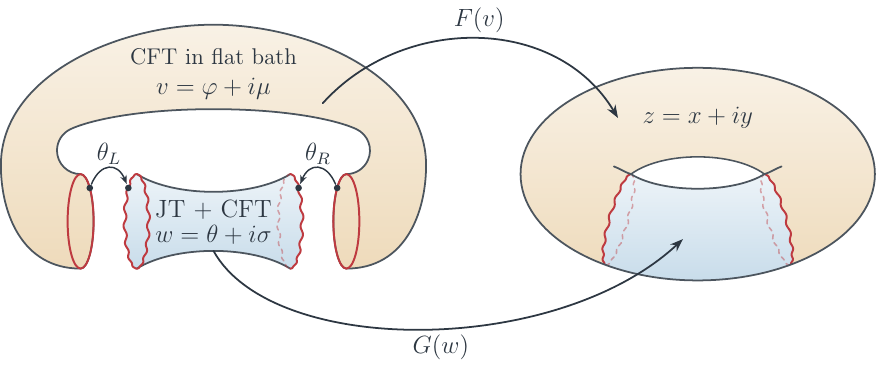}
  \caption{Conformal welding of a double trumpet to a single connected bath.
  The boundary maps $\theta_L$ and $\theta_R$ glue the two regions;
  $F(v)$ and $G(w)$ map the bath and gravitational region, respectively,
  to a flat reference torus.}
  \label{fig:torus-welding}
\end{figure}

In this appendix we exhibit how to derive the CFT action on the torus geometry in section~\ref{subsec:cylinder_torus} and evaluate the one loop path integral, which demonstrates the open-closed duality.
\paragraph{CFT action and conformal welding problem on the torus.}
Keep the gravitational region  $w=\theta+i\sigma$ with $-\kappa<\sigma<0$, and take the bath coordinate to be $v=\varphi+i\mu$, $0<\mu<\ell$. The cylinder geometry identifies two boundaries $\sigma = -\kappa$ and $ \mu = \ell$. Introduce two conformal maps $G$ and $F$, which map the gravitational region and the bath to the flat torus, as shown in figure~\ref{fig:torus-welding},
with the conformal welding condition
\begin{equation}
 \begin{aligned}
F(\varphi)=G(\theta_R(\varphi)),\quad  F(\varphi+\varpi+i\ell)
   =G(\theta_L(\varphi)-i\kappa)+\mathcal{T},
 \end{aligned}
 \label{eq:tor_sewing}
\end{equation}
where $\varpi = 2 \pi \tau_{\text{tw}}/b$ is the angular twist along the neck of the double trumpet. The reference torus has periods $z \sim z +2\pi$ and $z \sim z + \mathcal{T}$, where $\mathcal{T}=\mathcal{W} +i s$ is determined by the welding. Evaluation of the conformal anomaly~\eqref{ConfAn} is similar to what we discussed in the cylinder computation of appendix~\ref{ConformalAnomalyCylinder}. The reference partition function is
\begin{equation}
 \begin{aligned}
 Z_{\rm tor}(s,\mathcal{W})
 =\Tr_{\mathcal H_{\rm CFT}}
  e^{-s (L_0+\bar L_0-c/12)+i \mathcal{W}(L_0-\bar L_0)}=\sum_{h,\bar h}d_{h,\bar h}
   \chi_h(e^{- s+i \mathcal{W}})
   \chi_{\bar h}(e^{- s -i \mathcal{W}} ).
 \end{aligned}
 \label{eq:tor_reference}
\end{equation}
Subtracting the same local UV divergences as before gives the effective action
\begin{equation}
 \begin{aligned}
 -I_{\rm CFT}={}&-\frac{cb}{24} + \log Z_{\rm tor}(s,\mathcal{W})\\
 &+\frac{c}{24\pi}\int_0^{2\pi}d\varphi\left[
  \log\theta_L'\,\Im\frac{F''(\varphi+\varpi+i\ell)}
                              {F'(\varphi+\varpi+i\ell)}
 -\log\theta_R'\,\Im\frac{F''(\varphi)}{F'(\varphi)}\right].
 \end{aligned}
 \label{eq:tor_action}
\end{equation}
The interface orientations give the relative sign. At the round configuration $\theta_a(\varphi)=\varphi$, $F(v)=v$, and $G(w)=w$. The sewing conditions~\eqref{eq:tor_sewing} then give $\mathcal{W}=\varpi$ and $s = s_0 :=\kappa+\ell$, leaving
\begin{equation}
 Z_{\rm CFT}[b,\varpi;\ell]
 =e^{-cb/24}\,
 Z_{\rm tor}(s_0,\varpi).
 \label{eq:tor_round}
\end{equation}

\paragraph{Effective quadratic action.}
We expand around the round configuration as $\theta_a=\varphi+\varepsilon_a$, where $\varepsilon_a=\sum_{n\ne0}\varepsilon_n^a e^{in\varphi}$ and $\varepsilon_{-n}^a=\overline{\varepsilon_n^a}$.
The common rotation is quotiented out, while the relative rotation is
included in the twist $\tau_{\text{tw}}$.
Writing $F(v)=v+\sum_n f_ne^{inv}$ and
$G(w)=w+\sum_n g_ne^{inw}$, the linearized solution to the conformal welding condition~\eqref{eq:tor_sewing} is
\begin{equation}
 \begin{aligned}
 f_n=\frac{\varepsilon_n^R-e^{-n\kappa}\varepsilon_n^L}
                {1-e^{-n(s_0-i\varpi)}}, \quad
 g_n=\frac{e^{-n\kappa}
       [-\varepsilon_n^L+e^{-n\ell+in\varpi}\varepsilon_n^R]}
                {1-e^{-n(s_0-i\varpi)}},\quad n\ne0.
 \end{aligned}
 \label{eq:tor_linear_maps}
\end{equation}
The fluctuations of the reparametrizations change both components of the reference period. We write $\Delta\mathcal{T}=\mathcal{T}  -(\varpi+i s_0) =\Delta\mathcal{W}+i\Delta y$, where $\Delta\mathcal W=\mathcal W-\varpi$ and $\Delta y=s-s_0$. Using~\eqref{eq:tor_sewing}, we can write
\longeq{
f(\varphi) &= \vare_R (\varphi) + g(\varphi) + \vare_R (\varphi) g'(\varphi) + O(\vare^3), \\
f(\varphi+ \varpi+i \ell) &= \vare_L (\varphi) + g(\varphi-i \kappa) + \vare_L(\varphi) g'(\varphi-i \kappa) + \Delta \mathcal{T} + O(\vare^3) \, .
}
Using the holomorphic property we can take the angular mean of the above equations and eliminate the common additive constant in the maps. The first-order shift vanishes, $\Delta\mathcal{T}^{(1)}=0$, and at quadratic order we find
\begin{equation}
 \begin{aligned}
 \Delta\mathcal{T}^{(2)}
 &=\sum_{n\ne0}in g_n
   \left(\varepsilon_{-n}^R-e^{n\kappa}\varepsilon_{-n}^L\right) \, .
 \end{aligned}
 \label{eq:tor_modulus_response}
\end{equation}
    We are interested in the one loop partition function and assume the fluctuation to be small. To the quadratic order the change of the reference partition function is simply
\lceq{
\label{eq:tor_DetltalogZ}
 \Delta\log Z_{\rm tor}
 =-E\Delta y^{(2)}
  + J\,\Delta\mathcal{W}^{(2)},\quad E=-\partial_{s}\log Z_{\rm tor}, \quad J=\partial_{\mathcal{W}}\log Z_{\rm tor} \, .
}
Both derivatives are evaluated at $(s,\mathcal{W})=(s_0,\varpi)$. As in the cylinder calculation, the linearized maps suffice to determine the quadratic change in the reference partition function.

As we will discuss around \nref{eq:tor_classical_exponent} the saddle point will be at  $\varpi=0$. So we set $\varpi=0$ now.
Define $\boldsymbol{\vare}_n = \left(\vare_n^L, \vare_n^R\right)^T$ and the quadratic action is simply
\begin{equation}
 -I_{\rm CFT}^{(2)}
 =-\frac c{12}\sum_{n\ge1}n(n^2+\gamma_e^2) \boldsymbol{\vare}_n^\dagger K_n \boldsymbol{\vare}_n  \quad  K_n = \frac{1}{\sinh \frac{n s_0}{2}}
 \mx{
  \cosh \frac{n s_0}{2}&-\cosh \frac{n(\kappa-\ell)}{2}\\
  -\cosh \frac{n(\kappa-\ell)}{2} &\cosh\frac{n s_0}{2}
  },
 \label{eq:tor_quadratic}
\end{equation}
where $ \gamma_e^2=12E/c$.
\paragraph{The full integral.} The integration variables are the boundary fluctuations $\vare_n^L$, $\vare_n^R$ with $n\ge 1$, the neck length $b$, and the twist $\tau_{\rm tw}$. For the boundaries, we use the trumpet measure of~\eqref{eq:appG_dmuvarea} for each boundary. For the moduli we use the measure in~\eqref{eq:appG_dbdtau}. The full integral is
\lceq{
Z_{\text{one loop}} (\ell) = \int_0^\infty d b  \int_{-b/2}^{b/2} \frac{d \tau_{\rm tw}}{4 \pi} \int \prod_{n \ge 1} \left[n(n^2+\gamma^2)\right]^2 d^2 \vare_n^L d^2 \vare_n^R e^{-I_{\text{CFT}}} \, ,
}
where $I_{\text{CFT}}$ is the full action~\eqref{eq:tor_action}. At fixed moduli,
\lceq{
-I_{\text{CFT}} = - \frac{cb}{24} + \log Z_{\rm tor} (s_0, \varpi) - I_{\rm CFT}^{(2)} + O (\vare^3) \, .
}
The first two terms come from the partition function at the round configuration~\eqref{eq:tor_round}. At $\varpi = 0$, $-I_{\rm CFT}^{(2)}$ is what we derived in~\eqref{eq:tor_quadratic}. At $\varpi \neq 0$, the twist response~\eqref{eq:tor_DetltalogZ} also contributes. However, for our one loop calculation around the saddle $\varpi_* = 0$, using~\eqref{eq:tor_quadratic} is sufficient, as demonstrated in the saddle point analysis later. The Gaussian integral gives the fixed-moduli partition function
\lceq{
\label{eq:tor_fixedZ}
Z_{\rm CFT}^{(1)} [b,\varpi;\ell] = \frac{c}{12 \pi} e^{- \frac{cb}{24}} Z_{\rm tor} (s_0, \varpi) D(b, \varpi), \quad D(b, 0) = \prod_{n \ge 1} \frac{(n^2+\gamma^2)^2}{(n^2 + \gamma_e^2)^2 \det K_n} \, .
}
The constant $c/12 \pi$ comes from the infinite product $\prod_{n \ge 1} \left(12 \pi/c\right)^2$. The factor $D(b, \varpi)$ is the ratio of the measure factors to the determinant of the quadratic action. We display it at $\varpi=0$, the only value needed below. This gives the full integral
\lceq{
Z_{\text{one loop}} (\ell) \simeq  \int_0^\infty d b  \int_{-b/2}^{b/2} \frac{d \tau_{\rm tw}}{4 \pi} Z_{\rm CFT}^{(1)} [b,\varpi;\ell] \, .
}

\paragraph{The short-cylinder saddle.}
As explained in the main text, despite the integral being divergent as $b \rightarrow 0$, we can still evaluate the local contribution of the saddle at finite $b$ in the semiclassical approximation. We take $c\gg1$ and $\ell\ll1$. Under this assumption, the reference partition function is approximated by the unique vacuum module propagating in the open channel
\begin{equation}
 Z_{\rm tor}(s_0,0)\simeq
 \chi_0(e^{-4\pi^2/s_0})^2
 =e^{c\pi^2/(3s_0)}\mathcal V(e^{-4\pi^2/s_0})^2, \quad \mathcal V(q):=\prod_{n=2}^{\infty}(1-q^n)^{-1} \, .
 \label{eq:tor_modular_vacuum}
\end{equation}
The square $\mathcal{V}(q)^2$ accounts for the left- and right-moving vacuum modules of the same CFT. Restoring a small twist, the leading term in the exponent is
\begin{equation}
 -I_{\rm CFT}^{(0)}
 =-\frac{cb}{24}
  +\frac{c\pi^2}{6}\left(\frac1{s_0-i\varpi}
                         +\frac1{s_0+i\varpi}\right).
 \label{eq:tor_classical_exponent}
\end{equation}
The saddle point for $\varpi$ and $b$ based on this leading exponent is simply
\begin{equation}
 \varpi_*=0,\qquad \kappa_*= \frac{2\pi^2}{b_*} =  \ell,\qquad
 -I_{\rm CFT}^{(0)} \big|_*=\frac{c\pi^2}{12\ell},
 \label{eq:tor_saddle}
\end{equation}
which is the same as what we derived in~\eqref{saddlpk}. This is the static traversable-wormhole saddle discussed in the main text. Around this saddle $\gamma_e^2=\gamma_*^2+O(c^{-1})$ with $\gamma_*=\pi/\ell$. Combined with $\det K_n|_* = 1$, the factor $D(b_*,0)$ is simply $1$.
With a small twist, the two vacuum characters in~\eqref{eq:tor_modular_vacuum} depend on $s_0 \mp i \varpi$, so $\mathcal{V}(e^{-4\pi^2/s_0})^2$ becomes $\mathcal{V}(q_+) \mathcal{V}(q_-)$ with $q_\pm = e^{-4\pi^2/(s_0 \mp i \varpi)}$. The fixed-moduli partition function~\eqref{eq:tor_fixedZ} is then
\longeq{
 Z^{(1)}_{\rm CFT} [b,\varpi;\ell] &\simeq \frac{c}{12\pi} D(b,\varpi) \mathcal{V}(q_+) \mathcal{V} (q_-) \exp \left(-I_{\rm CFT}^{(0)}\right) \\
 &\simeq \frac{c}{12\pi} \mathcal{V}(q_o)^2 \exp \left[-\frac{cb}{24}
  +\frac{c\pi^2}{6}\left(\frac1{s_0-i\varpi}
                         +\frac1{s_0+i\varpi}\right)\right]
}
with $q_o := e^{-2\pi^2/\ell}$. In the first line, the exponent~\eqref{eq:tor_classical_exponent} is of order $c$, while the prefactors $D(b, \varpi)$ and $\mathcal{V}(q_\pm)$ are of order $c^0$.  The prefactors therefore shift the saddle~\eqref{eq:tor_saddle} only by $\delta b/b_* =O(c^{-1})$, and to this order it is enough to evaluate them at the saddle, where $D(b_*,0) = 1$ and $q_\pm = q_o$. Define $\Delta b = b - b_*$ and expand~\eqref{eq:tor_classical_exponent} to the quadratic order
\lceq{
 -I_{\rm CFT}^{(0)}=\frac{c\pi^2}{12\ell}
  -\frac{c}{48b_*}(\Delta b)^2
  -\frac{cb_*}{48\pi^2}\tau_{\text{tw}}^2+\cdots
}
which gives
\begin{equation}
 \begin{aligned}
 Z_{\text{one loop}}(\ell) = { 1 \over 4 \pi } \int_{\mathbb R^2}d(\Delta b)\,d\tau_{\text{tw}}\,
 Z^{(1)}_{\rm CFT} [b,\varpi;\ell]
 = q_o^{-c/24} \mathcal{V}(q_o)^2 \, .
 \end{aligned}
 \label{eq:tor_answer}
\end{equation}
The Gaussian prefactor $c/(12\pi)$ in
\eqref{eq:tor_fixedZ}
and the integral~\eqref{eq:tor_answer} multiply to one.

\paragraph{The closed-channel calculation.}
In the above derivation we assume the reference partition function is approximated by~\eqref{eq:tor_modular_vacuum}. We next recover~\eqref{eq:tor_answer} by summing heavy scalar Ishibashi sectors, using the interface projection~\eqref{IntProject}. For a nondegenerate scalar primary, $h=\bar h>c/24$, use the existing parameter $\gamma_h^2=24h/c-1$ and replace the energy response $E$ in~\eqref{eq:tor_DetltalogZ} by $E_h=c\gamma_h^2/12$. At nonzero twist, the determinant of the $K_n$ is
\begin{equation}
 \det K_n(\varpi)
 =\frac{(1-e^{-2n\kappa})(1-e^{-2n\ell})}
              {|1-e^{-n(s_0-i\varpi)}|^2}.
\end{equation}
Its inverse cancels the two independent descendant denominators:
\begin{equation}
 \prod_{n\ge1}\frac1{\det K_n(\varpi)}\,
 \chi_h(e^{-s_0+i\varpi})\chi_h(e^{-s_0-i\varpi})
 =\chi_h(e^{-2\kappa})\chi_h(e^{-2\ell}).
 \label{eq:tor_character_sewing}
\end{equation}
The resulting scalar integrand is independent of $\varpi$, so the twist integral reduces the modulus measure to $b\,db/(4\pi)$. We denote the remaining determinant ratio by
\begin{equation}
R_h(b):=\prod_{n\ge1}\frac{n^2+(b/2\pi)^2}{n^2+\gamma_h^2}
 =\frac{\gamma_h\sinh(b/2)}{(b/2\pi)\sinh(\pi\gamma_h)}.
\end{equation}
We recover the same connected coefficient as in the cylinder calculation:
\begin{equation}
 \begin{aligned}
 Z_{\text{one loop},h}(\ell)
 =A^{(2,0)}_{h\bar h}\,\chi_h(e^{-2\ell}),\quad
 A^{(2,0)}_{h\bar h}
 =\frac{c}{48\pi^2}\int_0^\infty b\,db\,
       e^{-cb/24}\chi_h(e^{-4\pi^2/b})R_h(b)^2.
 \end{aligned}
 \label{eq:tor_closed_coefficient}
\end{equation}
Here the superscript denotes two gravitational boundaries and genus
zero, as in~\eqref{eq:proposal_cylinder_match}.

For fixed $\gamma_h\gg1$ and large $c$, the exponent in this integral is
$-c[b/24+\pi^2\gamma_h^2/(6b)]$. Its saddle is
$b_h=2\pi\gamma_h$, with $R_h(b_h)=1$. The Gaussian integral of $b$ around saddle therefore gives
\begin{equation}
 A^{(2,0)}_{h\bar h}
 =\frac{\sqrt c\,\gamma_h^{3/2}}{2\sqrt3}\,
 \frac{e^{-c\pi\gamma_h/6}}
      {\prod_{n\ge1}(1-e^{-2\pi n/\gamma_h})}
 [1+O(c^{-1})].
 \label{eq:tor_heavy_gaussian}
\end{equation}
The oscillator product is essential for matching the prefactor and the order-one term in the exponent. The above result is the one loop partition function with a fixed primary state $h$ propagating along the closed channel. To give the full partition function we need to perform the primary sum.

Let us use the Liouville parameters introduced in the main text, $c=1+6Q^2$, $Q=\mathfrak b+\mathfrak b^{-1}$ with $0<\mathfrak b<1$, and $h_P=(c-1)/24+P^2$, $P\ge0$. The vacuum modular kernel $S_{0P}$ transforms the Virasoro character $\chi_{h_P}$ to the vacuum Virasoro character in the open channel
\begin{equation}
S_{0P}=4\sqrt2\sinh(2\pi\mathfrak b P)\sinh(2\pi P/\mathfrak b),
 \qquad
 \int_0^\infty dP\,S_{0P}\chi_{h_P}(e^{-2\ell})=\chi_0(q_o).
 \label{eq:tor_modular_kernel}
\end{equation}
This identity uses the positive-momentum normalization of~\cite{Zamolodchikov:2001ah}. We approximate the heavy-primary density by the Cardy formula~\cite{Collier:2019weq}
\lceq{
\rho(P,\bar P)\simeq S_{0P}S_{0\bar P} \,.
}
Let $\rho_{\rm sc}(P)$ denote the density of scalar Virasoro primaries with $h=\bar h=(c-1)/24+P^2$, for $P>0$. Writing the primary spin as $j_{\rm p}=P^2-\bar P^2$, the change of variables $dP\,d\bar P=dP\,dj_{\rm p}/(2\bar P)$ then gives
\begin{equation}
  \rho_{\rm sc}(P)\simeq\frac{S_{0P}^2}{2P}.
  \label{eq:tor_scalar_density}
\end{equation}
We now express the gravitational amplitude~\eqref{eq:tor_heavy_gaussian} in terms of $P$. Applying the modular transformation of the Dedekind eta function to the product
in~\eqref{eq:tor_heavy_gaussian}, and using
$P^2=(c\gamma_h^2+1)/24$, one can verify that
\begin{equation}
 A^{(2,0)}_{h_P\bar h_P}
 =\frac{2P}{S_{0P}}\,
   \mathcal V(e^{-b_h(P)})[1+O(c^{-1})].
 \label{eq:tor_inverse_density}
\end{equation}
Here $b_h (P)$ is the saddle point $b_h  = 2\pi \gamma_h = 2 \pi \sqrt{(24P^2-1)/c}$ of the $b$ integral in~\eqref{eq:tor_closed_coefficient}. The factor $2P/S_{0P} \simeq \sqrt{2} P e^{-2\pi Q P}$ is exponentially small in $c$ while the factor $\mathcal{V}(e^{-b_h(P)})$ is of order $c^0$, since $b_h(P) \simeq (4 \pi P/Q)$ at large $c$. Inserting~\eqref{eq:tor_inverse_density} and \eqref{eq:tor_scalar_density} into the closed-channel sum
\longeq{
\label{eq:tor_closed}
Z_{\text{one loop}} (\ell)  &\simeq
 \int_{\rm heavy}dP\,\rho_{\rm sc}(P)\,
 A^{(2,0)}_{h_P\bar h_P}\,
 \chi_{h_P}(e^{-2\ell}) \\
 & \simeq \int dP\,S_{0P}\chi_{h_P}(e^{-2\ell})
          \mathcal V(e^{-b_h(P)}) = q_o^{-c/24}\mathcal V(q_o)^2 \left[1 + O(c^{-1})\right]
}
The last step needs more explanation. The order-$c$ part is the weight $S_{0P} \chi_{h_P}(e^{-2\ell})$ with the saddle determined by the exponent
\lceq{
\label{eq:tor_P*}
\exp \left(2 \pi Q P - 2 \ell P^2\right)\quad \Rightarrow \quad P_* = \frac{\pi Q}{2 \ell}.
}
Since $\mathcal{V}(e^{-b_h(P)})$ is of order $c^0$ and almost a constant across the peak $P = P_*$, the factor $\mathcal{V}(e^{-b_h(P)})$ is then evaluated at the saddle $P = P_*$, giving $\mathcal{V}(q_o)$. The modular transformation identity~\eqref{eq:tor_modular_kernel} gives the rest of the factor $\chi_0(q_o) =q_o^{-c/24} \mathcal{V}(q_o)$.  Thus the heavy closed-channel sum reproduces~\eqref{eq:tor_answer} at leading order in $1/c$.

\paragraph{Comments on light non-vacuum states.}
  So far we kept only the vacuum module in~\eqref{eq:tor_modular_vacuum}. The other states of the CFT give the remaining terms in~\eqref{eq:sec4_Ztor}. If we include non vacuum states, say with $h_i, \bar h_i > 0$\footnote{If either $h_i$ or $\bar h_i$ is zero, there is a modification involving the identity character.}, we get
\lceq{
Z_{\rm tor} (s_0, \varpi) = \sum_i \chi_{h_i} (q_+) \chi_{\bar{h}_i} (q_-) = (q_+q_-)^{-c/24} \sum_i f_i (q_+, q_-) ,
}
where $f_i (q_+, q_-)$ has different expressions for the vacuum state $i=0$ and the non-vacuum states $i \neq 0$
\lceq{
f_0 (q_+, q_-) = \mathcal{V}(q_+) \mathcal{V} (q_-), \quad f_i (q_+, q_-) = q_+^{h_i} q_-^{\bar{h}_i} \prod_{n \ge 1} \left(1 - q_+^n\right)^{-1 }\left(1 - q_-^n\right)^{-1 } \, .
}
Every character contains the same factor $q_\pm^{-c/24}$, so the order-$c$ exponent~\eqref{eq:tor_classical_exponent} is the same for all Virasoro families, which leads to the same saddle~\eqref{eq:tor_saddle}, where $q_\pm = q_o$. To the order we consider, $f_i$ only shifts the saddle at the order $c^{-1}$. For each family, the energy and twist responses~\eqref{eq:tor_DetltalogZ} differ from those of the vacuum only at relative order $h_i/c$, which is negligible for light states. Therefore the determinant ratio in~\eqref{eq:tor_fixedZ} is again $1+ O(c^{-1})$ at the saddle. Combined with the Gaussian integral over the moduli we get the one loop partition function
\lceq{
\label{eq:tor_h_Ztor}
Z_{\text{one loop}} (\ell) \simeq q_o^{-c/24} \sum_i f_i (q_o, q_o) = q_o^{-c/24} \mathcal{V}(q_o)^2 \left[1 + \sum_{i \neq 0} \frac{q_o^{h_i + \bar{h}_i}}{(1-q_o)^2}\right] \, ,
}
as in~\eqref{eq:sec4_Ztor}.

The closed channel sum can also reproduce this result. Each light primary adds its contribution to the spectral density \cite{Collier:2019weq}
\lceq{
\rho(P, \bar{P}) \simeq \sum_i S_{h_i P} S_{\bar{h}_i \bar{P}}, \quad S_{h_i P} = 2 \sqrt{2} \cos \left(4 \pi P_i P\right),
}
where $h_i = (c-1)/24+P_i^2$ and $\bar{h}_i = (c-1)/24+\bar{P}_i^2$. This formula holds for $h_i \neq 0$; for the vacuum, $S_{0P}$ is given by~\eqref{eq:tor_modular_kernel}. Using the modular transformation identity
\lceq{
\int_0^\infty d P S_{h_i P} \chi_{h_P} (e^{-2\ell}) = \chi_{h_i} (q_o),
}
we can repeat the computation of~\eqref{eq:tor_closed}. For a given non-vacuum primary labelled by $i$, the change of variables that led to~\eqref{eq:tor_scalar_density} gives the scalar density $S_{h_i P} S_{\bar{h}_i P}/(2P)$, and summing over the heavy primaries gives the integral over $P$
\longeq{
\label{eq:tor_h_closed_int}
\int d P \frac{S_{h_i P} S_{\bar{h}_i P}}{2 P} A_{h_P \bar{h}_P}^{(2,0)} \chi_{h_P}(e^{-2\ell})& \simeq \mathcal{V}(q_o) \frac{q_o^{\bar{h}_i}}{1-q_o} \int_0^{\infty} d P S_{h_i P} \chi_{h_P} (e^{-2\ell}) \\
&=q_o^{-c/24} \mathcal{V}(q_o)^2 \frac{q_o^{h_i + \bar{h}_i}}{(1-q_o)^2} \,  .
}
In the first line we used the amplitude in~\eqref{eq:tor_inverse_density}, which cancels the factor $2P$ and leaves two slowly varying factors $\mathcal{V}(e^{-b_h(P)})$ and $S_{\bar{h}_i P}/S_{0P}$. Similar to the analysis below~\eqref{eq:tor_closed}, those should be evaluated at the saddle governed by the leading exponent in $S_{h_i P} \chi_{h_P} (e^{-2\ell})$. Using~\eqref{eq:sec3_chih}, we can write
\lceq{
S_{h_i P} \chi_{h_P} (e^{-2\ell}) \simeq \frac{\sqrt{2} e^{4 \pi |P_i| P - 2 \ell P^2}}{\eta (e^{-2\ell})}, \qquad \eta(q) = q^{1/24} \prod_{n \ge 1} (1-q^n) \, ,
}
giving the saddle $P = \tilde{P}_* = \pi |P_i|/\ell$. Notice that we have the expansion
\lceq{
\tilde{P}_* = P_* \sqrt{1 - \frac{4 h_i}{Q^2}} = P_* \left[1 - \frac{12 h_i}{c-1} +O(c^{-2})\right],
}
the shift compared to the saddle we used in~\eqref{eq:tor_P*} is of order $c^{-1/2}$ and approximately we can still use the saddle~\eqref{eq:tor_P*}. Therefore we can use the asymptotic behavior of the ratio evaluated at $P= P_*$
\lceq{
\frac{S_{\bar{h}_i P}}{S_{0P}} \simeq  \frac{e^{-2\pi  \left( Q-2 |\bar{P}_i|\right) P}}{1-e^{-4 \pi \mathfrak{b} P}} = \frac{q_o^{\bar{h}_i}}{1-q_o} \left[1+ O(c^{-1}) \right]\, , \quad \text{at} \quad P = P_* = \frac{\pi Q}{2 \ell} \, .
}
This, together with the $\mathcal{V}(e^{-b_h(P)})$ evaluated at the saddle $P_*$, provides the overall factor after the first equality in~\eqref{eq:tor_h_closed_int}. Summing over $i$ and adding the vacuum term~\eqref{eq:tor_closed} reproduces~\eqref{eq:tor_h_Ztor} as required by open-closed duality.

\section{A new operator of dimension two at the boundary}
\label{app:boundary_operator}

In this appendix, we study the extra dimension two primary operator that we expect in the BCFT describing the black hole, as discussed around \nref{NewPrimPar}.

 \paragraph{Identifying the expected primary.}
Use $w=\theta+i\sigma$, $v=\varphi+i\mu$ and the welding condition
\eqref{eq:sec2_welding}. Denote the holomorphic stresses in the flat
cylinder frames inside and outside by $t_{w} = - 2 \pi T_{ww} $ and $t_v = - 2\pi T_{vv}$.
The transformation used in~\eqref{eq:sec2_eom} gives
\begin{equation}
 t_v=F'^2t_z(F)+\frac c{12}\{F,v\},\qquad
 t_w=G'^2t_z(G)+\frac c{12}\{G,w\},\qquad
 \langle t_z\rangle=\frac c{24}.
 \label{eq:bop_stress_maps}
\end{equation}
For the disk metric~\eqref{Diskme}, the Weyl anomaly subtracts $c/24$. The physical interior component is $t_w- c/24$, whereas the exterior
component is $t_v$. Consider the two boundary insertions
\begin{equation}
 B(\varphi):=T(\varphi)-\frac c{24},\qquad
 S(\varphi):=\frac c{12}
 \left[\left\{\tan\frac{\theta(\varphi)}2,\varphi\right\}-\frac12\right].
 \label{eq:bop_ingredients}
\end{equation}
Promoting $\phi_r$ in~\eqref{eq:sec2_Sch} to a local source defines $S$
as a curvature response, even at zero background coupling. Note that in the pure Schwarzian theory we also have an operator like $S$, but in that case it is the conserved energy and its correlators are constant, so it does not give rise to a dimension two operator. However, we will see that in our context, we can use it to construct the extra operator.

The Virasoro-primary condition concerns conformal transformations of
the {\it{bath coordinate}} $\varphi$. Locally, consider
$v=p(\tilde v)$, with $p$ real at the interface, so that
$\varphi=p(\tilde\varphi)$. We keep the interior coordinate fixed;
the welding function in the new bath coordinate is therefore
$\tilde\theta(\tilde\varphi)=\theta(p(\tilde\varphi))$.
The stress transformation and the Schwarzian chain rule give
\begin{equation}
 \begin{aligned}
 \widetilde B(\tilde\varphi)
   &=p'^2 B(p)+\frac c{12}\{p,\tilde\varphi\}
       +\frac c{24}(p'^2-1),\\
 \widetilde S(\tilde\varphi)
   &=p'^2 S(p)+\frac c{12}\{p,\tilde\varphi\}
       +\frac c{24}(p'^2-1).
 \end{aligned}
 \label{eq:bop_anomaly}
\end{equation}
Here $p'=dp/d\tilde\varphi$, and the last terms arise from the cylinder
background subtractions in~\eqref{eq:bop_ingredients}.
These identical inhomogeneous terms cancel in the combination
\begin{equation}
 \hat T:=B-S
 =t_v-\frac c{12}\left\{\tan\frac{\theta}2,\varphi\right\}
 =\theta'^2 (t_w - { c \over 24 } )
 \label{eq:bop_definition}
\end{equation}
It has precisely the weight-two transformation law under a change of
$\varphi$:
\begin{equation}
 \widetilde{\hat T}(\tilde\varphi)
 =p'(\tilde\varphi)^2\hat T\bigl(p(\tilde\varphi)\bigr).
 \label{eq:bop_primary_transform}
\end{equation}
We therefore expect $\hat T$ to be the new primary.
Note that $\hat T $ is basically the stress tensor in the interior, $t_w$, but it is important that it is dressed to the exterior and viewed as an exterior operator, depending on the bath coordinate.  In fact, we wanted to find a second primary from the exterior point of view.

\paragraph{The Gaussian integral.}
The solution of conformal welding and the expression for the action are, from
\eqref{eq:sec2_disk_welding_sol} and~\eqref{eq:sec2_inf_I2},
\begin{equation}
 f(v)=\sum_{n\ge2}\varepsilon_n e^{inv},\qquad
 g(w)=-\sum_{n\ge2}\varepsilon_{-n}e^{-inw},\qquad
 I_2=\sum_{n\ge2}{ c \over 12} n (n^2-1)|\varepsilon_n|^2.
 \label{eq:bop_linear_welding}
\end{equation}
The expressions of the stress tensor expectation values at first order in the small fluctuations are
\begin{equation}
 \begin{aligned}
 \delta t_v&=\frac c{12}(f'''+f')
       =-i\sum_{n\ge2}a_n\varepsilon_n e^{inv}, ~~~~~~~a_n= { c\over 12 } n (n^2-1)\\
 \delta t_w&=\frac c{12}(g'''+g')
       =-i\sum_{n\ge2}a_n\varepsilon_{-n}e^{-inw}.
 \end{aligned}
 \label{eq:bop_means}
\end{equation}
  The first derivatives come from the cylinder Casimir term.

In addition, we also need to remember that the two point function of the stress tensor for the trivial reparametrization is $\langle t_v(\varphi) t_v(0) \rangle_{\rm con} = \frac{c}{2 ( 2 \sin{\varphi \over 2 } )^4}$. This then needs to be reparametrized, which leads to a contribution at quadratic order.

\paragraph{The matrix of inner products.}

Using \nref{eq:bop_means} and \nref{eq:bop_linear_welding} we can find the matrix of correlators
\begin{equation}
 \begin{gathered}
 \left\langle
 \begin{pmatrix}B(\varphi)\\S(\varphi)\end{pmatrix}
 \begin{pmatrix}B(0)&S(0)\end{pmatrix}\right\rangle_{\rm conn}
 =\frac c2  { 1 \over [ 2 \sin { \varphi \over 2 } ]^4 }\begin{pmatrix}1&1\\1&2\end{pmatrix}+O(1),\\
 \left\langle
 \begin{pmatrix}B(\varphi)\\\hat T(\varphi)\end{pmatrix}
 \begin{pmatrix}B(0)&\hat T(0)\end{pmatrix}\right\rangle_{\rm conn}
 =\frac c2  { 1 \over [ 2 \sin { \varphi \over 2 } ]^4 }\begin{pmatrix}1&0\\0&1\end{pmatrix}+O(1).
 \end{gathered}
 \label{eq:bop_orthogonality}
\end{equation}
We see that $\hat T$ has
nonzero norm and is orthogonal to the stress-tensor.

As a related check, for $\operatorname{Im}(v-w)>0$ the mixed bulk correlator of the stress tensor in the interior and the stress tensor in the bath vanishes at this order in the $1/c$ expansion.

\begin{equation}
 \langle t_w(w)t_v(v)\rangle_{\rm conn}
 =\frac{c}{32\sin^4[(v-w)/2]}
  -\frac c{12}\sum_{n\ge2}n (n^2-1)e^{in(v-w)}+O(1)=O(1),
 \label{eq:bop_cross_cancellation}
\end{equation}
using $\sum_{n\ge2}n (n^2-1)z^n=6z^2/(1-z)^4$ for $|z|<1$.
At this order, there is no connected correlation between the stress tensor inside and the one outside, which is related to the fact that the matter reflects at the boundary from both the inside and outside points of view.

Once we have the operator $\hat T $, and we notice that at large $c$ the interactions are small, then we can construct multiparticle states from it. The easy way to count them is as follows. First consider $\hat T $ and all its descendants. They give contributions like
\be
 q_o^2 + q_o^3 + q_o^4 + \cdots
\ee
from $\partial^n \hat T$. Now consider all possible ``multi-particle'' or ``multi-trace'' operators. We can count them viewing each of the above states as a harmonic oscillator, so that we get
\be
 { 1 \over \prod_{n=2}^\infty (1 - q^n ) } -1
 \ee
 This counts all the states that contain a non-zero number of particles.  If we want to count only the $\SL(2,\R)$ primaries, we simply multiply by a factor of $(1-q)$. This is precisely what we had in the bracket in \nref{NewPrimPar}.  These states can be viewed as fluctuations of the length of the wormhole, see appendix \ref{app:wormhole_length}.

 We should also note that the operator $\hat T $ has dimension 2 at leading order in the $1/c$ expansion. Using AI we computed the first correction to its dimension and found that it is $h= 2 - 24/c + O(1/c^2) $.  In particular the transformation law \nref{eq:bop_primary_transform}, which is only valid for $c=\infty$, should be changed appropriately for finite $c$.

 A curious feature of this operator is that it is not produced by the boundary OPE of a bulk operator, in the disk approximation. In other words, if we take  a bulk operator to the boundary, we do not produce it. The two point function of a non-identity bulk Virasoro primary operator $V$ and the boundary operator $\hat T$ vanishes exactly because it involves the one point function of the bulk operator $V$ on an infinite reference cylinder (or sphere), which vanishes by conformal symmetry. In other words, it vanishes before we do the integral over reparametrizations. If the bulk operator is the stress tensor, then we only produce identity descendants but not $\hat T$ which is a separate primary.

\section{The dimension-two operator and the wormhole length}
\label{app:wormhole_length}
We explain the relation between $\hat T$ and fluctuations of the wormhole
length at leading order in the large-$c$ expansion. We set the AdS radius
to one and initially keep $\gamma=\pi/\kappa=b/(2\pi)$ fixed.

Let $D(\varphi_L,\varphi_R)$ be the renormalized geodesic length between
the two regulated boundaries. Mapping the gravitational strip to a
half-plane by $z=e^{\gamma w}$ gives
\begin{equation}
 D=\log\frac{4\cosh^2\!\left[\frac{\gamma}{2}
 (\theta_L(\varphi_L)-\theta_R(\varphi_R))\right]}
 {\gamma^2\theta_L'(\varphi_L)\theta_R'(\varphi_R)} ,
 \label{eq:wh_ren_length}
\end{equation}
where we subtracted the cutoff divergence $-\log(\epsilon_L\epsilon_R)$.
Differentiating with respect to one boundary point $\varphi_a$ while holding the opposite endpoint fixed yields
\begin{equation}
 \partial_a^2D+\frac12(\partial_aD)^2
 =-\{e^{\gamma\theta_a},\varphi_a\},\qquad a=L,R.
 \label{eq:wh_length_schwarzian}
\end{equation}
The cylinder counterpart of the disk operator
\eqref{eq:bop_definition} is therefore
\begin{equation}
 \hat T_{\gamma,a}:=t_{v,a}
 -\frac c{12}\{e^{\gamma\theta_a},\varphi_a\}
 =t_{v,a}+\frac c{12}
 \left[\partial_a^2D+\frac12(\partial_aD)^2\right].
 \label{eq:wh_length_operator}
\end{equation}
Thus $\hat T_\gamma$ is related to endpoint derivatives of the length,
rather than to the length itself. At the round saddle,
$t_{v,a}=-c\gamma^2/24$ and $\partial_a^2D=\gamma^2/2$ for $\varphi_L = \varphi_R$, so the
background value of $\hat T_{\gamma,a}$ vanishes.

For $\theta_a=\varphi+\varepsilon_a$, the equal-time length fluctuation
is $\delta D=-\varepsilon_L'-\varepsilon_R'$. Writing
$B_a=\delta t_{v,a}$, we obtain
\begin{equation}
 \hat T_{\gamma,L}+\hat T_{\gamma,R}-B_L-B_R
 =\frac c{12}(\partial_\varphi^2-\gamma^2)\delta D .
 \label{eq:wh_linear_length}
\end{equation}
The exterior-stress terms must be retained unless we project onto
the sector orthogonal to the ordinary boundary stress tensor.

We can also identify the lowest length oscillation directly.
At the wormhole saddle $\kappa=\ell$, the quadratic action
\eqref{eq:tor_quadratic}, with
$\varepsilon_\pm=(\varepsilon_L\pm\varepsilon_R)/\sqrt2$, becomes
\begin{equation}
 I_2=\frac c{12}\sum_{n>0}n(n^2+\gamma^2)
 \left[\tanh\frac{n\ell}{2}|\varepsilon_{+,n}|^2
 +\coth\frac{n\ell}{2}|\varepsilon_{-,n}|^2\right].
\end{equation}
Since $\delta D_n=-\sqrt2\,in\varepsilon_{+,n}$, its Gaussian
two-point function is
\begin{equation}
 \langle\delta D_n\delta D_{-n}\rangle
 =\frac{24}{c}\frac{n\coth(n\ell/2)}{n^2+\gamma^2}.
\end{equation}
After continuation to Lorentzian frequency, the pole structure is
proportional to
$\omega\cot(\omega\ell/2)/(\gamma^2-\omega^2)$.
The apparent pole at $\omega=\gamma$ cancels, and the first
nonzero-frequency pole is $\omega=2\gamma$.
Since the open-channel energy spacing is $\gamma=\pi/\ell$,
this is precisely the energy of a dimension-two excitation.

These relations support interpreting the extra dimension-two sector
as contributing to the lowest dynamical fluctuation of the wormhole
length. The equal-time symmetric length probes the even-frequency
tower. The odd-frequency modes are instead probed by the relative
endpoint derivative $[(\partial_{\varphi_L}-\partial_{\varphi_R})D]_{
\varphi_L=\varphi_R}$, which couples to $\varepsilon_-$.
Its first surviving pole is at $\omega=3\gamma$, followed by
$5\gamma,7\gamma,\ldots$, consistent with the complementary
descendant levels of the dimension-two operator. This fluctuation is
distinct from varying the constant modulus $b$, which is an off-shell mode that does not correspond to any Lorentzian state.  The geometric identities above
are classical; a finite-$c$ operator statement would additionally
require a prescription for renormalizing the length and its
composite derivatives.

\bibliographystyle{unsrturl}

\bibliography{ref}

\end{document}